\documentclass[leqno, 12pt]{article}
\usepackage{kokoszka}
\usepackage{graphicx}
\usepackage{amsfonts}
\usepackage{amsmath}
\usepackage{amssymb}
\usepackage{enumitem}
\usepackage{color}
\usepackage{epsfig,subfigure}
\usepackage{epstopdf}
\usepackage{booktabs}
\usepackage{float, lscape,multirow}
\usepackage{tikz-cd}

\usepackage{algorithm}
\usepackage{algpseudocode}
\usepackage{capt-of}

\renewcommand{\baselinestretch}{1.02}
\newcommand\mbZ{{\mathbb Z}}

\begin{document}

\title{Detection of  a  structural break in intraday volatility pattern}

\author{
Piotr Kokoszka\footnote{Corresponding author:  Department of Statistics,
Colorado State University, Fort Collins, CO 80521-1877, USA \
Email: Piotr.Kokoszka@colostate.edu}\\
{\small Colorado State University}
\and
Tim Kutta \\
{\small Colorado State University}
\and
Neda Mohammadi \\
{\small Colorado State University}
\and
Haonan Wang \\
{\small Colorado State University}
\and
Shixuan Wang \\
{\small University of Reading}
}

\date{}
\maketitle

\begin{abstract}
We develop theory leading to
testing procedures for the presence of a change point in
the intraday volatility pattern. The new theory
is  developed in the framework of Functional Data Analysis.
It is  based on a model akin to the stochastic volatility model for
scalar point-to-point returns. In our context, we study
intraday curves, one curve per trading day. After postulating a
suitable model for such functional data, we present three tests
focusing, respectively, on changes in the shape, the magnitude
and arbitrary changes in the sequences of the curves of interest.
We justify the respective procedures by showing that they
have asymptotically correct size and by deriving consistency
rates for all tests. These rates  involve the sample size (the number of
trading days) and the grid size (the number of observations per day).
We also derive the corresponding change point estimators and  their
consistency rates. All procedures are additionally validated by
a simulation study and an application to US stocks.

\end{abstract}
MSC 2020 subject classifications:
62R10, 
62G10, 
62M10. 
\\
Keywords and phrases: Change point, Functional data,
Intraday volatility, It\^{o} integral.

\section{Introduction}\label{sec:intro}
\begin{figure}
	\centering
\includegraphics[width=0.9\linewidth]{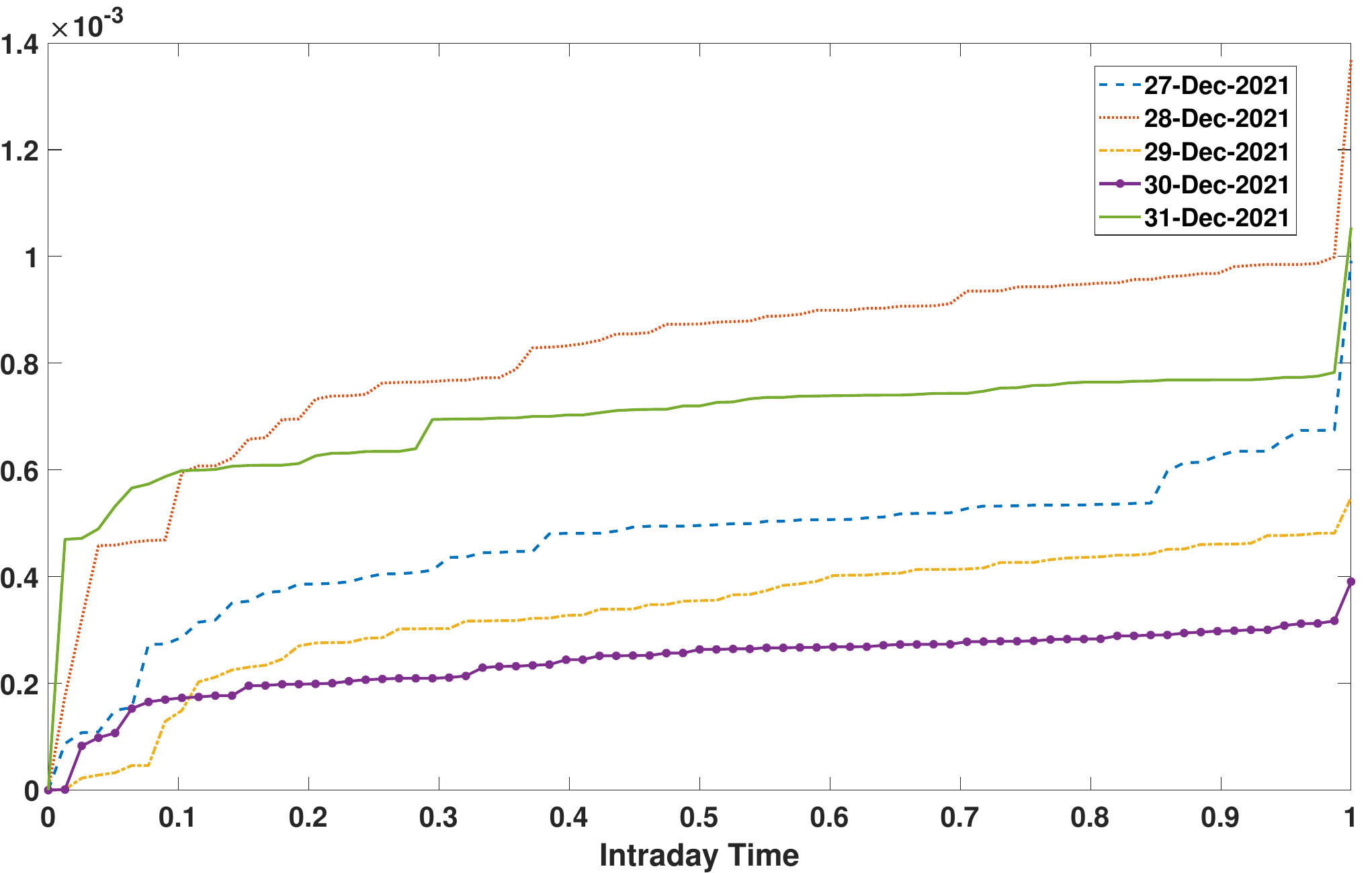}
	\caption{Five consecutive realized volatility
curves computed with $K = 78$ for Tesla Inc. from Dec 27 to Dec 31, 2021.
	\label{f:rv5}}
\end{figure}

Consider a sample  of intraday price
curves $\lbr P_i(t), t \in[0,1] \rbr$, $1 \le i \le N$,
where $i$ indexes the trading day
and $t$ is intraday time normalized to the standard unit
interval.  For each $i$, we study the  limits, as $\Dg \to 0$,
of cumulative intraday realized volatility curves
\begin{equation} \label{e:rvDgt}
{\rm RV}_i (\Dg)(t) =  \sum_{1 \le k\le Kt}
\left | \log[P_i( k \Dg)] - \log [P_i( (k-1)\Dg)]  \right |^2,
\ \ \ t\in [0,1].
\end{equation}
To illustrate, five
consecutive curves ${\rm RV}_i (\Dg)(\cdot)$ are shown in Figure \ref{f:rv5}.

Under suitable assumptions, see Section~\ref{sec:prelim},
for each $t\in [0,1]$,
\[
{\rm RV}_i (\Dg)(t)\convP \int_0^t \nu_i^2(u) du,
\ \ \ {\rm as } \ \Dg \to 0.
\]
The object of our study are basically the curves $\nu_i$, but a more precise
problem statement is needed. We represent the curves
$\nu_i$ as
\begin{equation} \label{e:nu-i}
\nu_i(u) = h_i \sg_i(u), \ \ \ u\in [0,1], \  i \in \mbZ,
\end{equation}
where the $h_i >  0$ describe the evolution
of the curves  from day  to day (between-day volatility),
while the functions $\sg_i$ quantify
the residual volatility after the between days volatility has been
accounted for by the sequence $\lbr h_i\rbr$. The identifiability
of the components in decomposition \refeq{nu-i} is addressed in
Lemma \ref{l:id}.  {\em We develop a
statistical framework to test if  the functions $\sg_i$ change
over a time period  of many  days.}
The volatilities $h_i$ typically exhibit persistent magnitude clusters.
We propose methodology,  and supporting theory,
that allows us to test the constancy of the functions $\sg_i$ in index
$i$. We emphasize that we do not test if each $\sg_i$ is a constant
function because this
is well-known not to be true. The functions $\sg_i$ are approximately
the derivatives of functions like those in Figure \ref{f:rv5},  divided by $h_i$.
The  shape of the functions $\sg_i$ can change, and such a change
will be undetectable if one focuses only on realized daily volatilities
${\rm RV}_i (\Dg)$ because they are dominated by the $h_i$.
In the following, we refer to the
$\sigma_i(\cdot)$ as volatility or diffusion
functions, while keeping in mind that the stochastic volatility functions
are  the products $h_i \sigma_i(\cdot)$.

Model \refeq{nu-i} is a useful approximation that allows us to construct
an effective test and justify it. Using a related  perspective,
\cite{christensenHP:2018} assume that
$\nu_i(u) = h_i(u) \sg_i(u)$ and propose a test of $H_0: h_i(u) = h_i$,
$\forall \ u\in [0,1]$. They apply it to over thousand days $i$ and to
30 stocks, about 30,000 tests in total. There are overall more rejections
than acceptances of their $H_0$,  the results depend on the
implementation of the test. The rejections dominate if the test
is implemented with a good estimate of the function $\sg^2$, under the
assumption that it does not change,  i.e.
\begin{equation} \label{e:H0-i}
\sg_i^2(\cdot)  = \sg^2(\cdot),  \ \ \ \forall \ i=1,2, \ldots, N.
\end{equation}
If condition \refeq{H0-i} does not hold, any estimate of the function
$\sg^2(\cdot)$ may be meaningless.
Our test is thus complementary to that of \cite{christensenHP:2018}
and impacts its implementation;
we test  assumption \refeq{H0-i} with an unknown $\sg^2(\cdot)$.
We evaluate and apply the test to five minute intraday returns, so we
do not need to be concerned with microstructure noise and price jumps.
Our objective is to provide and effective principled  way of
testing condition \refeq{H0-i} within a broad framework of
statistical change point detection and functional data analysis. Our  tests will be useful in any context that requires the verification that
an intraday volatility pattern remains constant over the period of many trading days.

Research on the detection and estimation of a change
point in various statistical models is over 70 years old and
forms a well-established subfield of statistics. Its importance stems
from the fact that most statistical models assume a single data generating
mechanism, so if this mechanism changes over the observational
period, their application will be meaningless.
There are several monographs
and thousands of research papers; the paper of \cite{horvath:rice:2014}
can serve as a concise and modern introduction to the general
framework of this paper.  It includes change point detection for 
different data structures, among them functional data. 
A broad review of  inference and estimation techniques is given 
in \cite{casini:perron:2019}, who also consider a range of 
different applications. The related literature on sequential (online) 
tests for structural changes is  reviewed in \cite{aue:kirch:2023}.
In the following two paragraphs,
we briefly  review the most closely related research.
An important point to note is that in the framework
of functional data analysis (FDA), the tests are $\sqrt{N}$-consistent
($N$ is the sample size),
whereas in our framework of replications of a diffusion process,
we can obtain $\sqrt{NK}$-consistency, where $K$ is the number
of sampling points for each replication. Moreover, these rates
depend on the type of local alternative, shape change vs. size change,
and arise because we explicitly use modeling though It{\^o} integrals.
A key starting point  is  new concentration results,
Proposition \ref{prop:unif:4} and its corollaries, that can be used in
other contexts that require information about the rates, in terms
of grid size, at which population volatility functions can be approximated
by realized volatility curves. As far as
we know,  the framework we study has  been considered neither in
change point research nor in intraday volatility research. We  hope
that the theoretical advances we make together with a comprehensive
application will motivate further research at the nexus of FDA and SDEs.

\noindent{\bf Related change point research in the framework of FDA} \
In FDA, the observations are
random elements in some function space, such the space
$L^2$ equipped with the canonical
$L^2$-norm, or  $\mathcal{C}$ (continuous functions)
equipped with the supremum norm.
The space $L^2$ has played a particularly important role in FDA
since, under weak assumptions,  it is a separable Hilbert space.
\cite{berkes:gabrys:horvath:kokoszka:2009}
proposed a test for a change in the mean function in an $L^2$
setting; extensions
were considered by  \cite{aston:kirch:2012},
\cite{horvath:kokoszka:rice:2014},
\cite{gromenko:kokoszka:reimherr:2017}
and \cite{aue:rice:sonmez:2018}, among others.
Structural breaks of time series in the space $\mathcal{C}([0,1])$
were studied in \cite{dette:kokot:aue}, see also
\cite{rackauskas:suquet:2006} for a more abstract context.
Changes in the covariance of a functional time series
were considered in \cite{stoehr:aston:kirch:2021}
and in the cross-covariance operator in \cite{rice:shum:2019}.
\cite{dette:kokot:2022} considered inference for the covariance kernel
of continuous data. More recently, \cite{horvath:rice:zhao:2022}
employed a weighted CUSUM statistic for the detection and localization
of changes in the covariance operator. Tests for the stability of
eigenvalues and principal components were presented in
\cite{aue:rice:sonmez:2020} and \cite{dette:kutta}.
A self-normalization approach, see \cite{shao:zhang:2010}
and \cite{shao:2015},
was used in the context of change point detection
in functional time series by \cite{dette:kokot:volgushev:2020}.
The cited works comprise only a small fraction of the relevant literature.

\noindent\textbf{Change point detection in It{\^o} semimartingales} \
We are not aware of any research that consideres a change point
problem in a sample of trajectories,  each of which is an
It{\^o} semimartingale. Research to date has focused on the detection
of a change point in a single continuous--time realization.
\cite{sahalia:jacod:2009} modified the commonly used CUSUM
approach to detect jumps in It{\^o} semimartingales.
In particular, in order to detect jumps in asset returns,
they proposed a  test statistic based on multiplicative difference
of realized truncated $p$--th variation.
\cite{bibinger2017nonparametric} used a similar approach to detect
structural changes in the volatility of It{\^o} semimartingales.
They  addressed detection of jumps, the so called local changes,
as well as changes in the roughness of  sample path,
the so called global changes. Other related papers are
\cite{Hoffmann:dette:2018}, \cite{hoffmann:dette:2019},
and \cite{bibinger:madensoy:2019} who provide further
references to the general area of  change point detection in  It{\^o}
semimartingales.

The remainder of the paper is organized as follows.
In Section \ref{sec:prelim},
we collect the minimum required background on It{\^o} integrals
and the FDA. Section \ref{s:mp} is dedicated to the precise
formulation of the problem outlined above. Testing and estimation
approaches are developed in Section \ref{sec:theor}. Their finite
sample properties are examined in Section \ref{sec:fsp}. Section \ref{s:app}
contains an application to sequences of intraday returns on US stocks.
Supplementary material contains proof of all results stated in Section
\ref{sec:theor},  details of practical implementation of
all  procedures, and some additional information.

\section{Mathematical preliminaries}\label{sec:prelim}
We begin by providing some mathematical background,
beginning with stochastic differential equations.
We first  recall the definition of the quadratic variation of
a stochastic process $\{X(t): t \in [0,1]\}$.
 We assume throughout that 
$0 = t_0 ^{(K)}< t_1^{(K)}<\cdots<t_K^{(K)}= 1$
is a  grid on the interval $[0,1]$ with $\max_k[t_k - t_{k-1}] \to 0$,
as $K \to \infty$.  Then, 
\begin{equation} \label{e:ito1}
\sum_{k=1}^{K}\left |X(t_k^{(K)})-X(t_{k-1}^{(K)})\right |^2
\mathbb{I}\{t_k^{(K)}\leq t\}, \convP  [ X,X ]_t \quad t \in [0,1],
\end{equation}
 The limit  $[ X,X ]_t$ exists for any semimartingale,  and is called the 
the quadratic variation at time $t$, see e.g. 
Theorem 1.14 and relation (3.23) in \cite{ait-sahalia_high-frequency_2014}. 
In this work, we assume that  the process
$X$ is given by the It{\^o} integral
\[
X(t) := \int_0^t \nu(u) d W(u),
\]
where  $W$ is a standard Wiener process and
$\nu: [0,1]\to (0,\infty)$ is a continuous  function
(for a detailed discussion of the existence and properties of this process,
see Theorem 5.2.1 in \cite{Oksendal03}).
It is well-known  that for  It{\^o} integrals
the quadratic variation is given by
\begin{equation}\label{e:ito2}
[ X,X] _t  := \int_0^t \nu^2(u) d u , \quad t \in [0,1], 
\end{equation}
 see e.g. equation (2.1) in  \cite{karatzas_brownian_1991}. 
Moreover, in this case,  \eqref{e:ito1} can be strengthened to
\begin{align}\label{e:emp->}
 \underset{0 \leq t \leq 1}{\sup}
 \left \vert \sum_{k}\left \vert X(t_k^{(K)})-X(t_{k-1}^{(K)}) \right \vert^2
 \mathbb{I}\{t_k^{(K)}\leq t\} - \int_0^t \nu^2 (u) du \right
 \vert \convP 0, \ \ K \to \infty. 
\end{align}
see again Theorem 1.14 and relation (3.23) in 
\cite{ait-sahalia_high-frequency_2014}. 
Next, we present a consequence of the Dambis--Dubins--Schwarz 
 theorem which states that
\textit{any} continuous local martingale can be expressed as a time
change of a Brownian motion,
see e.g. Section 5.3.2 in \cite{le2016brownian}
for a general statement. In our setting,
\begin{align}\label{e:tim:chng}
    \left\{ \int_0^t \nu(u) dW(u), \ t\in [0,1] \right\}
 \overset{d}{=}
 \left\{ W\left( \int_0^t \nu^2(u) du\right), \ t\in [0,1] \right\},
\end{align}
where the equality in distribution is in the space $C([0,1])$ of
continuous functions, equipped with the topology of uniform convergence.
Identity \eqref{e:tim:chng} entails
\begin{align}\label{e:tim:chng2}
 \mathbb{E}\left[\int_0^t \nu(u) dW(u)\right]^2
 = \int_0^t \nu^2(u) du,
\end{align}
which is a special case of the It{\^o} isometry for a deterministic,
square integrable integrand  $\nu(\cdot)$.

In identities  \eqref{e:tim:chng} and \eqref{e:tim:chng2},
the  It{\^o} process is treated as a random function in $C([0,1])$.
However, in the context of FDA, it is often useful to embed the smaller
space of continuous functions in the larger Hilbert space of square
integrable functions. More precisely, we define $ L^2([0,1])$ as
the space of measurable functions $f:[0,1]\to \mathbb{R}$
that satisfy $\int_0^1 f^2(x) dx <\infty$.
Equipped with the inner product
\[
\langle f, g \rangle := \int_0^1 f(t) g(t) dt, \qquad f,g \in L^2([0,1]),
\]
and the induced norm $\|\cdot \|_{L^2}$, $L^2([0,1])$
becomes a separable Hilbert space,
where we identify functions equal almost everywhere.
A random function $X$ in $L^2([0,1])$ is then a measurable map $X: (\Omega, \mathcal{A},\mathbb{P}) \to L^2([0,1])$, where $(\Omega, \mathcal{A},\mathbb{P})$ is a probability space. If the first moment of $X$ exists in the sense that $\mathbb{E}\|X\|<\infty$, we can define the expectation $\mu \in L^2([0,1])$ of $X$, which is characterized by the identity
\[
\mathbb{E} \langle X, f \rangle = \langle \mu, f \rangle \qquad \forall f \in L^2([0,1]).
\]
Similarly, if the second moment of $X$ exists,  $\mathbb{E}\|X\|^2<\infty$, we can define the  covariance operator $C_X: L^2([0,1]) \to L^2([0,1])$ of $X$ by the identity
\[
 \langle C_X[f], g \rangle := \mathbb{E}\big[\langle X-\mu, f \rangle\langle X-\mu, g \rangle\big]  \qquad \forall f,g \in L^2([0,1]).
\]
It is known  that $C_X$ is a self--adjoint, positive semi-definite,
Hilbert-Schmidt operator and as such it can be identified with
a square integrable kernel function $c_X: [0,1]^2 \to \mathbb{R}$ via
\[
C_X[f](x):= \int_0^1 c_X(x,y) f(y) dy \qquad \forall f \in L^2([0,1]).
\]
Chapters 10 and 11 of \cite{KRbook} provide a concise
introduction to the $L^2$ framework of FDA. For  a comprehensive
treatment see \cite{hsing:eubank:2015}.

\section{Statistical model and problem formulation}\label{s:mp}
Suppressing the superscript $(K)$,
consider the grid $0 = t_0< t_1<\cdots<t_K= 1$
introduced in Section \ref{sec:prelim},  and the cumulative returns
\[
R_i(t_k) = \log[P_i(t_k)] - \log[P_i(0)].
\]
The realized volatility curves \refeq{rvDgt}  for this  grid
can be written as
\[
{\rm RV}_i(t) = \sum_{k=1}^K
\left | R_i(t_k) - R_i(t_{k-1}) \right |^2 \mathbb{I}\{t_k \leq t\}.
\]
Setting $h_i = \exp(g_i)$ in \refeq{nu-i}, we postulate the model
\begin{equation} \label{e:R1}
R_i(t) = \exp(g_i)
\int_0^t \sg_i(u) d W_i(u), \quad t \in [0,1], \quad i \in \mathbb{Z}.
\end{equation}
The $W_i(\cdot)$ are independent standard Wiener processes.
The sequence $ g_i$
is a  centered real--valued, weakly stationary time series
independent of $(W_i)_{i \in \mathbb{Z}}$.  The following lemma
shows that for each $i$ the volatility function $\sg_i(\cdot)$ depends only 
on $R_i(\cdot)$, so $g_i$ and $\sg_i(\cdot)$ are identifiable.

\begin{lemma} \label{l:id}
Suppose $g$ satisfies $Eg =0$ and is independent of 
the Wiener process $W(\cdot)$. Setting
\[
R(t) = e^g \int_0^t \sg(u) d W(u), \quad t \in [0,1],
\] 
for a continuous function  $\sg(\cdot)$, we have 
\[
\int_0^t \sigma^2(u) du = \exp \lbr \mathbb{E} \log [R,R] _t \rbr. 
\] 
\end{lemma}
\noindent{\sc Proof.}
By \refeq{ito2}, 
$[ R,R] _t  = \exp(2g) \int_0^t \sg^2(u) d u$, so 
\[
\log [ R,R] _t = 2 g  + \log  \int_0^t \sigma^2(u) du.
\] 
Since $\mathbb{E}(g) = 0$, 
$ \mathbb{E} \log [ R,R] _t =  \log \int_0^t \sigma^2(u) du$, 
and the claim follows. 
\hfill \QED

To test for changes in the volatility functions $\sg_i(\cdot)$, we
propose the following change point model.
Let $\theta \in (0,1)$ be a parameter that locates a potential change
in the discrete time index $i$ and let
$\sigma_{(1)},\sigma_{(2)}: [0,1] \to (0,\infty)$
denote two continuous volatility functions.
We postulate that
\begin{align}\label{e:change}
\begin{cases}
\sigma_i(\cdot) = \sigma_{(1)}(\cdot), \qquad \textnormal{for}\,\, i \le \lfloor N \theta \rfloor,\\
\sigma_i(\cdot) = \sigma_{(2)}(\cdot), \qquad \textnormal{for}\,\, i > \lfloor N \theta \rfloor.
\end{cases}
\end{align}
A change occurs if $\sigma_{(1)}(\cdot) \neq \sigma_{(2)}(\cdot)$.
The testing problem is thus
\begin{align}\label{e:hyp}
H_0: \sigma_{(1)}(\cdot) = \sigma_{(2)}(\cdot), \qquad \textnormal{vs.}\qquad   H_A: \sigma_{(1)}(\cdot) \neq \sigma_{(2)}(\cdot).
\end{align}
A cornerstone of our statistical analysis is the translation of changes
in volatility to  changes of certain features in the  quadratic variation process.
This allows us to take advantage of regularities of
the  quadratic variation process compared to the
process $R_i(\cdot)$.
Indeed,  \eqref{e:ito2} directly entails
\begin{align}\label{e:QV}
 Q_i(t) :=   [ R_i,R_i]_t = \exp(2g_i) \int_0^t \sigma_i^2(u) du
 , \quad t \in [0,1], \quad
 i=1,2,\ldots, N.
\end{align}
This, together with the stationarity of the time series
$(g_i)_{i \in \mathbb{Z}}$, indicates that a change in
volatility corresponds to a change in the distribution of the
quadratic variation process $Q_i(\cdot)$, over index $i$.

Suppose we observe  a sample  $R_1,\ldots,R_N$.
Reflecting practically available data,
we assume that the functions $R_i$ are observed at $K+1$
equidistant points in  $[0,1]$.
This means  that inference will be based on the matrix of observations
\begin{equation} \label{e:data}
\{R_i({k/K}): i=1,\ldots,N, \,\, k=0,\ldots,K\}.
\end{equation}
In  our theory, we assume
that the number of grid points, $K+1$, as well as the number of
curves, $N$, tend to infinity.
In view of approximations \eqref{e:ito1} and \eqref{e:emp->},
we consider  the realized quadratic variation  processes
\begin{align}
\label{e:Qhat}
     \widehat{Q}_i(t)=&   \sum_{k=1}^{K}\vert R_i({k/K})-R_i({(k-1)/K}) \vert^2 \mathbb{I}\{{k/K}\leq t\} \\
    =& \exp(2g_i) \sum_{k=1}^K
    \left\vert  \int_{{(k-1)/K}}^{{k/K}}\sigma(u) dW_i(u)\right\vert^2
    \mathbb{I}\{{k/K}\leq t\}, \qquad t \in [0,1]. \nonumber
\end{align}
as  estimators of the $Q_i(\cdot)$ in  \refeq{QV}.
Observe that $\widehat{Q}_i(t)$ is equal to the  realized
volatility function \refeq{rvDgt}, with the second line
reflecting the assumed model.

Assuming the
$g_i$ have exponential moments, a test could be based on
the approximation
\begin{align} \label{e:approx}
    \mathbb{E}[\widehat Q_i(t)]\approx \mathbb{E}[Q_i(t)]
    = \mathbb{E}[\exp(2g_i)] \cdot \int_0^t \sigma_i^2(u) du
\end{align}
which  indicates that volatility function changes  translate
to mean changes in  the realized quadratic variation process.
Detecting changes in the mean of a functional time series is a
well-studied problem, as discussed in Section \ref{sec:intro}.
However, a test based on \refeq{approx},
requires the existence of exponential
moments of the $g_i$ and is  not robust against distributional changes
in $g_i$, which might be mistaken for changes in the volatility functions
$\sigma_i(\cdot)$. Moreover,  CUSUM based
FDA tests are  $\sqrt{N}$-consistent,
but we  demonstrate that
against large classes of common alternatives a much stronger
consistency rate of $\sqrt{NK}$ is attainable by some  tests we propose.
For these reasons, we present in this paper a different,
more effective method to test the hypotheses \eqref{e:hyp}.
Our approach does not require exponential moments of $g_i$,
is more stable against distributional changes (or spurious changes)
in the $g_i$,
and benefits from $\sqrt{NK}$-consistency under typical alternatives.
As a first step, we express the hypothesis $H_0$  in
\eqref{e:hyp} in terms of  two null hypotheses, $H_0^{(1)}$
and $H_0^{(2)}$, that are together equivalent to $H_0$:
\begin{align}
&H_0^{(1)}: \ \label{e:shape}
\frac{\int_0^t \sigma_{(1)}^2(u) du}{\int_0^1 \sigma_{(1)}^2(u) du}
= \frac{\int_0^t \sigma_{(2)}^2(u) du}{\int_0^1 \sigma_{(2)}^2(u) du},
\quad \forall \ t \in [0,1], \\
&H_0^{(2)}: \ \int_0^1 \sigma_{(1)}^2(u) du
=\int_0^1 \sigma_{(2)}^2(u) du. \label{e:total}
\end{align}
Heuristically, $H_0^{(1)}$ states that the volatility function
does not change its shape, while  $H_0^{(2)}$ states
that the total volatility stays the same.
 In Section \ref{sec:theor}, we formulate statistical tests of
 $ H_0^{(1)}$ and $ H_0^{(2)}$ separately, and then
combine them to test the $H_0$ in \refeq{hyp}.

\section{Change point tests}\label{sec:theor}
We begin by stating assumptions for our subsequent analysis.

\begin{assumption} \label{a:model}
\noindent{}
\medskip
\begin{enumerate}
\item \label{itm:sig} The volatility function $\sigma_{(1)}, \sigma_{(2)}: [0,1] \to (0, \infty)$ are continuous.
\item \label{itm:BM} The standard Wiener processes $W_i$, $i \in \mathbb{Z}$,  are independent.\item \label{itm:ind} The two sequences $(g_i)_{i \in \mathbb{Z}}$ and $(W_i)_{i \in \mathbb{Z}}$ are independent of each other.
\item \label{itm:g:inv} The time series $(g_i)_{i \in \mathbb{Z}}$ is centered, weakly stationary and satisfies a weak invariance principle of the form
        \[
        \Big\{\frac{1}{\sqrt{N}}\sum_{i=1}^{\lfloor Nx \rfloor}g_i: x \in [0,1] \Big \} \convd \{\lambda W(x): x \in [0,1]\},
        \]
        where $W$ is a standard Wiener process and $\lambda^2>0$ denotes the long-run variance.
\end{enumerate}
\end{assumption}
Assumption \ref{a:model}  is satisfied in many different scenarios.
It basically postulates a very general functional stochastic volatility model.
Condition \ref{itm:sig}  (before and after the change) is common  in
the literature on diffusion processes and intuitive in our setting.
Conditions \ref{itm:BM} and \ref{itm:ind} determine the dependence
structure along our functional time series,
which is moderated by the scaling factors
$e^{g_i}$. In \cite{kokoszka:mohammadi:wang:wang:2023},
the $g_i$  follow an AR($p$) model, but for our theory the precise
dependence structure is immaterial.  If the dependence is sufficiently
weak, the partial sum process on the left-hand side of condition 4
converges to a Wiener process. This is true under a multitude of
dependence conditions, see e.g. \cite{merlevede:peligrad:utev:2006},
so instead of choosing some of them, we postulate the general
condition 4.

\subsection{Inference for a shape change}\label{subsec:H^1_0}
 Recall the hypothesis $H_0^{(1)}$ in \refeq{shape}.
We begin with the simple observation that,
according to \eqref{e:QV}, the functions in \refeq{shape}
 can be represented by the \textit{standardized quadratic variation} as follows:
 \begin{equation} \label{e:F}
 F_i(t):=\frac{Q_i(t)}{Q_i(1)}
 =\frac{\int_0^t \sigma_{(j)}^2(u) du}{\int_0^1 \sigma_{(j)}^2(u) du},
 \,\quad \textnormal{with }
 j = \lbr
 \begin{array}{ll}
 1, \ {\rm for} \ & i=1,\ldots,\lfloor N\theta \rfloor, \\
 2, \ {\rm for} \ & i=\lfloor N\theta \rfloor+1,\ldots,N.
 \end{array}
 \right.
 \end{equation}
 This motivates using for statistical inference the empirical versions:
 \begin{align} \label{e:hatF}
 \qquad \widehat{F}_i(t):=&\frac{\widehat{Q}_i(t)}{\widehat{Q}_i(1)} :=
 \frac{\sum_{k=1}^K
\left | R_i(k/K) - R_i((k-1)/K) \right |^2 \mathbb{I}\{k/K \leq t\}}{\sum_{k=1}^K
\left | R_i(k/K) - R_i((k-1)/K) \right |^2 }\\[0.5ex]
   = &\frac{\sum_{k=1}^K  \left\vert  \int_{{(k-1)/K}}^{{k/K}}\sigma_i(u)dW_i(u)\right\vert^2 \mathbb{I}\{{k/K}\leq t\}}{\sum_{k=1}^K  \left\vert\int_{{(k-1)/K}}^{{k/K}}\sigma_i(u) dW_i(u)\right\vert^2}. \nonumber
 \end{align}

\begin{remark}\label{rem_1} We highlight two useful properties
of $\widehat F_i$:
\begin{itemize}
    \item[i)] $\widehat F_i$ is monotonically increasing with $\widehat{F}_i(0)=0$ and $\widehat{F}_i(1)=1$ and in particular it is a random cdf (and thus measurable). It can be interpreted as a random function, mapping into the space $L^2[0,1]$ of square integrable functions on the unit interval.
    \item[ii)] The functions $\widehat F_1, \ldots,\widehat F_N$ are
    independent,  and they do not involve  the $g_i$.
\end{itemize}
Property ii) implies that any test statistic based on the $\widehat F_i$s
will be unaffected by the structure of the $g_i$, or a potentially
misspecified model for them.
\end{remark}

\begin{lemma}\label{lem:FFhat}
Suppose that Conditions \ref{itm:sig} and \ref{itm:BM} of Assumption \ref{a:model}
hold. Then, each $\widehat F_i$ is a consistent estimator of the standardized quadratic variation $F_i$ (defined in \eqref{e:F})
and satisfies a functional central limit theorem of the form
\begin{align} \label{e:sqrtK}
\sqrt{K}\{\widehat{F}_i(\cdot)-F_i(\cdot)\} \convd Z(\cdot)
\end{align}
where $Z$ is a centered, Gaussian process in $L^2([0,1])$, with distribution depending on the volatility function $\sigma_i$.
\end{lemma}
The proof of Lemma \ref{lem:FFhat} follows by an application of
Theorem 5.3.6 in \cite{jacod2011discretization}  together
with the continuous mapping theorem.
In view of the convergence in \eqref{e:sqrtK}, we expect a test statistic based
on $\widehat F_1, \ldots,\widehat F_N$ to have variance of order $O(1/(NK))$,
or a corresponding test for $H_0^{(1)}$ to be $\sqrt{NK}$-consistent.

To test $H_0^{(1)}$, we  use  the  CUSUM statistic
\begin{align}\label{e:S1}
\widehat{S}^{(1)} := \frac{K}{N^2}\sum_{n=1}^N \int_0^1\Big(\sum_{i=1}^n \widehat{F}_i(u) - \frac{n}{N} \sum_{i=1}^N \widehat{F}_i(u) \Big)^2 du.
\end{align}
In the following result, the asymptotics ``$N, K \to \infty$" should
be understood in terms of a sequence $K=K_N$ of natural numbers that
diverges as $N \to \infty$. We do not impose any restrictions on the growth
rate of $K$ relative  to $N$,
making our method valid regardless of the interplay between $K$ and $N$.

\begin{theorem}\label{thm:S1}
Suppose that Conditions  \ref{itm:sig} and \ref{itm:BM} of Assumption hold
and that $N, K \to \infty$. Then, under $H_0^{(1)}$, the weak convergence
\begin{align}\label{e:def:S1}
\widehat{S}^{(1)} \convd S^{(1)}:= \sum_{\ell =1}^\infty \lambda_\ell \int_0^1 \mathbb{B}_\ell(u)^2 du
\end{align}
holds, where $(\mathbb{B}_\ell)_{\ell \in \mathbb{N}}$ is a sequence of i.i.d. Brownian bridges and $(\lambda_\ell)_{\ell \in \mathbb{N}}$ the collection of eigenvalues of the asymptotic covariance kernel
\begin{align} \label{e:covS1}
c_F(u,v):= & \lim_{K \to \infty} K \cdot \mathbb{E}\left[\{\widehat{F}_1(u)-\mathbb{E}[\widehat{F}_1(u)]\}\{\widehat{F}_1(v)-\mathbb{E}[\widehat{F}_1(v)]\}\right].
\end{align}
Moreover,  if $H_0^{(1)}$ is violated, $\widehat{S}^{(1)} \overset{\mathbb{P}}{\rightarrow} \infty$.
\end{theorem}
Denoting for any $\alpha \in (0,1)$ the upper
$\alpha$-quantile of $S^{(1)}$ by $q_{1-\alpha}^{(1)}$,
the decision
\[
{\rm reject \ if} \ \  \widehat{S}^{(1)}>q_{1-\alpha}^{(1)}
\]
yields a consistent asymptotic level $\alpha$ test of $H_0^{(1)}$.
While in practice, we do not know the distribution of
$S^{(1)}$, it is uniquely determined by the eigenvalues of $c_F$,
which can be estimated by off-the-shelf methods
(we provide details in  Appendix \ref{sec:imple}).
 An explicit formula for the kernel $c_F(u, v)$ is given
in Theorem \ref{thm:cov:F},  but it is not needed to estimate
the $\la_\ell$ because  \refeq{covS1}
is a limit of covariance kernels, and  many FDA packages
output their eigenvalues. The distribution of the
integral in  \refeq{def:S1} is easy to simulate, and it is fairly
well-known how to compute
the approximate quantiles of the right-hand side of  \refeq{def:S1}. 

In the next theorem, we demonstrate the consistency of our test procedure
against local alternatives.

\begin{theorem}\label{t:loc1}
Suppose Conditions  \ref{itm:sig} and \ref{itm:BM} of Assumption hold.
Let $\tilde \sigma:[0,1]\to (0,\infty)$ be a continuous function such that
$\tilde \sigma(\cdot)/\sigma_{(1)}(\cdot)$ is not constant and let
$(a_N)_{N \in \mathbb{N}}$ be  a bounded sequence of positive numbers.
Then, defining $\sigma_{(2)} := \sigma_{(1)}+ a_N  \tilde \sigma$ and imposing the growth conditions $ a_N \sqrt{NK} \to \infty$ and $a_N K \to \infty$, it follows that
    \[
    \lim_{N,K \to \infty}\mathbb{P}(\widehat S^{(1)}>c)=1
    \quad \forall \ c \geq 0.
    \]
\end{theorem}

Finally, we define the change point estimator
\begin{align} \label{e:theta1}
\hat \theta^{(1)} := \frac{1}{N}\underset{n\in \{1,\ldots,N\}}{\textnormal{argmax}}\int_0^1\Big(\sum_{i=1}^n \widehat{F}_i(u) - \frac{n}{N} \sum_{i=1}^N \widehat{F}_i(u) \Big)^2 du.
\end{align}
If the hypothesis of no change in the shape of volatility is violated, i.e.
\begin{align}\label{e:change1}
\begin{cases}
F_i(\cdot) = F_{(1)}(\cdot), \qquad \textnormal{for}\,\, i \le \lfloor N \theta \rfloor\\
F_i(\cdot) = F_{(2)}(\cdot), \qquad \textnormal{for}\,\, i > \lfloor N \theta \rfloor
\end{cases} , \quad F_{(1)}(\cdot) \neq F_{(2)}(\cdot)
\end{align}
for some  $\theta \in (0,1)$, we can show that the  estimator
$\hat \theta^{(1)}$  in \eqref{e:theta1} is consistent
under local alternatives.

\begin{theorem}\label{t:th1}  Under the assumptions of Theorem \ref{t:loc1},
\[
\hat \theta^{(1)}-\theta =\mathcal{O}_P\bigg(\max\Big\{\frac{a_N^{-2}}{NK}, \frac{1}{N} \Big\}\bigg),
\]
where $\theta\in (0,1) $ is the rescaled time of the change
in \eqref{e:change1}.
\end{theorem}
{The rate in Theorem \ref{t:th1} can be explained as follows:
For a change of size $a_N$ (potentially tending to $0$) it is well-known in change point estimation that an optimal approximation rate is given by
\[
\hat \theta^{(1)}-\theta = \mathcal{O}_P\Big(\frac{a_N^{-2}}{sample \, size}\Big).
\]
In our case the "sample size" is $NK$, yielding a rate of
$\mathcal{O}_P(a_N^{-2}/(NK))$. However, since the number
of curves in discrete time is $N$, it is also clear that a convergence
rate cannot be faster than $\mathcal{O}_P(1/N)$. This
limitation is simply due to the discretization of time in $N$ steps.
As a consequence, the best attainable rate is as specified in Theorem \ref{t:th1}.
Notice that in the special case of $a_N=O(1/\sqrt{K})$,
we obtain the optimal rate $O_P(1/N)$ on the right side, the same rate as
for fully observed functions, see e.g.
\cite{aue:gabrys:horvath:kokoszka:2009}.

\subsection{Inference for a change in total volatility} \label{subsec:H^2_0}
Recall the hypothesis $ H_0^{(2)}$ in \refeq{total}.
The integrals in  \refeq{total} are closely related to the total quadratic
variation $Q_i(1)$ and taking its logarithm, we obtain
\begin{align}\label{e:log:det}
\log(Q_i(1)) = 2 g_i + \log\Big(\int_0^1 \sigma_{(j)}^2(u) du\Big),
\,\quad \textnormal{for} \quad\begin{cases}
     i=1,\ldots,\lfloor N\theta \rfloor,\,\,\,\,\,\,\,\,\,\,\quad j=1,\\
     i=\lfloor N\theta \rfloor+1,\ldots,N,\,\,\, j=2.
 \end{cases}
\end{align}
Since the $g_i$ are centered, any change in total volatility translates
into a mean change of the real--valued time series $\{\log (Q_i(1))\}$.
An empirical analogue of \eqref{e:log:det} is
\begin{align}
\label{e:logQ(1)}
     \log(\widehat{Q}_i(1)) = & 2 g_i + w_i,
\end{align}
where
     \begin{align} \label{e:def:wi}
   w_i := & \log \Big(\sum_{k=1}^K \Big\vert  \int_{{(k-1)/K}}^{{k/K}}\sigma_i(u) dW_i(u)\Big\vert^2 \Big).
\end{align}
 This decomposition shows that the observations $\log(\widehat{Q}_1(1)),\ldots,\log(\widehat{Q}_N(1))$  form (for any fixed $K$) a dependent time series that is stationary before and after a potential change. For the purpose of statistical inference, we use the following CUSUM statistics:
\begin{align}\label{e:S2}
\widehat{S}^{(2)}:= \frac{1}{N^2} \sum_{n=1}^N \Big(\sum_{i=1}^n \log(\widehat{Q}_i(1)) - \frac{n}{N}\sum_{i=1}^N \log(\widehat{Q}_i(1))\Big)^2.
\end{align}

\begin{theorem}\label{thm:S2}
If  Assumption \ref{a:model}  holds and $N,K \to \infty$,
then, under $H_0^{(2)}$,
\begin{align}\label{e:def:S2}
\widehat{S}^{(2)} \convd S^{(2)}:=
(4\lambda) \cdot \int_0^1 \mathbb{B}(u)^2 du,
\end{align}
where $\mathbb{B}$ is a standard Brownian bridge and the long-run variance
$\lambda$ is defined as
\begin{align}\label{e:long:cov:g}
\lambda:= \sum_{i \in \mathbb{Z}} \mathbb{C}ov(g_0, g_i).
\end{align}
Moreover, if $H_0^{(2)}$ is violated, $\widehat{S}^{(2)}
\overset{\mathbb{P}}{\rightarrow} \infty$.
\end{theorem}
The fact that the long-run variance  $\lambda$ only depends on the $g_i$s
is not an accident. As we will see in the next section, \
the statistic $\widehat{S}^{(2)}$ is asymptotically only dependent on
the $g_i$s and independent of the $W_i$s.
This implies that $\widehat{S}^{(1)}$
(which does not depend on the $g_i$s) and $\widehat{S}^{(2)}$
are asymptotically independent.

Theorem \ref{thm:S2} implies that if  we denote by $q_{1-\alpha}^{(2)}$
the upper $\alpha$-quantile of the distribution $S^{(2)}$, then
the decision to
\[
{\rm reject \ if} \ \ \widehat{S}^{(2)}>q_{1-\alpha}^{(2)}
\]
yields a  consistent  asymptotic level $\alpha$ test of
the hypothesis $H_0^{(2)}$. Again, $q_{1-\alpha}^{(2)}$
cannot be directly computed, but it can be approximated,
if a consistent estimator for the long-run variance is given
(see Appendix \ref{sec:imple}).

We now show consistency of the test against local alternatives.
\begin{theorem} \label{t:loc2}
Suppose Assumption \ref{a:model} holds
and  $(a_N)_{N \in \mathbb{N}}$ is a bounded sequence of positive numbers.
Then, defining $\sigma_{(2)}:= (1+a_N)\sigma_{(1)}$ and imposing
the growth conditions $a_N\sqrt{N} \to \infty$ and $a_NK \to \infty$,
it follows that
    \[
    \lim_{N,K \to \infty}\mathbb{P}(\widehat S^{(2)}>c)=1,\quad \forall \ c \geq 0.
    \]
\end{theorem}
Finally, with the change point estimator
\begin{align}\label{e:theta2}
\hat \theta^{(2)} :=  \frac{1}{N}  \underset{n\in \{1,\ldots,N\}}{\textnormal{argmax}}\Big(\sum_{i=1}^n \log(\widehat{Q}_i(1)) - \frac{n}{N}\sum_{i=1}^N \log(\widehat{Q}_i(1))\Big)^2,
\end{align}
we can localize a change in total volatility.
If the hypothesis of no change in the total volatility is violated, i.e.
\begin{align}\label{e:change2}
\begin{cases}
\log({Q}_i(1)) = \log({Q}_{(1)}(1)), \qquad \textnormal{for}\,\, i \le \lfloor N \theta \rfloor\\
\log({Q}_i(1)) = \log({Q}_{(2)}(1)), \qquad \textnormal{for}\,\, i > \lfloor N \theta \rfloor
\end{cases} , \quad \log({Q}_{(1)}(1)) \neq \log({Q}_{(2)}(1)),
\end{align}
for some  $\theta \in (0,1)$, we obtain the following result.

\begin{theorem} \label{t:th2} Under the assumptions of Theorem \ref{t:loc2},
$
\hat \theta^{(2)}-\theta=O_P(a_N^{-2}/N)$.
\end{theorem}
Taking $a_N=1$, we obtain the optimal rate.

\subsection{Inference for an arbitrary change} \label{subsec:H_0}
In the previous subsections, we have developed test statistics
$\widehat{S}^{(1)}, \widehat{S}^{(2)}$ for the null hypotheses
$H_0^{(1)}, H_0^{(2)}$ in \refeq{shape} and \refeq{total},
respectively.  We now want to combine these two tests  to yield
a test for the global null hypothesis $H_0$  in \eqref{e:hyp}.
As a first step, we show  that as $N, K \to \infty$, t
he two test statistics \eqref{e:S1} and \eqref{e:S2} become independent
of each other.

\begin{proposition}\label{p:ind}
If Assumption \ref{a:model}  and  $H_0$ in \refeq{hyp} hold,
then, as $N, K \to \infty$,
    \[
    \lp \widehat{S}^{(1)}, \widehat{S}^{(2)}\rp  \convd
     \lp S^{(1)}, S^{(2)}\rp,
    \]
where $S^{(1)}, S^{(2)}$ are independent
and defined in \eqref{e:def:S1}, \eqref{e:def:S2}, respectively.
\end{proposition}
In order to combine the results from both test statistics, we employ their asymptotic $p$-values. To be precise, if $\Lambda^{(j)}$ is the (continuous) cumulative distribution function of $S^{(j)}$, we define the $p$-values
\begin{align} \label{e:pval}
    p^{(j)} = 1 - \Lambda^{(j)}\big(\widehat{S}^{(j)}\big), \quad j=1,2.
\end{align}
In practice, the $\Lambda^{(j)}$ are not known,
but can uniformly approximated,  yielding empirical $p$-values.
We discuss this issue in Appendix \ref{sec:imple}.
To combine our test statistics, we recall that under $H_0$,
both $p$-values $p^{(1)}, p^{(2)}$ are asymptotically uniformly
distributed on $[0,1]$ and according to Proposition \ref{p:ind}
asymptotically independent. Hence, using Fisher's method, see e.g.
\cite{quinn:keough:2002},  we can combine them to
\begin{align} \label{e:hatSp}
\widehat{S} := -2\{\log(p^{(1)})+ \log(p^{(2)})\},
\end{align}
which then converges under $H_0$ to a chi-squared distribution with
four degrees of freedom. Denoting the  upper $\alpha$-quantile of this
distribution by $q_{1-\alpha}$, gives us the test decision
\begin{equation} \label{e:tc}
{\rm reject  \ if} \ \ \widehat{S}>q_{1-\alpha}.
\end{equation}
We make this result precise in the following proposition.
\begin{proposition} \label{p:comb} Under the assumptions of
Proposition \ref{p:ind},
\[
\widehat{S} \convd \chi_4^2,
\]
where $\chi_4^2$ is a  chi-squared distribution with four degrees of freedom. \
If $H_0$ is violated, $\widehat{S} \convP  \infty$.
\end{proposition}

It is a simple consequence of  Theorems \ref{t:loc1} and \ref{t:loc2}
that the test  \refeq{tc} is consistent against local alternatives
of shape changes and  changes in total volatility, with the rates discussed
in those theorems.

\begin{remark}
 In view of Proposition \ref{p:ind} there are different ways of combining
the test statistics $\widehat{S}^{(1)}, \widehat{S}^{(2)}$ for a joint test,
apart from our choice of $\widehat{S}$. Such combinations correspond
to different rejection regions in
$\mathbb{R}_{\ge 0}^2$ for $(\widehat{S}^{(1)}, \widehat{S}^{(2)})$.
Generically we can define for a function
$f: \mathbb{R}_{\ge 0}^2 \to \mathbb{R}_{\ge 0}$
the combined statistic
$\widehat{S}_f=f(\widehat{S}^{(1)}, \widehat{S}^{(2)})$.
A simple choice might be a sum $f_{\rm sum}(x,y)=x+y$,
which has linear, downward sloping contour lines and thus triangular
rejection regions. Our choice
\[
f_{\rm Fisher}(x,y) = -2\Big\{\log\big(1 - \Lambda^{(1)}(x)\big)+
\log\big(1 - \Lambda^{(2)}(y)\big)\Big\}
\]
has astroid shaped contour lines (like a $p$-norm with $0<p<1$).
Accordingly rejection regions are shaped like ellipsoids. The precise
shape of the contour lines depends on the asymptotic distributions
$\Lambda^{(1)}, \Lambda^{(2)}$.  The function $f$ implies how
evidence against the null hypothesis is interpreted in different scenarios.
Roughly speaking, $f_{\rm sum}$ is indifferent between large
$x$, large $y$ or large $x+y$. This means that more evidence against
the null might come just as well from one statistic, or the other,
or their sum. In contrast $f_{\rm Fisher}$ is largest if both $x$ and $y$
are large, treating evidence against the null hypothesis as strongest,
when it comes from both statistics and weaker if it only comes from one.

\end{remark}

Finally, we discuss the problem of change point localization.
For this purpose, we introduce the pooled change point estimator
\begin{align}\label{e:theta}
 \hat\theta:= \frac{ p^{(1)}}{ p^{(1)}+ p^{(2)}} \hat \theta^{(2)}+\frac{ p^{(2)}}{ p^{(1)}+ p^{(2)}} \hat \theta^{(1)}.
\end{align}
Intuitively, $\hat \theta$ combines information from both estimators
$\hat\theta^{(1)},\hat\theta^{(2)}$, putting priority on the one
where the change is more pronounced (smaller $p$-value).
Our proof rests on a careful investigation of the tail behavior of the distributions $\Lambda^{(1)}, \Lambda^{(2)}$, see Theorem \ref{thm:thp} in the Appendix. The tail behavior of these distributions determines the relative size of the $p$-values $p^{(1)}, p^{(2)}$ in the above weights.

\begin{proposition} \label{prop:th:fin}
Suppose Assumption \ref{a:model} holds, $K \to \infty$, $K/N \to 0$, and the continuous  function $\tilde \sigma:[0,1]\to (0,\infty)$ is such that $\tilde \sigma(\cdot)/\sigma_{(1)}(\cdot)$ is not constant.

\noindent (i) If only $H_0^{(2)}$ is violated with $\sigma_{(2)}=(1+1/\sqrt{K}) \sigma_{(1)}$, then
\[
|\hat \theta - \theta| = \mathcal{O}_P\left( \frac{K}{N}\right).
\]

\noindent (ii)
If, in addition,  $H^{(1)}$ is violated with $\sigma_{(2)}=  (1+1/\sqrt{K})\sigma_{(1)}+\tilde \sigma/\sqrt{K}$,  then
\[
|\hat \theta - \theta| = \mathcal{O}_P\left( \frac{1}{N}\right).
\]
\end{proposition}

\section{Finite sample properties} \label{sec:fsp}

\subsection{Empirical size} \label{ss:es}
We generate data under the null hypothesis according
to the  Functional Stochastic
Volatility Model of  \cite{kokoszka:mohammadi:wang:wang:2023}:
\begin{align*}
	R_i(t) &= \exp(g_i) \int_{0}^{t} \sigma (u) d W_i (u), \qquad t \in [0,1], \qquad i = 1,...,N,\\
	g_i &= \varphi g_{i-1} + \varepsilon_i, \qquad \varepsilon_i \sim i.i.d. \ \mathcal{N}(0, \sigma_\varepsilon^2).
\end{align*}

There are a number of settings to be carefully chosen:
\begin{itemize}\setlength\itemsep{0em}
\item Following \cite{kokoszka:mohammadi:wang:wang:2023},
   we set $\varphi = 0.55$ and $\sigma_\varepsilon^2 = 0.25$
   in order to reflect real-world data.
\item We have four settings of $\sigma (\cdot)$
	\begin{itemize}
		\item {\bf Flat}: $\sigma (u) = 0.2$. This is a simple case that we have the same intraday volatility throughout the day.
		\item {\bf Slope}: $\sigma (u) = 0.1 + 0.2u$. The is a case that the intraday volatility is increasing in a linear manner.
		\item {\bf Sine}: $\sigma (u) = 0.1\sin(2\pi u) + 0.2$. This the case we have higher volatility in the morning, but lower volatility in the afternoon.
		\item {\bf U-shape}: $\sigma (u) = (u-0.5)^2 + 0.1145299$.
This choice is the most relevant one because it reflects the stylized fact that volatility is typically higher at the beginning and the end of a trading day.
	\end{itemize}
The coefficients in $\sigma (\cdot)$ are set to ensure that
the above four $\sigma (\cdot)$ have a similar scale.
	\item The continuous time $t$ in $[0,1]$ is discretized as $\left[t_0, t_1, ..., t_K \right] $, where $t_k=  k \Delta$ and $k=1,...,K$. This is the same for all random curves.
	\item The number of intraday observations is $K=26, 39, 78$,
which corresponds to 15-min, 10-min 5-min sampling intervals in our data analysis respectively.
Their corresponding stepsizes are $\Delta=1/26, 1/39, 1/78$.
\item The sample size is $N = 100, 200, 500$.
\end{itemize}

Details on the computation of $\int_{0}^{t} \sigma (u) d W (u)$
and $\int_{0}^{1} \mathbb{B}^2 (u) d u$, both use special approaches,
are presented in Section \ref{s:dc}, which also contains step-by-step
formulas for the computation of the three test statistics.
The long-run variance of the $\log \widehat{Q}_i (1)$ was computed
using
the Bartlett  kernel with bandwidth selected
by the procedure of \cite{newey1994automatic} with  prewhitening.

Table \ref{tab:size_FDE} provides the empirical sizes
of the three tests under four different shapes of $\sigma(\cdot)$.
We see that the test performs very well, even for fairly small
sample sizes $N$ and low resolution $K$.

\begin{table}[htbp]
	\centering
	\caption{Empirical size}
	\resizebox{0.9\columnwidth}{!}{\begin{tabular}{llrrrrrrrrrrrr}
		\toprule
		\toprule
		&       &       & \multicolumn{3}{c}{\textbf{\underline{Shape of Volatility}}} &       & \multicolumn{3}{c}{\textbf{\underline{Total Volatility}}} &       & \multicolumn{3}{c}{\textbf{\underline{Global}}} \\
		\textbf{Flat} &       &       & \multicolumn{1}{c}{10\%} & \multicolumn{1}{c}{5\%} & \multicolumn{1}{c}{1\%} &       & \multicolumn{1}{c}{10\%} & \multicolumn{1}{c}{5\%} & \multicolumn{1}{c}{1\%} &       & \multicolumn{1}{c}{10\%} & \multicolumn{1}{c}{5\%} & \multicolumn{1}{c}{1\%} \\
		\midrule
		$N=100$ & $K=26$ &       & 11.4\% & 5.9\% & 1.4\% &       & 10.5\% & 5.1\% & 0.5\% &       & 11.4\% & 5.4\% & 1.2\% \\
		& $K=39$ &       & 10.9\% & 5.8\% & 1.2\% &       & 10.5\% & 4.7\% & 0.4\% &       & 10.6\% & 5.0\% & 1.0\% \\
		& $K=78$ &       & 11.9\% & 5.9\% & 0.9\% &       & 9.8\% & 4.3\% & 0.4\% &       & 11.0\% & 5.0\% & 0.9\% \\
		\midrule
		$N=200$ & $K=26$ &       & 10.6\% & 5.3\% & 1.1\% &       & 10.5\% & 5.2\% & 0.9\% &       & 10.8\% & 5.6\% & 1.2\% \\
		& $K=39$ &       & 10.8\% & 5.5\% & 1.3\% &       & 10.6\% & 5.1\% & 0.7\% &       & 11.3\% & 5.3\% & 1.0\% \\
		& $K=78$ &       & 11.8\% & 5.4\% & 1.2\% &       & 10.1\% & 5.0\% & 0.9\% &       & 11.2\% & 5.4\% & 0.9\% \\
		\midrule
		$N=500$ & $K=26$ &       & 11.0\% & 5.6\% & 1.0\% &       & 10.4\% & 5.6\% & 1.1\% &       & 11.0\% & 5.9\% & 1.1\% \\
		& $K=39$ &       & 11.2\% & 5.5\% & 1.2\% &       & 11.0\% & 5.1\% & 0.8\% &       & 11.5\% & 5.6\% & 1.0\% \\
		& $K=78$ &       & 11.1\% & 5.6\% & 1.3\% &       & 10.2\% & 4.7\% & 0.9\% &       & 10.7\% & 5.5\% & 1.2\% \\
		\midrule
		\midrule
		\textbf{Slope} &       &       &       &       &       &       &       &       &       &       &       &       &  \\
		\midrule
		$N=100$ & $K=26$ &       & 11.3\% & 5.9\% & 1.4\% &       & 10.4\% & 4.6\% & 0.4\% &       & 11.1\% & 5.3\% & 0.9\% \\
		& $K=39$ &       & 10.9\% & 5.3\% & 1.2\% &       & 10.0\% & 4.3\% & 0.4\% &       & 10.7\% & 5.2\% & 1.0\% \\
		& $K=78$ &       & 10.5\% & 5.6\% & 1.4\% &       & 9.5\% & 4.0\% & 0.6\% &       & 10.5\% & 5.2\% & 0.8\% \\
		\midrule
		$N=200$ & $K=26$ &       & 10.3\% & 5.4\% & 1.1\% &       & 10.9\% & 5.4\% & 1.1\% &       & 11.3\% & 5.9\% & 1.2\% \\
		& $K=39$ &       & 11.7\% & 5.9\% & 1.3\% &       & 9.8\% & 4.8\% & 0.7\% &       & 11.1\% & 5.4\% & 1.0\% \\
		& $K=78$ &       & 11.2\% & 6.1\% & 1.2\% &       & 10.7\% & 5.1\% & 0.5\% &       & 11.2\% & 5.8\% & 1.1\% \\
		\midrule
		$N=500$ & $K=26$ &       & 10.7\% & 5.1\% & 0.9\% &       & 10.7\% & 5.5\% & 1.2\% &       & 11.1\% & 5.6\% & 0.8\% \\
		& $K=39$ &       & 11.3\% & 5.6\% & 1.2\% &       & 10.3\% & 5.0\% & 1.0\% &       & 11.3\% & 5.6\% & 1.1\% \\
		& $K=78$ &       & 11.2\% & 5.6\% & 1.1\% &       & 10.2\% & 5.1\% & 0.9\% &       & 10.9\% & 5.4\% & 1.1\% \\
		\midrule
		\midrule
		\textbf{Sine} &       &       &       &       &       &       &       &       &       &       &       &       &  \\
		\midrule
		$N=100$ & $K=26$ &       & 11.3\% & 5.6\% & 1.1\% &       & 11.0\% & 5.5\% & 0.7\% &       & 11.6\% & 5.3\% & 0.8\% \\
		& $K=39$ &       & 11.3\% & 5.9\% & 1.2\% &       & 10.8\% & 5.1\% & 0.7\% &       & 11.6\% & 5.7\% & 1.0\% \\
		& $K=78$ &       & 11.8\% & 6.4\% & 1.5\% &       & 9.6\% & 4.6\% & 0.6\% &       & 11.5\% & 5.3\% & 0.9\% \\
		\midrule
		$N=200$ & $K=26$ &       & 10.7\% & 5.1\% & 1.3\% &       & 11.6\% & 6.2\% & 1.2\% &       & 11.5\% & 6.2\% & 1.4\% \\
		& $K=39$ &       & 11.1\% & 5.8\% & 1.3\% &       & 10.3\% & 4.8\% & 0.8\% &       & 11.5\% & 5.5\% & 1.1\% \\
		& $K=78$ &       & 10.8\% & 5.5\% & 1.0\% &       & 10.3\% & 5.0\% & 0.8\% &       & 11.3\% & 5.4\% & 0.8\% \\
		\midrule
		$N=500$ & $K=26$ &       & 10.6\% & 5.2\% & 1.0\% &       & 11.0\% & 5.5\% & 1.0\% &       & 11.2\% & 5.8\% & 0.9\% \\
		& $K=39$ &       & 11.4\% & 5.3\% & 1.1\% &       & 9.8\% & 4.9\% & 0.8\% &       & 11.1\% & 5.7\% & 1.2\% \\
		& $K=78$ &       & 11.1\% & 5.8\% & 1.0\% &       & 10.0\% & 5.0\% & 0.6\% &       & 11.1\% & 4.9\% & 0.8\% \\
		\midrule
		\midrule
		\textbf{U-Shape} &       &       &       &       &       &       &       &       &       &       &       &       &  \\
		\midrule
		$N=100$ & $K=26$ &       & 11.0\% & 5.7\% & 1.2\% &       & 10.6\% & 4.7\% & 0.5\% &       & 11.0\% & 5.8\% & 1.0\% \\
		& $K=39$ &       & 11.3\% & 6.1\% & 1.3\% &       & 11.0\% & 4.9\% & 0.4\% &       & 11.4\% & 5.2\% & 1.1\% \\
		& $K=78$ &       & 10.9\% & 5.8\% & 1.4\% &       & 10.1\% & 4.3\% & 0.4\% &       & 10.7\% & 5.4\% & 1.1\% \\
		\midrule
		$N=200$ & $K=26$ &       & 11.0\% & 5.9\% & 1.3\% &       & 11.0\% & 5.4\% & 0.9\% &       & 11.4\% & 6.1\% & 1.2\% \\
		& $K=39$ &       & 11.2\% & 6.0\% & 1.1\% &       & 10.5\% & 5.4\% & 1.0\% &       & 11.2\% & 5.8\% & 1.1\% \\
		& $K=78$ &       & 11.4\% & 6.0\% & 1.4\% &       & 10.4\% & 5.0\% & 0.8\% &       & 11.2\% & 6.3\% & 1.2\% \\
		\midrule
		$N=500$ & $K=26$ &       & 11.0\% & 5.5\% & 1.1\% &       & 11.3\% & 5.2\% & 0.8\% &       & 11.0\% & 5.8\% & 1.0\% \\
		& $K=39$ &       & 9.7\% & 4.9\% & 0.8\% &       & 10.4\% & 5.4\% & 1.0\% &       & 10.3\% & 4.9\% & 1.0\% \\
		& $K=78$ &       & 10.6\% & 5.2\% & 1.1\% &       & 10.6\% & 5.2\% & 1.1\% &       & 11.0\% & 5.7\% & 1.0\% \\
		\bottomrule
		\bottomrule
	\end{tabular}}
	\label{tab:size_FDE}
\end{table}

 One advantage of using our tests is that it is robust against changes in $g_i$,
which should not be mistaken as changes in the volatility function
$\sigma_i(\cdot)$. To verify this property, we consider
\begin{equation*}
	g_i = \begin{cases}
		0.45 g_{i-1} + \varepsilon_i, \quad \epsilon_i \sim i.i.d. \ \mathcal{N}(0, \sigma_\varepsilon^2), \qquad  i=1,\ldots,\lfloor N/2\theta \rfloor,\\
		0.65 g_{i-1} + \varepsilon_i, \quad \epsilon_i \sim i.i.d. \ \mathcal{N}(0, \sigma_\varepsilon^2), \qquad  i=\lfloor N/2\theta \rfloor + 1, \ldots, N,
	\end{cases}
\end{equation*}
and all other settings are the same as before.
Table \ref{tab:size_change_gi} presents the empirical sizes
of the three tests under the U-Shaped  $\sigma_i(\cdot)$.
The other three shapes yield similar results. As can be seen,
the empirical sizes of our three tests are not affected by
the change in $g_i$ and match their theoretical levels reasonably well.

\begin{table}[htbp]
	\centering
	\caption{Empirical size under a change in $g_i$}
	\resizebox{0.9\columnwidth}{!}{\begin{tabular}{llrrrrrrrrrrrr}
			\toprule
			\toprule
			&       &       & \multicolumn{3}{c}{\textbf{\underline{Shape of Volatility}}} &       & \multicolumn{3}{c}{\textbf{\underline{Total Volatility}}} &       & \multicolumn{3}{c}{\textbf{\underline{Global}}} \\
			&       &       & \multicolumn{1}{c}{10\%} & \multicolumn{1}{c}{5\%} & \multicolumn{1}{c}{1\%} &       & \multicolumn{1}{c}{10\%} & \multicolumn{1}{c}{5\%} & \multicolumn{1}{c}{1\%} &       & \multicolumn{1}{c}{10\%} & \multicolumn{1}{c}{5\%} & \multicolumn{1}{c}{1\%} \\
			\midrule
			$N=100$ & $K=26$ &       & 12.0\% & 6.0\% & 1.4\% &       & 11.9\% & 5.8\% & 0.8\% &       & 12.2\% & 6.0\% & 1.3\% \\
			& $K=39$ &       & 11.4\% & 6.0\% & 1.2\% &       & 10.9\% & 5.3\% & 0.5\% &       & 11.9\% & 5.4\% & 1.0\% \\
			& $K=78$ &       & 10.7\% & 5.6\% & 1.1\% &       & 11.2\% & 4.8\% & 0.5\% &       & 11.0\% & 5.1\% & 0.8\% \\
			\midrule
			$N=200$ & $K=26$ &       & 11.6\% & 6.3\% & 1.2\% &       & 12.8\% & 6.7\% & 1.2\% &       & 13.3\% & 6.7\% & 1.2\% \\
			& $K=39$ &       & 10.4\% & 5.1\% & 1.0\% &       & 12.0\% & 6.0\% & 1.2\% &       & 11.6\% & 5.7\% & 1.3\% \\
			& $K=78$ &       & 10.6\% & 5.0\% & 0.9\% &       & 11.2\% & 5.3\% & 0.9\% &       & 10.7\% & 5.5\% & 1.1\% \\
			\midrule
			$N=500$ & $K=26$ &       & 10.8\% & 5.3\% & 1.0\% &       & 12.0\% & 6.2\% & 1.2\% &       & 11.5\% & 5.8\% & 1.3\% \\
			& $K=39$ &       & 11.6\% & 5.9\% & 1.2\% &       & 11.6\% & 6.5\% & 1.4\% &       & 12.9\% & 7.2\% & 1.3\% \\
			& $K=78$ &       & 11.3\% & 5.6\% & 1.3\% &       & 11.6\% & 6.3\% & 1.1\% &       & 12.3\% & 6.3\% & 1.3\% \\
			\bottomrule
			\bottomrule
	\end{tabular}}%
	\label{tab:size_change_gi}%
\end{table}

\subsection{Empirical power} \label{ss:ep}
We set the time of the change at $\theta = 0.25, 0.5, 0.75$
and consider $N=250$ and $N=500$.
All other settings are the same as  under the null.

There are unlimited possibilities for a change in $\sigma_i(\cdot)$.
To focus on the scenarios emphasized in this paper,
we consider the following three alternative hypotheses:
\begin{enumerate}\setlength\itemsep{0em}
	\item $H_{A,1}$: a shape change in volatility, but no change in total volatility,
	\item $H_{A,2}$: a change in total volatility, but no change in the
shape of volatility,
	\item $H_{A,3}$: a simultaneous change
in the shape of volatility and total volatility.
\end{enumerate}

Under $H_{A,1}$, we have a change in $\sigma_i(\cdot)$
from the flat shape to a sine shape.
We  have noticed that our test is very effective
in detecting changes in the shape of volatility,
and it  can easily get empirical power of 100\%.
That is why we deliberately choose a very small change
in the shape
in order to show the convergence of the empirical power
with respect to $N$ and $K$. Specifically, we set
$$
	\sigma_i (u) = \begin{cases}
		0.2, \qquad &\mbox{for } i = 1, ..., \lfloor N \theta \rfloor,\\
		0.02\sin(2\pi u) + \sqrt{199/5000}, \qquad &\mbox{for } i = \lfloor N \theta \rfloor + 1, ..., N.
	\end{cases}
$$
The constant in the sine function is to ensure that the total volatility
before and after the change is the same,
i.e. $\int_{0}^{1} 0.2^2 du = \int_{0}^{1} \left[ 0.02\sin(2\pi u)
+ \sqrt{199/5000}\right]^2  du = 0.04$.
Thus,  there is a change in the shape of volatility, but
no change  in the total volatility.

Under $H_{A,2}$, we introduce an upward parallel shift
of  the flat shape:
$$
\sigma_i (u) = \begin{cases}
	0.2, \qquad &\mbox{for } i = 1, ..., \lfloor N \theta \rfloor,\\
	0.4, \qquad &\mbox{for } i = \lfloor N \theta \rfloor + 1, ..., N.
\end{cases}
$$
Note that an upward parallel shift in the other three shapes
(slope, sine, U-shape) will cause a change in total volatility
as well as in the shape of volatility. This is because the other
three shapes are actually ``compressed'' due to a
higher total volatility.

Under $H_{A,3}$, we have a simultaneous change  in shape and total volatility:
$$
\sigma_i (u) = \begin{cases}
	0.2, \qquad &\mbox{for } i = 1, ..., \lfloor N \theta \rfloor,\\
	(u-0.5)^2 + 0.4, \qquad &\mbox{for } i = \lfloor N \theta \rfloor + 1, ..., N.
\end{cases}
$$
The shape of $\sigma_i(\cdot)$ is changed from flat to U-shape, and total volatility is changed from $\int_{0}^{1} 0.2^2 du = 0.04$ to $\int_{0}^{1} \left[  (u-0.5)^2 + 0.3\right] ^2 du = 0.1525$.

Table \ref{tab:power_FDE} reports  the empirical power. The conclusions
can be summarized as follows.
\begin{enumerate}\setlength\itemsep{0em}
	\item Under  $H_{A,1}$, the empirical power of the shape test
and the global test increases with  of $N$ and $K$,
in agreement with the  $\sqrt{NK}$-consistency we established theoretically.
The rejection rate of  total volatility test
is always around 5\%, as expected since  there is no change in total volatility in $H_{A,1}$.
\item Under $H_{A,2}$, the empirical power of the total volatility
test increases with the growth of $N$, not with $K$.
This is exactly what we expected because the test on total volatility
is $\sqrt{N}$-consistent. Additionally, the rejection rate
of testing the shape is typically around 5\%,
since there is no change in the shape in $H_{A,2}$.
\item Under $H_{A,3}$, the empirical power of the shape test and
the global test increases with the growth of $N$ and $K$,
and empirical power of the volatility test increases with the
growth of $N$, but not with $K$, again as predicted by our theory.
\end{enumerate}

\begin{table}[htbp]
	\centering
	\caption{Empirical power}
	\resizebox{\columnwidth}{!}{
	\begin{tabular}{llllllllllllll}
		\toprule
		\toprule
		&       &       & \multicolumn{3}{c}{$\theta=0.25$} &       & \multicolumn{3}{c}{$\theta=0.5$} &       & \multicolumn{3}{c}{$\theta=0.75$} \\
		\cmidrule{4-6}\cmidrule{8-10}\cmidrule{12-14}    $H_{A,1}$ &       &       & Shape & Total & Global &       & Shape & Total & Global &       & Shape & Total & Global \\
		\midrule
		$N=250$ & $K=26$ &       & 62.6\% & 5.2\% & 51.1\% &       & 86.8\% & 5.3\% & 78.2\% &       & 62.9\% & 5.6\% & 51.3\% \\
		& $K=39$ &       & 82.6\% & 5.0\% & 72.0\% &       & 96.9\% & 5.1\% & 93.3\% &       & 83.5\% & 5.0\% & 72.9\% \\
		& $K=78$ &       & 98.7\% & 5.1\% & 96.8\% &       & 100.0\% & 4.7\% & 99.9\% &       & 99.2\% & 4.5\% & 97.1\% \\
		\midrule
		$N=500$ & $K=26$ &       & 92.0\% & 5.5\% & 84.5\% &       & 99.2\% & 5.1\% & 97.6\% &       & 91.7\% & 5.5\% & 83.4\% \\
		& $K=39$ &       & 99.1\% & 4.6\% & 97.2\% &       & 100.0\% & 5.1\% & 99.9\% &       & 98.8\% & 5.2\% & 96.2\% \\
		& $K=78$ &       & 100.0\% & 5.5\% & 100.0\% &       & 100.0\% & 4.9\% & 100.0\% &       & 100.0\% & 5.1\% & 100.0\% \\
		\midrule
		$H_{A,2}$ &       &       & \multicolumn{3}{l}{}  &       & \multicolumn{3}{l}{}  &       & \multicolumn{3}{l}{} \\
		\midrule
		$N=250$ & $K=26$ &       & 5.8\% & 86.8\% & 74.5\% &       & 5.5\% & 98.6\% & 95.2\% &       & 5.7\% & 86.8\% & 73.2\% \\
		& $K=39$ &       & 5.9\% & 86.1\% & 72.7\% &       & 6.0\% & 98.5\% & 95.1\% &       & 6.0\% & 86.6\% & 73.0\% \\
		& $K=78$ &       & 5.9\% & 86.0\% & 72.3\% &       & 5.3\% & 98.3\% & 94.9\% &       & 5.5\% & 85.7\% & 71.6\% \\
		\midrule
		$N=500$ & $K=26$ &       & 5.6\% & 99.6\% & 97.9\% &       & 6.1\% & 100.0\% & 100.0\% &       & 5.2\% & 99.5\% & 98.1\% \\
		& $K=39$ &       & 5.4\% & 99.7\% & 98.4\% &       & 5.9\% & 100.0\% & 100.0\% &       & 5.3\% & 99.7\% & 98.3\% \\
		& $K=78$ &       & 5.7\% & 99.6\% & 98.3\% &       & 5.9\% & 100.0\% & 100.0\% &       & 5.3\% & 99.6\% & 98.0\% \\
		\midrule
		$H_{A,3}$ &       &       & \multicolumn{3}{l}{}  &       & \multicolumn{3}{l}{}  &       & \multicolumn{3}{l}{} \\
		\midrule
		$N=250$ & $K=26$ &       & 89.2\% & 97.4\% & 99.9\% &       & 99.9\% & 100.0\% & 100.0\% &       & 95.2\% & 97.4\% & 99.9\% \\
		& $K=39$ &       & 99.3\% & 97.3\% & 100.0\% &       & 100.0\% & 100.0\% & 100.0\% &       & 99.9\% & 97.7\% & 100.0\% \\
		& $K=78$ &       & 100.0\% & 97.1\% & 100.0\% &       & 100.0\% & 99.9\% & 100.0\% &       & 100.0\% & 97.3\% & 100.0\% \\
		\midrule
		$N=500$ & $K=26$ &       & 100.0\% & 100.0\% & 100.0\% &       & 100.0\% & 100.0\% & 100.0\% &       & 100.0\% & 100.0\% & 100.0\% \\
		& $K=39$ &       & 100.0\% & 100.0\% & 100.0\% &       & 100.0\% & 100.0\% & 100.0\% &       & 100.0\% & 100.0\% & 100.0\% \\
		& $K=78$ &       & 100.0\% & 100.0\% & 100.0\% &       & 100.0\% & 100.0\% & 100.0\% &       & 100.0\% & 100.0\% & 100.0\% \\
		\bottomrule
		\bottomrule
	\end{tabular}}%
	\label{tab:power_FDE}%
\end{table}

In Section \ref{ss:est}, we show that the change point estimators under the
three alternatives inherit the properties of the corresponding tests:
the performance of $\theta_1$ and $\theta$ improves with increasing
$N$ and $K$, $\theta_2$ improves with increasing $N$.

\section{Application to US stocks} \label{s:app}
We begin with  an individual stock as
a prototype analysis to showcase our developed tests.
Then, there are two ways to use the developed tests on a larger scale.
First, since there could be multiple changes, we use the binary segmentation
to explore all changes for one stock during a sample period.
Second, we apply our test procedure to a large number of stocks and present
the summary of first detected changes (without the binary segmentation).

For the purpose of demonstration, we  focus on Tesla Inc.
(Permno: 93436)  for our prototype analysis.
We consider  5-min intraday prices,
 the sample period is from Jun 29, 2010 (the IPO date) to Dec 31, 2021,
corresponding to $N=2891$ trading days. In each trading day $i$,
we have the opening price $P_i(t_0)$ and the subsequent 78
5-min intraday prices $P_i(t_k)$, $k=1,...,78$,
with the last trading price in every 5-min time interval.
Thus, the equidistant grid on the unit interval is
$t_k=k\Delta$, $k=0,1,...,K$, where $K=78$ and
the step size $\Delta=1/78$.

Based on the intraday price data,
we calculate the cumulative intraday return (CIDR) curves as
$$
R_i(t_k) = \log(P_i(t_k)) - \log(P_i(t_0)), \qquad k=1,...,K, \ i = 1,...,N.
$$
By definition, the CIDR curves always start from zero, i.e. $R_i(t_0)=0$,
and are scale invariant.
We also compute the cumulative intraday realized volatility (CIDRV) curves as
$$
	RV_i(t_k) = \sum_{k=1}^{K} \left|R_i(t_k) - R_i(t_{k-1}) \right|^2 \mathbb{I}\left\lbrace t_k <t\right\rbrace ,  \qquad k=1,...,K, \ i = 1,...,N.
$$
In order to visualize the important functional objects,
Figure \ref{fig:tesla} plots the intraday Price $P_i(t_k)$,
CIDRs $R_i(t_k)$, and CIDRVs $RV_i(t_k)$ in the upper,
middle, and lower panels, respectively.

\begin{figure}
	\centering
	\includegraphics[width=1\linewidth]{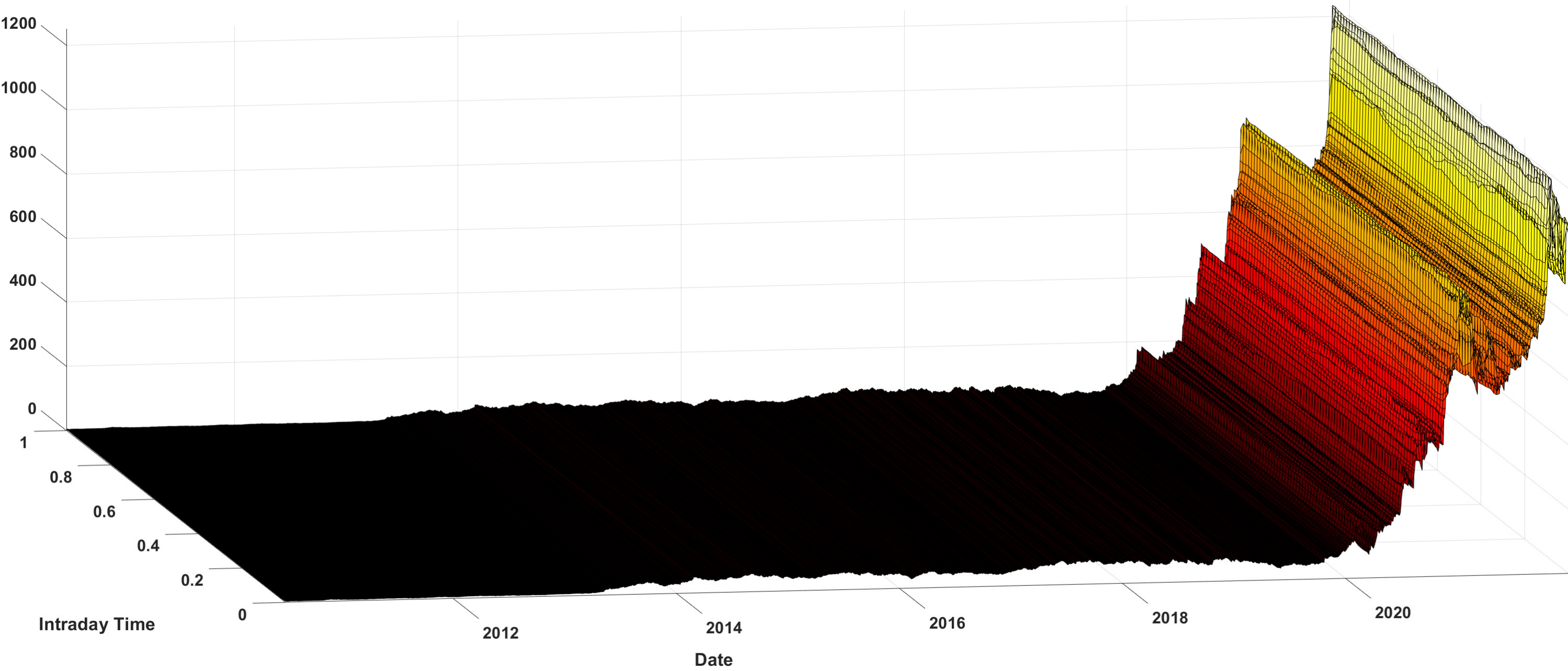}
	\includegraphics[width=1\linewidth]{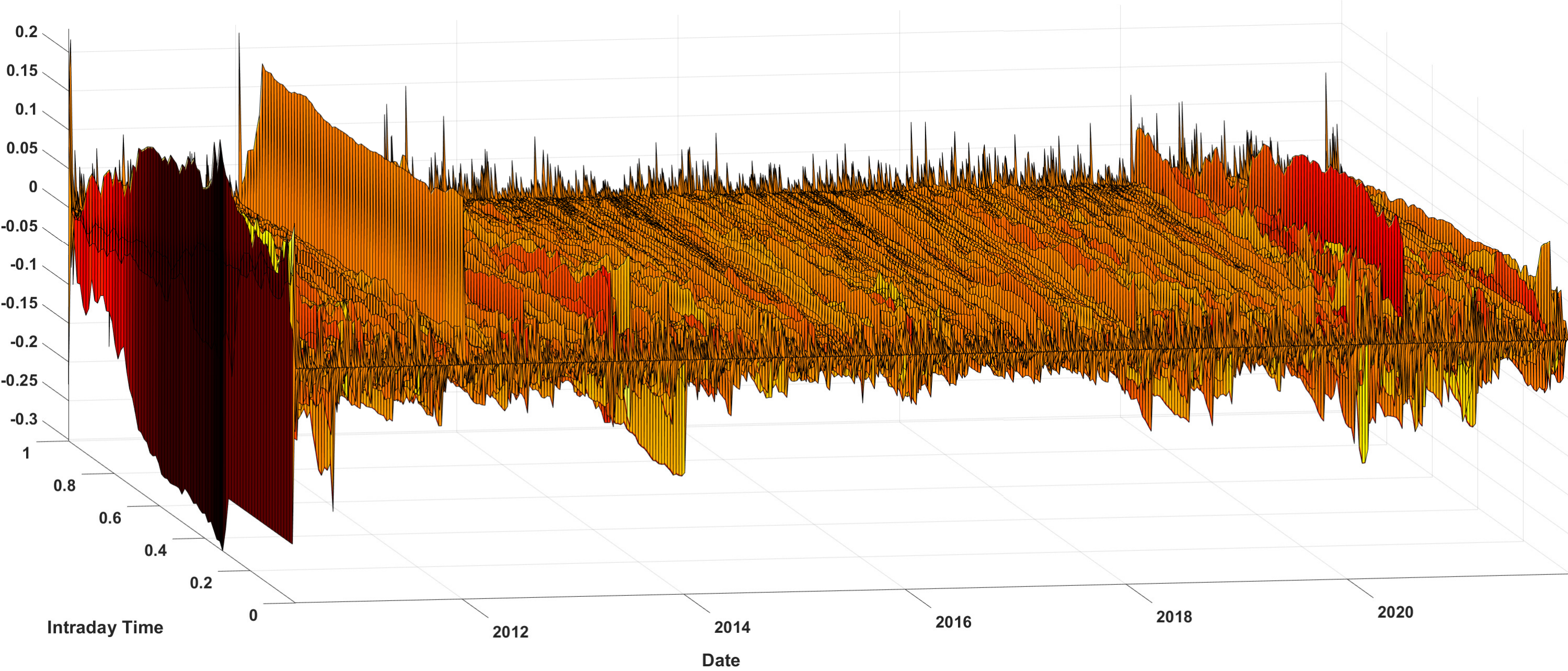}
	\includegraphics[width=1\linewidth]{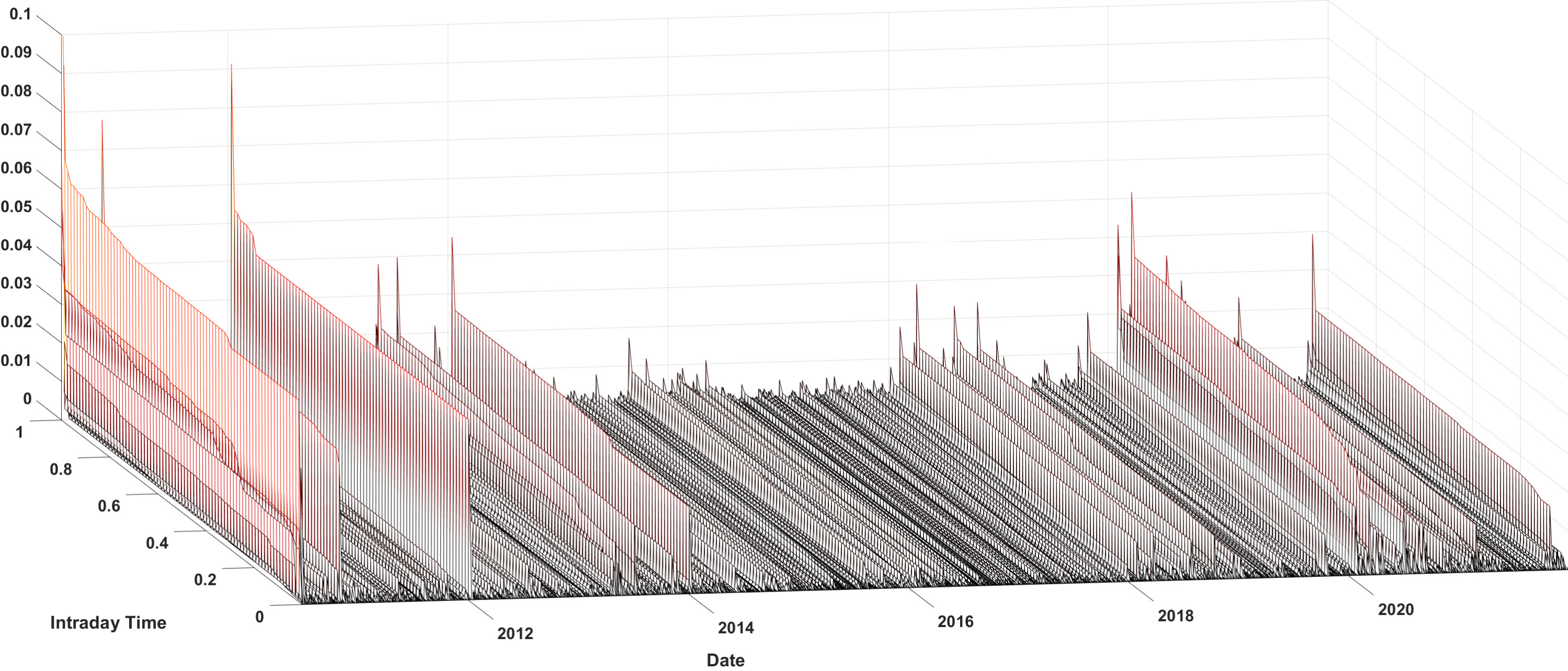}
	\caption{Time series of functional objects derived from intraday Tesla prices.
Upper Panel: Intraday Price $P_i(t_k)$; Middle Panel: CIDRs $R_i(t_k) $;
 Lower Panel: CIDRVs $RV_i(t_k)$.}
	\label{fig:tesla}
\end{figure}

We apply the tests for  the whole sample period,
in order to detect 1) a shape change, 2) a change in total volatility, and
3) an arbitrary change. Table \ref{tab:test_Tesla}
presents the test results. The $p$-value of testing $H_0^{1}$ is 0.02\%,
providing strong evidence of a shape change.
The change point estimator $\hat{\theta}_1$ is 0.26,
indicating the shape change occurred on Jul 1, 2013.
As for testing $H_0^{2}$, we find strong evidence of a
change in total volatility with $p$-value of 0.96\%.
The date of change in total volatility is May 12, 2014,
as suggested by the $\hat{\theta}_2 = 0.34$.
Combing the two tests, we have the $p$-value of 0.00\%
for the global null hypothesis $H_0$, with the pooled change
point estimator $\hat{\theta} = 0.26$, implying that the date of
arbitrary change in intraday volatility pattern is July 8, 2013.

\begin{table}[htbp]
	\centering
	\caption{Test results of Tesla (note that the $p$-values are in percent).}
\medskip
	\begin{tabular}{lp{2cm}p{5cm}p{3cm}}
		\toprule
		\toprule
		& \multicolumn{1}{l}{$p$-value} & \multicolumn{1}{l}{Change Point Estimator} & \multicolumn{1}{l}{Date of Change} \\
		\midrule
		Shape of Volatility ($H_0^{1}$) & 0.02\% & 0.26  & Jul 1, 2013 \\
		Total Volatility ($H_0^{2}$) & 0.96\% & 0.34  & May 12, 2014 \\
		Global ($H_0$) & 0.00\% & 0.26  & Jul 8, 2013 \\
		\bottomrule
		\bottomrule
	\end{tabular}%
	\label{tab:test_Tesla}%
\end{table}%

As the sample period of the Tesla analysis covers more than a decade,
there could be multiple changes in the intraday volatility pattern.
Thus, we use the standard binary segmentation
based on the global test
at the 5\% significance level and the pooled
change point estimator ($\hat{\theta}$). Table \ref{tab:binary}
presents the result with some associated events that could
be used to validate the identified change points.

\begin{table}[htbp]
	\centering
	\caption{Result of binary segmentation to test multiple changes for Tesla}
	\resizebox{\columnwidth}{!}{ \begin{tabular}{lll}
		\toprule
		\toprule
		$p$-value & Date of Change & Related News \\
		\midrule
		0.00\% & Jul 8, 2013 & Tesla joined the Nasdaq 100 index on Jul 15, 2013 \\
		0.00\% & Jul 16, 2014 & Tesla announced new smaller electric vehicle named Model 3 on Jul 16, 2014 \\
		0.08\% & Feb 6, 2018 & Elon Musk made history launching a car into space on Feb 6, 2018 \\
		0.09\% & Jan 23, 2019 & Tesla posted back-to-back profits for the first time \\
		4.79\% & Dec 20, 2019 & Tesla's Chinese factory delivered its first cars \\
		0.48\% & Mar 31, 2021 & NHTSA confirmed no violation of Tesla's touchscreen drive selector \\
		\bottomrule
		\bottomrule
	\end{tabular}}%
	\label{tab:binary}%
\end{table}%

It is also interesting to examine change in the intraday volatility pattern
for other stocks. Thus, we apply our test procedure to 7293 stocks in
the US stock markets. To preserve space, we focus on the first
change detected by the global test  at 5\% significance level,
without using binary segmentation to find additional changes.
The stocks used and the data cleaning procedure are  the same as in
\cite{kokoszka:mohammadi:wang:wang:2023}.
Their sample period varies in length from 2 to 25 years.
Shorter sample periods could be due to
IPO dates later than Jan 3, 2006 or stocks delisted before Dec 31, 2021.

\begin{figure}[H]
	\centering
	\includegraphics[width=1\linewidth]{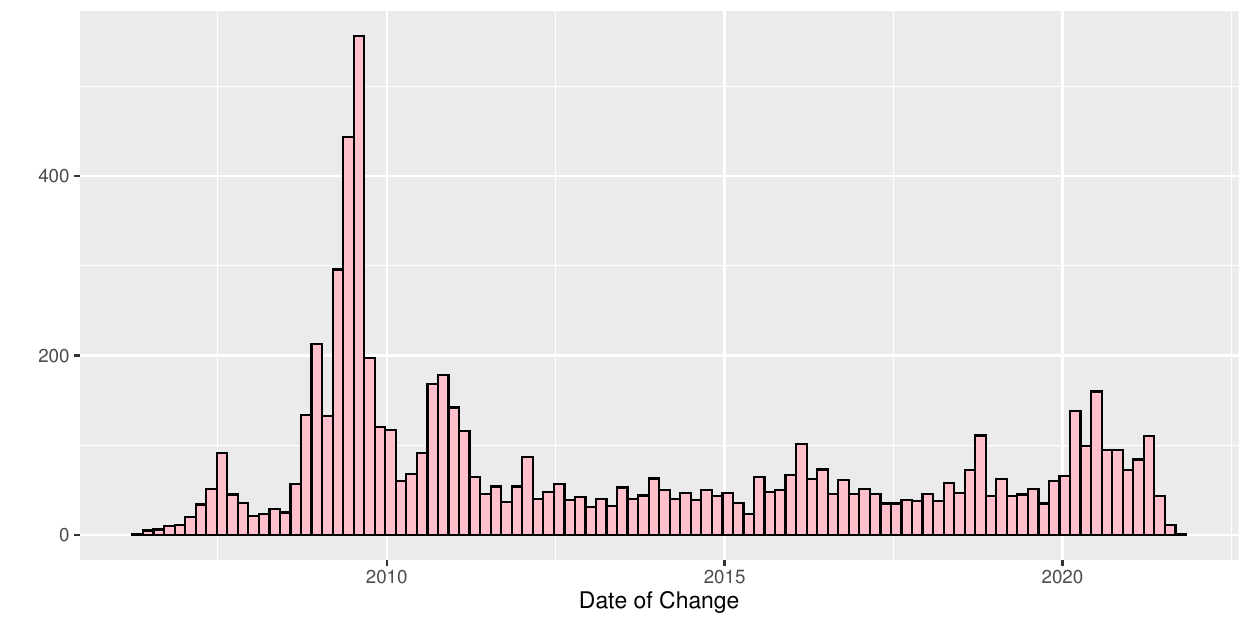}
	\caption{Dates of first change in the intraday volatility pattern
for 7168 US stocks.
	\label{fig:cpdist}}
\end{figure}

Our test indicates that 7168 out of 7293 stocks (98.3\%)
underwent at least one change in the intraday volatility pattern.
This provides the evidence that change in the intraday volatility
pattern is a common issue in the US stocks.
To provide further insights, we plots the histogram of the
first detected changes in Figure \ref{fig:cpdist}.
We can clearly see that 1) the highest frequent changes happen
during the subprime mortgage crisis in 2008,
2) the second highest frequent changes occur around the
European debt crisis in the 2010s, 3) the third highest
frequent changes appear after COVID in 2020.
These results show that our  test is able to detect
change points that are consistent with well-known market events,
providing additional validation on a very large data set.

\bigskip 

\noindent{\bf Acknowledgements} \ We thank two referees for 
insightful comments that helped us improve the paper. 
Piotr Kokoszka and Haonan Wang
were  partially supported  by  NSF grant DMS--2123761.

\bigskip
\bibliographystyle{plain}
\renewcommand{\baselinestretch}{0.9}
\small
\bibliography{neda}

\renewcommand{\baselinestretch}{1}
\normalsize

\newpage
\appendix
\centerline{\LARGE  \bf Proofs and additional information}

\bigskip

This supplementary material
contains  the proofs of our theoretical results and information
about numerical implementation.
In the following derivations, $c$ denotes a generic, positive constant that
may change from one line to another and is  always independent of $N,K$
and $i$. When needed,  we use a sequence of  positive constants,
which we denote by $c_1,c_2,\ldots$.
The  supplementary material consists of three parts:
In Appendix \ref{App_A},
we gather concentration results for the realized quadratic
variation process (denoted by $\widehat{V}(\cdot)$)
in the absence of  the random factor $\exp (g_i)$.
These results are of independent interest and
formulated for $K\to \infty$, since they do not require functional replications.
In Appendix \ref{App_B},  we prove our main results on the convergence
of the test statistics $\widehat{S}^{(i)}$, for $i=1,2$,
under the corresponding null hypotheses $H_0^{(i)}$
and local alternatives. Finally, Appendices \ref{sec:imple} and \ref{s:dc}
provide additional details, as well as pseudocode for all of our procedures.

\section{Uniform limits for the realized quadratic variation} \label{App_A}
In this section, we present some fundamental results concerning
the limiting behavior of the realized quadratic variation
process of the model \eqref{e:R1} in absence of the stochastic
coefficient $\exp (g_i)$. For the sake of simplicity we drop index $i$,
and define
\begin{equation}\label{e:V}
    V(t) := \int_0^t \sigma^2(u) du, \ \ \ t \in [0,1].
\end{equation}
    The empirical counterpart of $V(\cdot)$ is
\begin{align} \label{e:hat:V}
    \widehat{V} (t) := \sum_{k=1}^K \left \vert \int_{{(k-1)/K}}^{{k/K}}\sigma(u) dW(u)\right\vert^2 \mathbb{I}\{{k/K}  \leq t \}.
\end{align}
Most of the results stated in this section are more general
than necessary for our later investigation.
Yet, we consider them to be of independent interest.
The bounds are formulated with constants $c$ that are independent on
$\sigma$. We make the dependence on  $\sigma$ explicit,
as it is needed for our subsequent theory. It is convenient
to state the following assumption.

\begin{assumption} \label{a:A}
For  the It{\^o} process
$\int_0^t\sigma(u)dW (u)$,   Condition \ref{itm:sig} of
Assumption \ref{a:model}
holds, i.e. the volatility function
$\sigma: [0,1] \to (0, \infty)$ is continuous
\end{assumption}

\begin{lemma}\label{lem:4th:moment} Under Assumption \ref{a:A},
for any $M \in \mathbb{N}$,
     \begin{align}
     \label{e:V^2M}
        \mathbb{E}\left[ \widehat{V}(1) - V(1)\right]^{2M} \le &  \frac{c\|\sigma\|_\infty^{4M}}{K^M}.
     \end{align}
 \end{lemma}
\noindent{\sc Proof.}  Recall \eqref{e:V} and \eqref{e:hat:V}. Then we have
\begin{align}
    \label{e:V_sum}
    & \mathbb{E}\left[ \widehat{V}(1) - V(1)\right]^{2M} =  \mathbb{E}\left[ \sum_{k=1}^K \left \vert \int_{{(k-1)/K}}^{{k/K}}\sigma(u)dW(u)\right\vert^2 - V(1) \right]^{2M}\\ \nonumber
    = & \mathbb{E}\left[ \sum_{k=1}^K \left( \left \vert \int_{{(k-1)/K}}^{{k/K}}\sigma(u)dW(u)\right\vert^2 - [V({k/K})- V({(k-1)/K})] \right) \right]^{2M}\\ \nonumber
=: & \mathbb{E}\left[ \sum_{k=1}^K  Y_k [V({k/K})- V({(k-1)/K})] \right]^{2M},
  \end{align}
  where
  \begin{align}\label{e:Yk}
      Y_k =& \frac{\left \vert \int_{{(k-1)/K}}^{{k/K}}\sigma(u)dW(u)\right\vert^2 }{V({k/K})- V({(k-1)/K})}-1, \quad k=1,2,\ldots,K,
  \end{align}
are independent  centered random variables with law
\begin{align} \label{e:Y_k}
   Y_k + 1  \sim \chi^2_1.
\end{align}
We can now apply Theorem 1 in \cite{yoshihara:1978}
to the right-hand  side of \eqref{e:V_sum}, which yields
\begin{align*}
     \mathbb{E}\left[ \sum_{k=1}^K  Y_k [V({k/K})- V({(k-1)/K})] \right]^{2M} \le  &c \Big( \sum_{k=1}^K [V({k/K})- V({(k-1)/K})]^2 \Big)^M\\
    \le & c \big( \|\sigma\|_{\infty}^4/K\big)^M.
\end{align*}
This completes the proof of \eqref{e:V^2M}.
\hfill \QED

\medskip

Proposition \ref{prop:unif:4} below extends the result of Proposition
A.1 in \cite{kokoszka:mohammadi:wang:wang:2023} to all
even moments $2M$.

\begin{proposition}\label{prop:unif:4}
Under Assumption \ref{a:A},
\begin{align*}
          \mathbb{E}\left[ \underset{t \in [0,1]}{\sup} \left \vert \widehat{V}(t) -  V(t)\right \vert\right]^{2M}
    \le &  \frac{ c\|\sigma\|_\infty^{4M}}{K^M}.
     \end{align*}
(The constant $c$ depends on $M$.)
\end{proposition}

\noindent{\sc Proof.}
For  $t \in [0,1]$, we have
\begin{align}\label{e:V-G}
    \left \vert \widehat{V}(t) -  V(t)\right \vert^{2M} \leq & c  \left \vert  \widehat{V}(t) -  V(t) - \widehat{V}(t^{\ast}(t)) + V(t^{\ast}(t))\right \vert^{2M} \\
    & + c  \left \vert \widehat{V}(t^{\ast}(t)) - V(t^{\ast}(t)) \right \vert^{2M}, \nonumber
\end{align}
where $t^{\ast}(t) = \frac{1}{K}\lfloor t K \rfloor $,  and $c$ depends on the integer $M$ only.
Definition \eqref{e:hat:V} implies $\widehat{V}(t^{\ast}(t)) = \widehat{V}(t)$. This leads   to
\begin{align*}
 \left \vert \widehat{V}(t) -  V(t)\right \vert^{2M} \leq & c \left \vert     V(t) -V(t^{\ast}(t)) \right \vert^{2M} + c  \left \vert \widehat{V}(t^{\ast}(t)) - V(t^{\ast}(t)) \right \vert^{2M}.
\end{align*}
Continuity of the volatility function $\sigma (\cdot)$   implies
\begin{align}
\label{e:hatV:4}
 \underset{t \in [0,1]}{\sup} \left \vert \widehat{V}(t) -  V(t)\right \vert^{2M} \leq & c K^{-2M} \Vert\sigma \Vert_{\infty}^{4M} + c \underset{j \in \{1,2,\ldots,K\}}{\max} \left \vert \widehat{V}(j/K) - V(j/K) \right \vert^{2M}.
\end{align}
So, it is enough to focus on the discrete  index set $\{j/K\}_{j=0}^K$.
Observe that the sequence
\begin{align*}
    \widehat{V}(j/K) - V(j/K) =& \sum_{k=1}^j \left \vert \int_{{(k-1)/K}}^{{k/K}}\sigma(u)dW(u)\right\vert^2 - \int_0^{j/K}\sg^2(u) du\\
    =& \sum_{k=1}^j \left (\left \vert \int_{{(k-1)/K}}^{{k/K}}\sigma(u)dW(u)\right\vert^2 - \int_{{(k-1)/K}}^{{k/K}}\sg^2(u) du \right), \,\,\, j = 1,2 , \ldots K,
\end{align*}
forms a  martingale, see the proof of Proposition A.1
in \cite{kokoszka:mohammadi:wang:wang:2023} for details.
We now apply Doob's maximal inequality, see  {\cite{dasgupta:2011} Theorem 14.7}, to obtain
\begin{align*}
    \mathbb{E}\left[ \underset{j \in \{1,2,\ldots,K\}}{\sup} \left \vert \widehat{V}(j/K) -  V(j/K)\right \vert\right]^{2M} \leq &
\Big(\frac{2M}{2M-1} \Big)^{2M}\, \mathbb{E} \left \vert \widehat{V}(1) -  V(1)\right \vert^{2M}.
\end{align*}
This,  together with  inequality \eqref{e:hatV:4},  implies
\begin{align*}
    \mathbb{E}\left[ \underset{t \in [0,1]}{\sup} \left \vert \widehat{V}(t) -  V(t)\right \vert\right]^{2M} \leq &
    c K^{-2M} \Vert\sigma \Vert_{\infty}^{4M}
    + c \mathbb{E} \left \vert \widehat{V}(1) -  V(1)\right \vert^{2M},
\end{align*}
where, again, the constant $c$  depends only on $M$.   Applying Lemma \ref{lem:4th:moment} gives
\begin{align*}
    \mathbb{E}\left[ \underset{t \in [0,1]}{\sup} \left \vert \widehat{V}(t) -  V(t)\right \vert\right]^{2M} \leq &
    c K^{-2M} \Vert\sigma \Vert_{\infty}^{4M} + c  K^{-M}  \|\sigma\|_{\infty}^{4M}  .
\end{align*}
This completes the proof. \hfill \QED

\medskip

\begin{lemma} \label{lemma_1}
Under Assumption \ref{a:A},
there exists an absolute constant $c_1>0$
such that for any $\varepsilon \in (0,1)$,
    \[
    \mathbb{P}(| \widehat{V}(1)-V(1)|>\varepsilon)\le 2\exp\Big(-\frac{c_1 \varepsilon^2 K}{\|\sigma\|_\infty^4} \Big).
    \]
\end{lemma}

\noindent{\sc Proof.}
    We begin by rewriting the difference
    \begin{align*}
    \widehat{V} (1)-V (1) = & \sum_{k=1}^K \Big\{\Big \vert \int_{(k-1)/K}^{k/K}\sigma (u)dW (u)\Big\vert^2- V(1)\Big\}\\
= &  \sum_{k=1}^K \left( \left \vert \int_{{(k-1)/K}}^{{k/K}}\sigma(u)dW(u)\right\vert^2 - [V({k/K})- V({(k-1)/K})] \right)
    \\
    =: & \sum_{k=1}^K Y_k  [V({k/K})- V({(k-1)/K})] ,
    \end{align*}
    where $Y_1,\ldots,Y_K$ are independent centered random variables defined in \eqref{e:Yk} satisfying \eqref{e:Y_k}.
    We now define the  $\psi_1$-norm for a subexponential random variable $Y$ as
    \[
    \|Y\|_{\psi_1} := \inf\{t>0:\mathbb{E}[\exp(|Y|/t)] \le 2\}.
    \]
    For details on $\|\cdot\|_{\psi_1}$ as well as the fact that it
    indeed constitutes a norm see, for example,  \cite{vershynin:2018}.
    Due to  homogeneity of the norm, we can write
    \[
    \Big\|Y_k [V({k/K})- V({(k-1)/K})]\Big\|_{\psi_1}
    = \|Y_k\|_{\psi_1}[V({k/K})- V({(k-1)/K})] \le \frac{\|Y_k\|_{\psi_1}
    \|\sigma \|_\infty^2}{K},
    \]
    for any $K$. Moreover, $\|Y_k\|_{\psi_1} = c_2 <\infty$,
    which follows by Lemma 2.7.6 and Exercise 2.7.10
    in \cite{vershynin:2018}. Now, applying Bernstein's inequality for subexponential random variables (Theorem 2.8.1 in \cite{vershynin:2018}) proves the Lemma.
    \hfill \QED

\medskip

\begin{lemma}\label{lem:K:cov}
    Define the kernel function
\begin{align}\label{e:ker:c}
c_V (s,t) = & \int_0^{\min(s,t)} \sigma^4(x)dx,\quad s,t \in [0,1],
\end{align}
and the sequence (indexed by the grid size $K$) of  kernels
\begin{align*}
    &  \mathbb{E} \left(\left\{ \widehat{V}(s) -V(s)  \right\}\left\{ \widehat{V}(t)-V(t) \right\} \right),\quad s,t \in [0,1].
\end{align*}
Then, under Assumption \ref{a:A},
\begin{align} \label{e:CK->C}
 \underset{0 \leq s,t \leq 1}{\sup} \left \vert  K    \mathbb{E}
 \left(\left\{ \widehat{V}(s) -V(s)  \right\}\left\{ \widehat{V}(t)-V(t) \right\}
 \right)  - c_V (s,t)   \right \vert \to 0,\quad \mathrm{as}\; K \to \infty.
\end{align}
\end{lemma}

\noindent{\sc Proof.}
We start with
\begin{align*}
    &
    \mathbb{E} \left[\left\{ \widehat{V}(s)-V (s) \right\} \left\{ \widehat{V}(t) -V(t)  \right\} \right] \\
   = &   \mathbb{E} \left[\left\{ \widehat{V}(s)-V(t^{\ast}(s)) +V(t^{\ast}(s))-V (s) \right\} \left\{ \widehat{V}(t) -V(t^{\ast}(t))+ V(t^{\ast}(t))-V (t)  \right\} \right]
    \end{align*}
   where $t^{\ast}(t) = \frac{1}{K}\lfloor t K \rfloor $, for $t \in [0,1]$. This entails
   \begin{align*}
    K  \mathbb{E} \left[\left\{ \widehat{V}(s)-V (s) \right\} \left\{ \widehat{V}(t) -V(t)  \right\} \right]=& K \mathbb{E} \left[ \left\{\widehat{V}(s)- V(t^{\ast}(s))\right\}\left\{\widehat{V}(t) - V(t^{\ast}(t)\right\}\right]\\
    &+ K \mathbb{E} \left\{\widehat{V}(s)- V(t^{\ast}(s))\right\}\left\{V(t^{\ast}(t)) - V(t)\right\}\\
    &+ K \left\{V(t^{\ast}(s)) - V(s)\right\} \mathbb{E} \left\{\widehat{V}(t) - V(t^{\ast}(t))\right\}\\
    &+ K\left\{V(t^{\ast}(s)) - V(s)\right\} \left\{V(t^{\ast}(t)) - V(t)\right\},
\end{align*}
According to Proposition A.1 in
\cite{kokoszka:mohammadi:wang:wang:2023}, we have
\[
\mathbb{E} \left\{\underset{ s \in [0,1]}{\sup}\vert\widehat{V}(s)- V(t^{\ast}(s))\vert\right\} = O\left(K^{-\frac{1}{2}}\right).
\]
Moreover, continuity of the volatility function $\sigma(\cdot)$ implies
\[
\underset{ s \in [0,1]}{\sup}\left\vert V(t^{\ast}(s)) - V(s)\right\vert = O\left(K^{-1}\right).
\]
This gives
\begin{align}
\nonumber
 & K  \mathbb{E} \left[\left\{ \widehat{V}(s)-V (s) \right\} \left\{ \widehat{V}(t) -V(t)  \right\} \right]   \\
    =& K \mathbb{E} \left[ \left\{\widehat{V}(s)- V(t^{\ast}(s))\right\}\left\{\widehat{V}(t) - V(t^{\ast}(t))\right\}\right]+ K O\left(K^{-\frac{3}{2}}\right)\nonumber\\ \label{e:cov:V}
    =& K \mathbb{E} \left[ \left\{\widehat{V}(s)- V(t^{\ast}(s))\right\}\left\{\widehat{V}_i(t) - V(t^{\ast}(t))\right\}\right]+  O\left(K^{-\frac{1}{2}}\right).
\end{align}
We now investigate the first term in \eqref{e:cov:V}. Observe that
\begin{align*}
    \widehat{V}(t) - V(t^{\ast}(t))
    = \sum_{k=1}^K\left\{ \left \vert \int_{{(k-1)/K}}^{{k/K}}
    \sigma(x)dW(x)\right\vert^2  -  \int_{{(k-1)/K}}^{{k/K}}
    \sigma^2(x)dx \right\}\mathbb{I}\{{k/K}  \leq t \}.
\end{align*}
Therefore,
\begin{align*}
&K \mathbb{E} \left[ \left\{\widehat{V}(s)- V(t^{\ast}(s))\right\}\left\{\widehat{V}(t) - V(t^{\ast}(t))\right\}\right] \\
=& K \mathbb{E} \Bigg[ \left( \sum_{k=1}^K\left\{ \left \vert \int_{{(k-1)/K}}^{{k/K}}\sigma(x)dW(x)\right\vert^2  - [V({k/K})- V({(k-1)/K})]\right\}\right)^2 \\
&\qquad  \times \mathbb{I}\{{k/K}  \leq s \} \mathbb{I}\{{k/K}  \leq t \}\Bigg]\\
=: & K \mathbb{E} \left[ \left( \sum_{k=1}^K Y_k [V({k/K})- V({(k-1)/K})]\right)^2 \mathbb{I}\{{k/K}  \leq \min(s,t)\} \right],
\end{align*}
where the $Y_k$ are independent,
centered random variables defined in \eqref{e:Yk}
 with the law specified in  \eqref{e:Y_k}. Consequently,
\begin{align}
\nonumber
    &K \mathbb{E} \left[ \left\{\widehat{V}(s)- V(t^{\ast}(s))\right\}\left\{\widehat{V}(t) - V(t^{\ast}(t))\right\}\right] \\ \nonumber
    = &K  \mathbb{E}   \left[ \sum_{k=1}^K Y_k^2
    [V({k/K})- V({(k-1)/K})]^2
    \mathbb{I}\{{k/K}  \leq \min(s,t) \} \right]\\ \nonumber
    =  &  K \sum_{k=1}^K  \left( \int_{{(k-1)/K}}^{{k/K}} \sigma^2(x)dx\right)^2 \mathbb{I}\{{k/K}  \leq \min(s,t) \}\\ \label{e:rieman:4}
    =& K \sum_{k=1}^K \left( K^{-1} \sigma^2(\tilde{t}_k)\right)^2 \mathbb{I}\{{k/K}  \leq \min(s,t) \}
\end{align}
for some $\tilde{t}_k \in [{(k-1)/K} , {k/K}]$, $k=1,2,\ldots,K$. So,
it is enough to prove that the Riemann--type sum \eqref{e:rieman:4}
 converges to the kernel  \eqref{e:ker:c} uniformly. To do so, observe that
\begin{align}
\nonumber
   &\underset{0 \leq s,t \leq 1}{\sup} \left \vert K^{-1} \sum_{k=1}^K   \sigma^4(\tilde{t}_k) \mathbb{I}\{{k/K}  \leq \min(s,t) \} - \int_0^{\min(s,t)} \sigma^4(x)dx \right \vert\\ \nonumber
    = & \underset{0 \leq s,t \leq 1}{\sup}
    \left \vert \sum_{k=1}^{\lfloor K \cdot \min(s,t)\rfloor}
\int_{\frac{k-1}{K}}^{\frac{k}{K}} \left(  \sigma^4(\tilde{t}_k) - \sigma^4(x) \right)dx
    \right \vert + O(K^{-1})\\ \nonumber
    \leq & \underset{0 \leq s,t \leq 1}{\sup}  \lfloor K \cdot \min(s,t)\rfloor K^{-1} \underset{\vert x- y \vert \leq K^{-1}}{\sup}  \left \vert  \sigma^4(x)  - \sigma^4(y)   \right \vert + O(K^{-1})
   \\
   \label{e:cov:V:t*}
    =&    o(1),
\end{align}
where \eqref{e:cov:V:t*} is a consequence of the
uniform continuity of the function $\sigma(\cdot)$.
Combining \eqref{e:cov:V} and \eqref{e:cov:V:t*}
gives the desired convergence result \eqref{e:CK->C}.
\hfill \QED

\section{Proofs of the results of Section \ref{sec:theor}} \label{App_B}

\subsection{Proofs of Theorems \ref{thm:S1},   \ref{thm:S2} and
Propositions \ref{p:ind},  \ref{p:comb} (behavior under the null hypotheses)}
This section is dedicated to the analysis of the
test statistics $\widehat{S}^{(i)}$, for $i=1,2$.
We focus first on $\widehat{S}^{(1)}$
and subsequently turn to the   test statistic $\widehat{S}^{(2)}$, concluding
with their joint behavior.

To show weak convergence of $\widehat{S}^{(1)}$,
we show a weak invariance principle for the functional partial sum process
\begin{align}\label{e:PN}
    P_N^{(1)}(x,t):= \frac{1}{\sqrt{N}} \sum_{i=1}^{\lfloor xN \rfloor} \sqrt{K}\left \{\widehat{F}_i(t) - \mathbb{E}[\widehat{F}_i(t)]\right\}, \quad t \in [0,1]. 
\end{align}
Notice that thus defined, $P_N^{(1)}(x)=P_N^{(1)}(x,\cdot)$ is for every $x \in [0,1]$ a random function.

Our first theorem,  in conjunction with Lemma \ref{lemma_mean}
following it, establishes an explicit formula for the kernel
$c_F(\cdot, \cdot)$ in Theorem \ref{thm:S1} as well as the uniform
convergence needed in subsequent proofs.

\begin{theorem}\label{thm:cov:F}
Under Assumption \ref{a:A},
\begin{align*}
    \underset{0 \leq s,t \leq 1}{\sup} \left \vert K \mathbb{E}
    \left[   \left(\widehat{F}(s) -F(s) \right)
    \left( \widehat{F}(t) -F(t) \right)    \right]- c_F(s,t)\right\vert
    \to 0 , \quad \mathrm{as} \;  K \to 0,
    \end{align*}
where
\begin{align*}
    c_F(s,t)= \frac{1}{V^2(1)}c_V(s,t)- \frac{V(t)}{V^3(1)}c_V(s,1) - \frac{V(s)}{V^3(1)}c_V(1,t) + \frac{V(s)V(t)}{V^4(1)}c_V(1,1),\,\, s,t \in [0,1],
\end{align*}
and where $c_V(\cdot,\cdot)$ is defined in \eqref{e:ker:c} and $V(\cdot)$
in \refeq{V}.
\end{theorem}

\noindent{\sc Proof.} First define the event $A_{K,\varepsilon} = \left\{ \left\vert \widehat{V}(1) - V(1)\right \vert < \varepsilon\right\}$, for sufficiently small positive $\varepsilon \in (0, V(1))$. Then, we have
\begin{align*}
    & K \mathbb{E} \left[   \left(\widehat{F}(s) -F(s) \right)   \left( \widehat{F}(t) -F(t) \right)    \right] \\
    =& K \mathbb{E} \left[   \left(\widehat{F}(s) -F(s) \right)   \left( \widehat{F}(t) -F(t) \right)  \mathbb{I} \left\{A_{K,\varepsilon}\right\}   \right]\\
    &+ K \mathbb{E} \left[   \left(\widehat{F}(s) -F(s) \right)   \left( \widehat{F}(t) -F(t) \right)  \mathbb{I} \left\{ A_{K,\varepsilon}^c\right\}   \right]\\
    = &B_1 (s,t)+ B_2 (s,t).
    \end{align*}
    Since $F$, $\widehat{F}$ are cdfs, their difference is absolutely bounded by $1$.
    Boundedness of the difference  $\left(\widehat{F}(\cdot) -F(\cdot) \right) $ and  Lemma \ref{lem:4th:moment} entail
\begin{align}\label{e:B2}
    \underset{0 \leq s,t \leq 1}{\sup} \vert  B_2  (s,t) \vert \leq c K \mathbb{P}\left(A_{K,\varepsilon}^c\right) \leq cK \frac{\mathbb{E} \left\vert \widehat{V}(1) - V(1)\right \vert^4 }{\varepsilon^4}=O\left(K^{-1}\right).
\end{align}
    We now investigate $B_1(\cdot,\cdot)$. Define
    \begin{align*}
        U(s) := & \frac{\left[ \widehat{V}(s)- V(s)\right]V(1) - V(s) \left[ \widehat{V}(1)- V(1)\right]}{\widehat{V} (1)V(1)} .
    \end{align*}
A  simple calculation entails
    \begin{align*}
        B_1  (s,t) = & K \mathbb{E}  \left[U(s) U(t) \mathbb{I} \left\{ A_{K,\varepsilon}\right\}\right] \\
        = & K \mathbb{E}  \left[
        \frac{\left[ \widehat{V}(s)- V(s)\right] \left[ \widehat{V}(t)- V(t)\right]}{\widehat{V}^2 (1)} \mathbb{I} \left\{ A_{K,\varepsilon}\right\}
       \right]\\
        & - K \mathbb{E} \left[
        \frac{\left[ \widehat{V}(s)- V(s)\right] V(t) \left[ \widehat{V}(1)- V(1)\right] }{\widehat{V}^2 (1)V(1) }
        \mathbb{I} \left\{ A_{K,\varepsilon}\right\}
       \right]\\
        &- K \mathbb{E}
        \left[
        \frac{ V(s) \left[ \widehat{V}(1)- V(1)\right] \left[ \widehat{V}(t)- V(t)\right] }{\widehat{V}^2 (1)V(1) }
        \mathbb{I} \left\{ A_{K,\varepsilon}\right\}
       \right]
        \\
        &+ K \mathbb{E}
        \left[
        \frac{ V(s) V(t) \left[ \widehat{V}(1)- V(1)\right]^2   }{\widehat{V}^2 (1)V^2(1) }
        \mathbb{I} \left\{ A_{K,\varepsilon}\right\}
       \right]\\
       =: & B_{11} (s,t)-B_{12} (s,t)-B_{13} (s,t)+B_{14} (s,t).
    \end{align*}
The term $B_{11}$  satisfies
\begin{align*}
    B_{11}  (s,t) =&  K \mathbb{E}  \left[
        \frac{\left[ \widehat{V}(s)- V(s)\right] \left[ \widehat{V}(t)- V(t)\right]}{V^2 (1)}
        \mathbb{I} \left\{ A_{K,\varepsilon}\right\}
       \right]\\
       &+  K \mathbb{E}  \left[
\left[ \widehat{V}(s)- V(s)\right] \left[ \widehat{V}(t)- V(t)\right]
       \left[ \frac{1}{\widehat{V}^2 (1)} - \frac{1}{V^2 (1)} \right]
        \mathbb{I} \left\{ A_{K,\varepsilon}\right\}
       \right]\\
       =:& B_{111} (s,t)+ B_{112} (s,t) .
\end{align*}
We first explore $B_{111}(\cdot,\cdot)$. Doing so observe that
\begin{align*}
   &\left \vert B_{111} (s,t) -  K \mathbb{E}  \left[
        \frac{\left[ \widehat{V}(s)- V(s)\right] \left[ \widehat{V}(t)- V(t)\right]}{V^2 (1)}
       \right] \right \vert
       \\= & K \left \vert  \mathbb{E}  \left[
        \frac{\left[ \widehat{V}(s)- V(s)\right] \left[ \widehat{V}(t)- V(t)\right]}{V^2 (1)}
        \mathbb{I} \left\{ A^c_{K,\varepsilon}\right\}
       \right] \right \vert\\
       \leq &  \frac{K}{V^2 (1)}
       \mathbb{E}  \left\{
      \underset{0\leq t \leq 1}{\sup}  \left \vert \widehat{V}(t)- V(t)\right \vert^2
       \right\}
       \mathbb{P}  \left(  A^c_{K,\varepsilon}\right)\\
       \leq & \frac{K}{V^2 (1)}
       \mathbb{E}  \left\{
      \underset{0\leq t \leq 1}{\sup}  \left \vert \widehat{V}(t)- V(t)\right \vert^2
       \right\} \frac{ \mathbb{E}
        \left \vert \widehat{V}(1)- V(1)\right \vert^2
       }{\varepsilon^2}\\
       = & O\left( K^{-1}\right) .
\end{align*}
On the other hand, Lemma \ref{lem:K:cov} implies
\begin{align*}
 \underset{0 \leq s,t \leq 1}{\sup}  \left \vert  K \mathbb{E}  \left[
        \frac{\left[ \widehat{V}(s)- V(s)\right] \left[ \widehat{V}(t)- V(t)\right]}{V^2 (1)}
       \right] - \frac{1}{V^2(1)}c_V(s,t) \right \vert \to 0.
\end{align*}
This gives
\begin{align}\label{e:B111}
 \underset{0 \leq s,t \leq 1}{\sup} \vert B_{111} (s,t) - \frac{1}{V^2(1)}c_V(s,t)\vert
  \to 0, \quad \mathrm{as}\; K \to 0.
\end{align}
We now investigate $B_{112}(\cdot , \cdot) $. Using Lipschitz continuity of the map $x \mapsto x^{-1}$ on $[\varepsilon,\infty)$, we obtain
\begin{align*}
  & \underset{0 \leq s,t \leq 1}{\sup} \vert B_{112} (s,t) \vert \\
\leq & c K \mathbb{E}  \left[
\left[ \widehat{V}(s)- V(s)\right] \left[ \widehat{V}(t)- V(t)\right]
       \left[ \widehat{V}^2 (1)- V^2 (1) \right] \mathbb{I} \left\{ A_{K,\varepsilon}\right\}
              \right]\\
              \leq &  c K \mathbb{E}  \left[
 \underset{0 \leq t \leq 1}{\sup} \vert \widehat{V}(t)- V(t)\vert^2
        \vert \widehat{V}^2 (1)- V^2 (1) \vert \mathbb{I} \left\{ A_{K,\varepsilon}\right\}
              \right]
  \end{align*}
Applying  Cauchy-Schwartz inequality, Proposition \ref{prop:unif:4} and Lemma \ref{lem:4th:moment} we have
\begin{align}\label{e:B112}
  & \underset{0 \leq s,t \leq 1}{\sup} \vert B_{112} (s,t) \vert = K O\left(K^{-1}\right)O\left(K^{-1/2}\right) = O\left(K^{-1/2}\right).
  \end{align}
Combining \eqref{e:B111} and \eqref{e:B112},  implies
\begin{align}\label{e:B11}
    \underset{0 \leq s,t \leq 1}{\sup} \vert B_{11} (s,t) - \frac{1}{V^2(1)}c_V(s,t)\vert
  \to 0, \quad \mathrm{as}\; K \to 0.
\end{align}
A similar argument implies
\begin{align}\label{e:B12}
    \underset{0 \leq s,t \leq 1}{\sup} \vert B_{12} (s,t) - \frac{V(t)}{V^3(1)}c_V(s,1)\vert
  \to 0, \quad \mathrm{as}\; K \to 0,
\end{align}
\begin{align}\label{e:B13}
    \underset{0 \leq s,t \leq 1}{\sup} \vert B_{13} (s,t) - \frac{V(s)}{V^3(1)}c_V(1,t)\vert
  \to 0, \quad \mathrm{as}\; K \to 0,
\end{align}
\begin{align}\label{e:B14}
    \underset{0 \leq s,t \leq 1}{\sup} \vert B_{14} (s,t) - \frac{V(s)V(t)}{V^4(1)}c_V(1,1)\vert
  \to 0, \quad \mathrm{as}\; K \to 0.
\end{align}
Combining \eqref{e:B11}, \eqref{e:B12}, \eqref{e:B13} and \eqref{e:B14}, we have
\begin{align}\label{e:B1}
       \underset{0 \leq s,t \leq 1}{\sup} \vert B_{1} (s,t) - c_F(s,t)\vert
  \to 0, \quad \mathrm{as}\; K \to 0.
\end{align}
The limiting result \eqref{e:B1} together with  \eqref{e:B2} completes the proof.
\hfill \QED

The next lemma relates
$\mathbb{E}[\widehat{F}_i]$
to the deterministic function $F_i:= V_i/V_i(1)$.
As before, we drop dependence on the index $i$ for this result.

\begin{lemma} \label{lemma_mean}
Under Assumption \ref{a:A},
\[
        \|\mathbb{E}[\widehat{F} ] - F \|_\infty=c_\sg \,\,O(K^{-1}),
\]
where $c_\sg$, but not the $O$-term, depends on $\sigma (\cdot)$.
More precisely,   $c_\sg=g(\|\sigma\|_\infty,\|\sigma^{-1}\|_\infty )$,
for a function
$g: (0, \infty) \times (0, \infty) \to (0, \infty)$ that is
increasing in each component.
(Notice that  $\|\sigma\|_\infty$
and $\|\sigma^{-1}\|_\infty$ are finite by Assumption \ref{a:model}.)
\end{lemma}
\noindent{\sc Proof.}
 Let $0<\varepsilon<V (1)$ and define the event
\[
A_{K,\varepsilon}:= \{| \widehat{V} (1)-V (1)|\le \varepsilon\}.
\]
Then we can decompose
\begin{align*}
    & \|\mathbb{E}[\widehat{F}  ] - F \|_\infty
    \le  \|\mathbb{E}[(\widehat{F}   - F )\mathbb{I} \left\{ A_{K,\varepsilon}\right\}]\|_\infty+\|\mathbb{E}[(\widehat{F}   - F )\mathbb{I}\{A_{K,\varepsilon}^c\}]\|_\infty\\
    \le & \|\mathbb{E}[(\widehat{F}   - F )\mathbb{I} \left\{ A_{K,\varepsilon}\right\}]\|_\infty + \mathbb{P}(A_{K,\varepsilon}^c) \le \|\mathbb{E}[(\widehat{F}   - F )\mathbb{I} \left\{ A_{K,\varepsilon}\right\}]\|_\infty + 2\exp\Big(-\frac{c_1 \varepsilon^2 K}{\|\sigma\|_\infty^4} \Big).
\end{align*}
Here, we have used the fact that $F $, $\widehat{F} $ are cdfs
and hence their difference is absolutely bounded by $1$.
Moreover, in the last step, we have employed Lemma \ref{lemma_1}
to bound from above the probability of the event $A_{K,\varepsilon}^c$.
We can now focus on the first term on the right.
A straightforward calculation shows
\begin{align*}
    & \|(\mathbb{E}[\widehat{F}  ] - F )\mathbb{I}
    \left\{ A_{K,\varepsilon}\right\}\|_\infty = \bigg\|
    \mathbb{E}\bigg[\frac{[\widehat{V}  -V ]V (1)-V [\widehat{V}  (1)-V (1)]}{\widehat{V}  (1)V (1)}\mathbb{I} \left\{ A_{K,\varepsilon}\right\}\bigg]\bigg\|_\infty\\
    \le & \bigg\| \mathbb{E}\bigg[\frac{\widehat{V}  -V }{\widehat{V}  (1)}
    \mathbb{I} \left\{ A_{K,\varepsilon}\right\}\bigg]\bigg\|_\infty
    + \bigg\| \mathbb{E}\bigg[\frac{V [\widehat{V}  (1)-V (1)]}{\widehat{V}
    (1)V (1)}\mathbb{I} \left\{ A_{K,\varepsilon}\right\}\bigg]\bigg\|_\infty=:
R_1+R_2,
\end{align*}
where $R_1, R_2$ are defined in the obvious way.
For $R_2$ we furthermore observe the bound
\begin{align*}
    R_2 \le \frac{\|V \|_\infty}{V (1)}\bigg| \mathbb{E}\bigg[\frac{\widehat{V}  (1)-V (1)}{\widehat{V}  (1)}\mathbb{I} \left\{ A_{K,\varepsilon}\right\}\bigg]\bigg|=  \bigg| \mathbb{E}\bigg[\frac{\widehat{V}  (1)-V (1)}{\widehat{V}  (1)}\mathbb{I} \left\{ A_{K,\varepsilon}\right\}\bigg]\bigg| \le  R_1.
\end{align*}
Here we have used that $\max_t |V(t)|=V(1)$.
We can thus focus our further analysis on $R_1$,
which can be  bounded from above by
\begin{align*}
    &\bigg\| \mathbb{E}\bigg[\frac{\widehat{V}  -V }{V (1)}\mathbb{I} \left\{ A_{K,\varepsilon}\right\}+[\widehat{V}  -V ]\bigg(\frac{1}{\widehat{V}  (1)}-\frac{1}{V (1)}\bigg)\mathbb{I} \left\{ A_{K,\varepsilon}\right\}\bigg]\bigg\|_\infty\\
    \le& \bigg\| \mathbb{E}\bigg[\frac{\widehat{V}  -V }{V (1)}\mathbb{I} \left\{ A_{K,\varepsilon}\right\}\bigg]\bigg\|_\infty+\bigg\|\mathbb{E}\bigg[[\widehat{V}  -V ]\bigg(\frac{1}{\widehat{V}  (1)}-\frac{1}{V (1)}\bigg)\mathbb{I} \left\{ A_{K,\varepsilon}\right\}\bigg]\bigg\|_\infty=: R_{1,1}+R_{1,2}.
\end{align*}
Focusing on $R_{1,1}$ first, we observe that it can be bounded by
\begin{align*}
R_{1,1} \le &  \frac{\|\mathbb{E} [(\widehat{V} -V )\mathbb{I} \left\{ A_{K,\varepsilon}\right\}]\|_\infty}{V(1)}\\
\le &  \frac{\|\mathbb{E} [(\widehat{V} -V )]\|_\infty}{V(1)}+\frac{\mathbb{E} \|\widehat{V} -V
\|_\infty \mathbb{P}({A^c_{K,\varepsilon}})}{V(1)}=:R_{1,1,1}+R_{1,1,2}.
\end{align*}
Jensen's inequality implies that
\[
R_{1,1,2} \le \frac{\{\mathbb{E} \|\widehat{V} -V
\|_\infty^2\}^{1/2} \mathbb{P}({A^c_{K,\varepsilon}})}{V(1)}.
\]
As before, we can use Lemma \ref{lemma_1} to bound the probability in $R_{1,1,2}$, which combined with the bound in Proposition \ref{prop:unif:4} implies
\[
R_{1,1,2}\le \frac{2c^{1/2}\|\sigma\|_\infty^{2}}{K^{1/2} V(1)}\exp\Big(-\frac{c_1 \varepsilon^2 K}{\|\sigma\|_\infty^4} \Big).
\]
The constant $c$ here stems from Proposition \ref{prop:unif:4} (for $M=1$)  and is independent of $\sigma$.
Now, we turn to $R_{1,1,1}$, for which we observe
\[
R_{1,1,1} \le \frac{\max_{u \in [0,1]}|\mathbb{E} \widehat{V}  (u)-V (u)|}{V(1)}
\leq \frac{\max_{k=1,\ldots,K} \int_{(k-1)/K}^{k/K} \sigma ^2(u) du }{V(1)}\le \frac{\|\sigma ^2\|_\infty }{V(1)K}.
\]
\noindent This shows the desired rate for $R_{1,1}$. For $R_{1,2}$, we observe that it is upper bounded by
\[
R_{1,2} \le \frac{\mathbb{E}\Big[\|\widehat{V}  -V \|_\infty |\widehat{V}  (1)-V (1)| \mathbb{I} \left\{ A_{K,\varepsilon}\right\} \Big]}{\varepsilon^2},
\]
where we have used Lipschitz continuity of the map $x \mapsto x^{-1}$ on $[\varepsilon,\infty)$, with constant $1/\varepsilon^2$. Notice that here $\varepsilon$ has to be sufficiently small, and we can choose it as  $V(1)/2$ (so that $\widehat{V}(1)$ is also bounded away from 0). Now, employing the Cauchy-Schwartz inequality, together with the moment bound from Proposition \ref{prop:unif:4} yields a rate of $O(K^{-1})$ for the right side. Together, these considerations demonstrate that
\[
\|\mathbb{E}[\widehat{F} ] - F \|_\infty \le \frac{c}{K}
\]
for a sufficiently large constant $c>0$.

The bounds that we have employed so far,
namely from Proposition \ref{prop:unif:4} and Lemma \ref{lemma_1},
depend only on $\|\sigma\|_\infty$ and are monotonically increasing
in it. Moreover, $c_\sg$ depends on the factor $1/V(1)$
(explicitly and via $1/\varepsilon=2/V(1)$). Since
\[
\frac{1}{V(1)} \le \frac{1}{\min_{t \in [0,1]}\sigma^2(t)} = \|\sigma^{-1}\|_\infty^2,
\]
the constant $c_\sg \in (0, \infty)$  depends only
on $\sigma$ via the values $\|\sigma\|_\infty$ and $\|\sigma^{-1}\|_\infty$
and this dependence is monotone in  each norm.
\hfill  \QED
\medskip

 \noindent Lemma \ref{lemma_mean} has two important implications for our analysis. In the context of this section, it establishes for the covariance operator, that $F $ can be used as a centering term instead of $\mathbb{E} \widehat{F}  $, i.e. that
\begin{align}\label{e:rep:m}
    &K \mathbb{E}\big(\{\widehat{F}(u)  - \mathbb{E} \widehat{F}(u)  \}\cdot \{\widehat{F}(v)   - \mathbb{E} \widehat{F}  (v)\}\big) \\
    =& K \mathbb{E}\big( \{\widehat{F}(u)   - F(u) \}\cdot \{\widehat{F}(v)   - F(v) \} \big) + o(1).\nonumber
\end{align}
Here the $o$-term vanishes w.r.t. to the sup-norm and the non-negligible part on the right converges according to Theorem \ref{thm:cov:F}.
A second use of Lemma \ref{lemma_mean} discussed later will be that any change in the functions $F_i$
transpires through to the functions $\widehat{F}_i  $ as a (detectable) mean change.

\begin{lemma}\label{lem:Fhat}
Under Assumption \ref{a:A},
\begin{align}\label{e:Fhat:EF}
K \mathbb{E}\left \Vert\widehat{F}-\mathbb{E}(\widehat{ F})\right \Vert^2
\le c_\sg.
\end{align}
The constant $c_\sg$ in \refeq{Fhat:EF} is different from the $c_\sg$ in Lemma
\ref{lemma_mean},  but admits the same representation in terms of
the two norms.
\end{lemma}
\noindent{\sc Proof.}
The left hand side of \eqref{e:Fhat:EF} satisfies
\begin{align*}
    K \mathbb{E} \left \Vert \widehat{F}-\mathbb{E}(\widehat{ F}) \right \Vert^2 \leq&
    2K \mathbb{E} \left \Vert \widehat{F}-F \right \Vert^2 +
    2K   \left \Vert F-\mathbb{E}(\widehat{ F}) \right \Vert^2 \\
    =: B_1 + B_2.
\end{align*}
According to Lemma \ref{lemma_mean},
$B_2 = c_\sg \,O(K^{-1})$.
We now investigate $B_1$. Define the event
$A_{K,\varepsilon} = \left\{ \left\vert \widehat{V}(1)
- V(1)\right \vert < \varepsilon\right\}$, for $\varepsilon = V(1)/2$
and observe that
\begin{align*}
\nonumber
B_1 =& 2K \mathbb{E} \left[ \int_0^1 \left(\widehat{F}(s)-F(s)\right)^2 ds
    \right]\\ \nonumber
=& 2K \mathbb{E} \left[ \int_0^1 \left(\widehat{F}(s)-F(s)\right)^2
\mathbb{I} \left\{ A^c_{K,\varepsilon}\right\}ds\right]
+ 2K \mathbb{E} \left[ \int_0^1 \left(\widehat{F}(s)-F(s)\right)^2 \mathbb{I}
\left\{ A_{K,\varepsilon}\right\}ds\right].\\
\end{align*}
Therefore,
\begin{align*}
 B_1 \leq & 2K \mathbb{P}\left(A_{K,\varepsilon}^c\right)+2K \mathbb{E}
\left[ \int_0^1 \left(\widehat{F}(s)-F(s)\right)^2 \mathbb{I} \left\{ A_{K,\varepsilon}\right\}ds\right]
\\ \nonumber
=& 2K \mathbb{P}\left(A_{K,\varepsilon}^c\right)+2K \mathbb{E} \int_0^1 \left(\frac{ \widehat{V}(s)}{ \widehat{V}(1)} - \frac{ \widehat{V}(s)}{ {V}(1)}+ \frac{ \widehat{V}(s)}{ {V}(1)} -  \frac{ V(s)}{V(1)}\right)^2 \mathbb{I} \left\{ A_{K,\varepsilon}\right\}ds\\ \nonumber
\leq & 2K \mathbb{P}\left(A_{K,\varepsilon}^c\right)
    +4K \mathbb{E} \int_0^1 \left(\frac{ \widehat{V}(s)}{ \widehat{V}(1)}
    - \frac{ \widehat{V}(s)}{ {V}(1)} \right)^2 \mathbb{I}
    \left\{ A_{K,\varepsilon}\right\}ds\\
&    +4K \mathbb{E} \int_0^1 \left( \frac{ \widehat{V}(s)}{ {V}(1)}
    -  \frac{ V(s)}{V(1)}\right)^2 \mathbb{I}
    \left\{ A_{K,\varepsilon}\right\}ds \nn
    =: B_{11}+B_{12}+B_{13}.
\end{align*}
An argument similar to \eqref{e:B2} implies
\begin{align}\label{e:b_11}
    B_{11} \leq \Vert\sigma\Vert_\infty^{8} O\left(K^{-1}\right).
\end{align}
Regarding $B_{12}$, we have
\begin{align}
\nonumber
     B_{12} =&    4K \mathbb{E} \left[\int_0^1  \widehat{V}^2(s)\left(\frac{1}{ \widehat{V}(1)} - \frac{ 1}{ {V}(1)} \right)^2 \mathbb{I} \left\{ A_{K,\varepsilon}\right\}ds\right]\\ \label{e:vhat:mon}
     \leq & 4K \mathbb{E}  \left[\widehat{V}^2(1)\left(\frac{1}{ \widehat{V}(1)} - \frac{ 1}{ {V}(1)} \right)^2 \mathbb{I} \left\{ A_{K,\varepsilon}\right\}\right]\\ \label{e:vhat:2v}
      \leq & 16 K V^2(1) \mathbb{E}   \left[\left(\frac{1}{ \widehat{V}(1)} - \frac{ 1}{ {V}(1)} \right)^2 \mathbb{I} \left\{ A_{K,\varepsilon}\right\}\right]\\ \label{e:vhat:lip}
      \leq &  16 K V^2(1) \mathbb{E}   \left[\frac{4}{V^2(1)} \left( \widehat{V}(1) - {V}(1) \right)^2 \mathbb{I} \left\{ A_{K,\varepsilon}\right\}\right]\\ \nonumber
      =& 64 K \mathbb{E}   \left[ \left( \widehat{V}(1) - {V}(1) \right)^2 \mathbb{I} \left\{ A_{K,\varepsilon}\right\}\right]\\ \label{e:v1}
      \leq &  64 K \mathbb{E}   \left[ \left( \widehat{V}(1) - {V}(1) \right)^2 \right] \le c \Vert\sigma\Vert_\infty^{4}.
\end{align}
Inequality \eqref{e:vhat:mon}
 is a consequence of monotonicity of the empirical quadratic variation process $\widehat{V}(\cdot)$. Inequality \eqref{e:vhat:2v} is a result of restriction to the event $A_{K,\varepsilon}$. Inequality \eqref{e:vhat:lip} is a consequence of Lipschitz continuity of the function $x \mapsto x^{-1}$ on $[V(1)/2,\infty)$, with Lipschitz constant $4/V(1)^2$. Finally, Lemma \ref{lem:4th:moment} implies \eqref{e:v1}.
   We now turn to $B_{13}$. Observe that
\begin{align}
\nonumber
     B_{13} =&  4K \frac{ 1}{V^2(1)} \mathbb{E} \left[ \int_0^1 \left(  \widehat{V}(s) -   V(s)\right)^2 \mathbb{I} \left\{ A_{K,\varepsilon}\right\}ds \right]\\ \nonumber
     =& 4K \frac{ 1}{V^2(1)} \mathbb{E} \left[ \int_0^1 \left(  \widehat{V}(s) -   V(s)\right)^2 ds \right]\\ \label{e:vhat-v}
     =& 4K \frac{ 1}{V^2(1)} \mathbb{E} \left[\underset{0 \leq s \leq 1}{\sup}\left\vert  \widehat{V}(s) -   V(s)\right\vert^2\right] \le c \Vert\sigma\Vert_\infty^{4} \|\sigma^{-1}\|_\infty^2,
\end{align}
where \eqref{e:vhat-v} is a consequence of Proposition \ref{prop:unif:4}. Pooling  \eqref{e:b_11}, \eqref{e:v1}
 and \eqref{e:vhat-v} completes the proof.
 \hfill \QED

We are now in the position to demonstrate weak convergence of the partial
sum process $P_N^{(1)}$, see \eqref{e:PN}. In the following theorem,
we invoke the notion of a Brownian motion $\mathbb{G}$ in  a Hilbert space.
For the  definition and details, we refer to \cite{kuelbs:1973}.

\begin{theorem} \label{thm_inv}
Suppose that  Conditions  \ref{itm:sig}, \ref{itm:BM} of Assumption
\ref{a:model} hold and $N, K \to \infty$.
Then, under  $H_0^{(1)}$ \refeq{shape},
there exists a functional Brownian motion $\mathbb{G}$
in the Hilbert space $L^2[0,1]$ such that
    \[
    \{P_N^{(1)}(x, \cdot)\}_{x \in [0,1]} \convd
     \{\mathbb{G}(x, \cdot)\}_{x \in [0,1]},
    \]
    where $P_N^{(1)}$ is the partial sum process defined in \refeq{PN}. The process $\mathbb{G}$ is centered and characterized by the covariance in \refeq{covS1}.
\end{theorem}


\noindent{\sc Proof.}
We apply Theorem 1 in \cite{kuelbs:1973}
(a weak invariance principle for triangular arrays of i.i.d.
 random elements  in a Banach space).
We validate the following three conditions of this theorem.
\begin{itemize}
    \item[(K.1)] The array $\sqrt{K}[\widehat{F}_i-\mathbb{E}[\widehat{F}_i]]$, $i=1, \ldots, N,$
    consists of i.i.d. random elements for any fixed $N$ (and $K$).
 \item[(K.2)] $P_N^{(1)}(1)$ satisfies a central limit theorem in
    the space $L^2[0,1]$.
    \item[(K.3)] For any $\epsilon>0$, there exists an
    $h>0$ (sufficiently small) such that
    \[
    \limsup_{N} \mathbb{P}\Big(\|P_N^{(1)}(xh)\|\ge \epsilon \Big) <1.
    \]
\end{itemize}

Condition (K.1) clearly holds because the Wiener processes $W_i$ are
i.i.d. Condition (K.3) follows directly from Markov's inequality and the fact that
\[
    \mathbb{E}\|P_N^{(1)}(xh)\|^2 \le \frac{1}{N} \sum_{i=1}^{\lfloor hN \rfloor}\mathbb{E}\|\sqrt{K}[\widehat{F}_i-\mathbb{E}(\widehat{ F}_i)]\|^2 \le ch,
\]
where we have used the moment bound from Lemma \ref{lem:Fhat}
in the last step.

We now proceed to the proof of the above condition (K.2).
For this purpose we apply a central limit theorem for triangular arrays
in Hilbert spaces, Theorem 1.1 in \cite{kundu:majumdar:mukherjee:2000}.
This result has three conditions, where i)-ii) are implied by the fact that our random  functions $\sqrt{K}[\widehat{F}_i-\mathbb{E}(\widehat{F}_i)]$ have a covariance  that converges w.r.t. to the supremum norm to the limiting covariance and thus in particular w.r.t. the trace norm (see Theorem \ref{thm:cov:F}, together with Lemma \ref{lemma_mean}). The third condition is a Linderberg-type condition, which is implied by the fact that for any $f \in L^2[0,1]$ and $\delta>0$ it holds that
\[
\mathbb{E}\Big[\langle f, \sqrt{K}[\widehat{F}_i-\mathbb{E}(\widehat{F}_i)]\rangle^2 \cdot  \mathbb{I}\{\|\sqrt{K/N}[\widehat{F}_i-\mathbb{E}(\widehat{F}_i)]\|>\delta\} \Big]\to 0.
\]
 This fact follows directly using the Cauchy-Schwarz inequality (essentially the Lyapunov argument) and recalling the moment bound from Lemma \ref{lem:Fhat}, since
\begin{align*}
    & \mathbb{E}\Big[\langle f, \sqrt{K}[\widehat{F}_i-\mathbb{E}(\widehat{F}_i)]\rangle^2 \mathbb{I}\{\|\sqrt{K/N}[\widehat{F}_i-\mathbb{E}(\widehat{F}_i)]\|>\delta\} \Big]\\
    \le & \Big\{\mathbb{E}\Big[\langle f, \sqrt{K}[\widehat{F}_i-\mathbb{E}(\widehat{F}_i)]\rangle^2\Big] \Big\}^{1/2} \mathbb{P}(\|\sqrt{K/N}[\widehat{F}_i-\mathbb{E}(\widehat{F}_i)]\|>\delta)^{1/2}.
\end{align*}
Again using the Markov inequality and the second moment bound for the variable $\sqrt{K}[\widehat{F}_i-\mathbb{E}(\widehat{F}_i)]$ (see Lemma \ref{lem:Fhat}) we observe that
\[
\mathbb{P}(|\sqrt{K/N}[\widehat{F}_i-\mathbb{E}(\widehat{F}_i)]|>\delta) \le c /N =o(1).
\]
Thus, by Theorem 1.1 in \cite{kundu:majumdar:mukherjee:2000}
weak convergence of  $P_N^{(1)}(1)$ to a centered Gaussian variable
in $L^2[0,1]$ follows, and then by Theorem 1 in \cite{kuelbs:1973}
the weak invariance principle for $P_N^{(1)}$ of this theorem.
\hfill \QED

\medskip

\noindent{\sc Proof of Theorem \ref{thm:S1}:}
The proof is based on
Theorem \ref{thm_inv}. To demonstrate the weak convergence
of $\widehat{S}^{(1)}$ (under $H_0^{(1)}$), we rewrite the statistic as
\begin{align*}
    \widehat{S}^{(1)} =& \frac{K}{N^2}\sum_{n=1}^N \int_0^1\Big(\sum_{i=1}^n \widehat{F}_i(u) - \frac{n}{N} \sum_{i=1}^N \widehat{F}_i(u) \Big)^2 du\\
    =&\frac{1}{N}\sum_{n=1}^N \int_0^1\Big(\frac{1}{\sqrt{N}}\sum_{i=1}^n \sqrt{K}[\widehat{F}_i(u)-\mathbb{E}[\widehat{F}_i(u)]] - \frac{n}{N\sqrt N} \sum_{i=1}^N \sqrt{K}[\widehat{F}_i(u)-\mathbb{E}[\widehat{F}_i(u)]] \Big)^2 du\\
    = & \int_0^1 \|P_N^{(1)}(x) - (\lfloor xN \rfloor/N) \cdot  P_N^{(1)}(1) \|^2 dx\\
    \convd & \int_0^1 \|\mathbb{G}(x)-x\mathbb{G}(1)\|^2 dx.
\end{align*}
Here, $\mathbb{G}$ denotes a functional Brownian motion
(in the space $L^2[0,1]$) that characterizes
the limit of $P_N^{(1)}$ and is defined in  Theorem \ref{thm_inv}.
It is known that
\[
  \int_0^1 \|\mathbb{G}(x)-x\mathbb{G}(1)\|^2 dx
  \overset{d}{=} \sum_{\ell =1}^\infty \lambda_\ell \int_0^1
  B_\ell(u)^2 du,
\]
see, e.g. Theorem 1 in \cite{gromenko:kokoszka:reimherr:2017}.
The consistency under a fixed alternative follows from Theorem \ref{t:loc1}
that will be proven in Section \ref{s:alt}.
\hfill \QED

\medskip

\noindent{\sc Proof of Theorem \ref{thm:S2}:}
Recall the definition of the random variable $w_i$ in \refeq{def:wi}. We can now define the process
\begin{align*}
P_N^{(2)}(x) :=&  \frac{1}{\sqrt{N}}
\sum_{i=1}^{\lfloor xN \rfloor} \log(\widehat{Q}_i(1))
-\mathbb{E}[\log(\widehat{Q}_i(1))] =  \frac{1}{\sqrt{N}}
\sum_{i=1}^{\lfloor xN \rfloor} 2 g_i + \frac{1}{\sqrt{N}}
\sum_{i=1}^{\lfloor xN \rfloor} w_i-\mathbb{E}[w_i]\\
=: &\widetilde{P}_N^{(2)}(x) + R(x),
\end{align*}
where we have used  decomposition \refeq{logQ(1)} in the second equality. Both terms on the right are defined in the obvious way. We first show that
\begin{equation} \label{e:ws}
\sup_{x \in [0,1]}|R(x)|=o_P(1), \ \ {\rm as}  \  N\to \infty.
\end{equation}
Corollary A.1 in \cite{kokoszka:mohammadi:wang:wang:2023} implies that
\[
\mathbb{E}|w_i-\mathbb{E}w_i|^2 = O(1/K).
\]
Since the $w_i$'s are independent,
by Doob's martingale inequality, for any $\delta>0$,
\[
\mathbb{P}\Big( \sup_x| R(x)|>\delta\Big) \le \frac{\mathbb{E}|R(1)|^2}{\delta^2} \le \frac{c}{K}.\
\]
As a consequence, it follows as $N,K \to \infty$ that $P_N^{(2)} = \widetilde{P}_N^{(2)}+o_P(1)$ and hence that
\begin{equation} \label{e:cS2}
\widehat{S}^{(2)} = \frac{1}{N} \sum_{n=1}^N \bigg(\sum_{i=1}^n
\big(2g_i+\mathbb{E}[w_i]\big)- \frac{n}{N}\sum_{i=1}^N \big(2g_i
+\mathbb{E}[w_i]\big)\bigg)^2+o_P(1)=:\widetilde{S}^{(2)}+o_P(1).
\end{equation}
Here, the $o$-term is the same on both sides of the equality, which defines $\tilde{S}^{(2)}$.

It remains  to show that $\widetilde{S}^{(2)} \to S^{(2)}$.
This follows from  Condition \ref{itm:g:inv} of Assumption \ref{a:model} and a continuous mapping argument.

The claim that $\widehat{S}^{(2)} \convP \infty$ if $H_0^{(2)}$ is
violated follows from Theorem \ref{t:loc2} proven in Section \ref{s:alt}.

\hfill \QED

\medskip

\noindent{\sc Proofs of Propositions \ref{p:ind} and \ref{p:comb}:} \
By \refeq{cS2},
\[
(\widehat{S}^{(1)},\widehat{S}^{(2)})
= (\widehat{S}^{(1)},\tilde{S}^{(2)})+o_P(1).
\]
Recall that $\widehat{S}^{(1)},\tilde{S}^{(2)}$ are independent of each other, since $\widehat{S}^{(1)}$ is only a functions of the Brownian motions $W_1, W_2,...$ (see eq. \refeq{hatF} and \refeq{S1}) and $\tilde{S}^{(2)}$ only of the random factors $g_1, g_2,...$ (see \refeq{cS2}). As a consequence,
their limits
 ${S}^{(1)},{S}^{(2)}$ are independent as well.
Recalling that $\Lambda^{(i)}$ is the cdf of $S^{(i)}$, which is continuous, it follows that
\[
(p^{(1)}, p^{(2)}) = (1-\Lambda^{(1)}(\widehat{S}^{(1)}),1-\Lambda^{(2)}(\widehat{S}^{(2)}))  \convd (U_1, U_2),
\]
where $U_1, U_2$ are independent uniformly distributed random variables on the unit interval. It is then standard to show that
\[
2(\log(p^{(1)})+\log(p^{(2)}))\convd 2(\log(U_1)+\log(U_2)) \overset{d}{=}\chi_4^2.
\]
If $H_0$ is violated $H_0^{(i)}$ is violated for some $i \in \{1,2\}$. In this case $\widehat{S}^{(i)} \convP \infty $, hence $\Lambda^{(i)}(\widehat{S}^{(i)}) \convP 1$ and
\[
\log(p^{(i)}) = -\log(1-\Lambda^{(i)}(\widehat{S}^{(i)}) )\convP \infty
\]
implying that $\widehat{S}\convP \infty $ leading to rejection with asymptotic probability $1$. \hfill \QED

\subsection{Proofs of Theorems \ref{t:loc1}, \ref{t:th1}, \ref{t:loc2},
\ref{t:th2} and \ref{thm:thp}
(behavior under the alternative hypotheses)}\label{s:alt}

\noindent{\sc Proof of Theorem \ref{t:loc1}:} \
We begin rewriting  $\widehat{S}^{(1)}$ as
\begin{align} \label{e:SNdec}
\widehat{S}^{(1)} = &  \frac{K}{N^2}\sum_{n=1}^N \Big\|\sum_{i=1}^n \{\widehat{F}_i-\mathbb{E}[\widehat{F}_i]\}  - \frac{n}{N} \sum_{i=1}^N \{\widehat{F}_i-\mathbb{E}[\widehat{F}_i]\} \\
&\qquad -\Big(\frac{n}{N} \sum_{i=1}^N\mathbb{E}[\widehat{F}_i]-\sum_{i=1}^n\mathbb{E}[\widehat{F}_i]\Big)\Big\|^2\\
= & \int_0^1 \|P_N^{(1)}(x) - (\lfloor xN \rfloor/N) \cdot  P_N^{(1)}(1)+h_N(x) \|^2 dx. \nonumber
\end{align}
 Here,  $P_N^{(1)}$ denotes the partial sum process defined in \refeq{PN} and
 \[
h_N(x) := -\sqrt{\frac{K}{N}}\Big(\frac{\lfloor Nx \rfloor}{N} \sum_{i=1}^N\mathbb{E}[\widehat{F}_i]-\sum_{i=1}^{\lfloor Nx \rfloor}\mathbb{E}[\widehat{F}_i]\Big).
 \]
 We decompose $P_N^{(1)}$ as
\begin{align} \label{e:dec:P}
    P_N^{(1)}(x) = P_N^{(1)}(x\land \theta) + [P_N^{(1)}(x)- P_N^{(1)}(\theta)]  \mathbb{I}\{x>\theta\}.
\end{align}
The two processes on the right-hand side
 involve only functions $\widehat{F}_i$ before (left term)
 or after  the change point (right term).
 Moreover, they are stochastically independent of each other,
 since the $\widehat{F}_i$ are independent along $i$.
Consequently, using exactly the same arguments
as in the proof of Theorem \ref{thm_inv}, we conclude that
\begin{align*}
      \{P_N^{(1)}(x)\}_{x \in [0, \theta]} \convd& \{\mathbb{G}_1(x)\}_{x \in [0, \theta]}\\
     \{[P_N^{(1)}(x)- P_N^{(1)}(\theta)] \}_{x \in [ \theta,1]}\convd
     & \{\mathbb{G}_2(x-\theta)\}_{x \in [ \theta,1]},
\end{align*}
where $\mathbb{G}_1, \mathbb{G}_2$ are independent functional
Brownian motions characterized by the covariance functions
\[
c_{F,1}(u,v):= \lim_{K \to \infty} K \cdot \mathbb{E}
\left[\{\widehat{F}_1(u)-\mathbb{E}[\widehat{F}_1(u)]\}
\{\widehat{F}_1(v)-\mathbb{E}[\widehat{F}_1(v)]\}\right],
\]
\[
c_{F,2}(u,v):= \lim_{K \to \infty} K \cdot
\mathbb{E}\left[\{\widehat{F}_N(u)-\mathbb{E}[\widehat{F}_N(u)]\}
\{\widehat{F}_N(v)-\mathbb{E}[\widehat{F}_N(v)]\}\right],
\]
respectively (covariance before and after the change).
These limits exist, as demonstrated  in Theorem \ref{thm:cov:F}.
Decomposition \refeq{dec:P},
 together with the continuous mapping theorem,  implies
\begin{align} \label{e:PNweak}
\{P_N^{(1)}(x)\}_{x \in [0,1]} \convd \mathbb{G}^{(1)}(x\land \theta) + [\mathbb{G}^{(2)}(x-\theta)] \mathbb{I}\{x>\theta\}.
\end{align}
Next, we turn to the analysis of the deterministic function $h_N(x)$
that can be written as
\[
h_N(x)= \sqrt{NK}\begin{cases}
x(1-\theta)(\mathbb{E}[\widehat{F}_1]-\mathbb{E}[\widehat{F}_N]) \quad \textnormal{if} \quad x \le \theta,\\
\theta(1-x)(\mathbb{E}[\widehat{F}_1]-\mathbb{E}[\widehat{F}_N]) \quad \textnormal{if} \quad x > \theta.
\end{cases}
\]
According to Lemma \ref{lemma_mean},
$\|\mathbb{E}[\widehat F_i]- F_i\|_\infty = O(1/K)$,
where the $O$-term is independent of the volatility functions
$\sigma_{(j)}, \sigma_{(j)}^{-1}$. Consequently, setting
\[
\tilde h_N(x):= \begin{cases}
x(1-\theta)(F_1-F_N)/a_N \quad \textnormal{if} \quad x \le \theta,\\
\theta(1-x)(F_1-F_N)/a_N \quad \textnormal{if} \quad x > \theta,
\end{cases}
\]
we get the identity
\begin{equation}\label{e:def_hN}
    h_N(x) = O\Big( \sqrt{\frac{N}{K}}\Big) + a_N \sqrt{NK}\cdot \tilde h_N.
\end{equation}
Now, a simple calculation shows that
\begin{align*}
    & \frac{F_1-F_N}{a_N} =  \frac{\int_0^t \sigma_{(1)}^2(u) du}{a_N \int_0^1 \sigma_{(1)}^2(u) du}-  \frac{\int_0^t [\sigma_{(1)}(u)+a_N \tilde \sigma(u)]^2 du}{a_N \int_0^1 [\sigma_{(1)}(u)+a_N \tilde \sigma(u)]^2 du}\\
    = & \frac{\int_0^t \sigma_{(1)}^2(u) du \int_0^1 [\sigma_{(1)}(u)+a_N \tilde \sigma(u)]^2 du-\int_0^t [\sigma_{(1)}(u)+a_N \tilde \sigma(u)]^2 du\int_0^1 \sigma_{(1)}^2(u) du}{a_N \int_0^1 \sigma_{(1)}^2(u) du\int_0^1 [\sigma_{(1)}(u)+a_N \tilde \sigma(u)]^2 du}\\
    = &  \Bigg(2\int_0^t \sigma_{(1)}^2(u) du \int_0^1 \sigma_{(1)}(u) \tilde \sigma(u) du + a_N\int_0^t \sigma_{(1)}^2(u) du  \int_0^1 \tilde \sigma^2(u) du
    \\
    & -2\int_0^t \sigma_{(1)}(u) \tilde \sigma(u) du
    \int_0^1 \sigma_{(1)}^2 du
    -a_N  \int_0^t \tilde \sigma^2(u) du\int_0^1 \sigma_{(1)}^2(u) du\Bigg)\\
    & \Bigg(\int_0^1 \sigma_{(1)}^2(u) du\int_0^1 [\sigma_{(1)}(u)+a_N \tilde \sigma(u)]^2 du\Bigg)^{-1}\\
    \to & 2\frac{\int_0^t \sigma_{(1)}^2(u) du \int_0^1 \sigma_{(1)}(u) \tilde \sigma(u) du -\int_0^t \sigma_{(1)}(u) \tilde \sigma(u) du
    \int_0^1 \sigma_{(1)}^2(u) du}{\Big(\int_0^1 \sigma_{(1)}^2(u) du\Big)^2}=:\bar h(t).
\end{align*}
The function in the numerator of $\bar h$ is not identically
equal to $0$. To see this, let us define
\[
f(t):= \int_0^t \sigma_{(1)}^2(u) du, \qquad g(t) = \int_0^t \sigma_{(1)}(u) \tilde \sigma(u) du.
\]
Notice that by assumption  $\tilde \sigma/\sigma_{(1)}$ is not constant. Now, the numerator being $0$ would mean that
$
f(t)g(1) = g(t) f(1),
$
which is equivalent to $g/f$ being constant.
But this means, that $g'/f'=\tilde \sigma/\sigma_{(1)}$ must be constant too, which contradicts our assumption. \\
This means, that
\[
\tilde h_N \to h:=\begin{cases}
x(1-\theta) \bar h \quad \textnormal{if} \quad x \le \theta,\\
\theta(1-x)\bar h \quad \textnormal{if} \quad x > \theta,
\end{cases}
\]
where $h$ is a nonzero function. Notice that by assumption
$a_NK \to \infty$, which implies $\sqrt{N/K}=o(a_N \sqrt{NK})$.
This fact, together with the convergence  $\tilde h_N \to h$
and  identity \refeq{def_hN},  implies that
\[
{h_N = a_N \sqrt{NK} h + o(a_N \sqrt{NK}).}
\]

Now, recalling the representation of $\widehat{S}^{(1)}$
in \refeq{SNdec}, the weak convergence \refeq{PNweak}
and the fact that $a_N \sqrt{NK} \to \infty$, we obtain
\begin{align} \label{e:S1k}
\frac{\widehat{S}^{(1)}}{a_N^2 NK} \convd \int_0^1 \|h(x)\|^2dx>0,
\end{align}
concluding our proof. \hfill \QED

\bigskip

\noindent{\sc Proof of Theorem \ref{t:th1}:} \
Recall the definition of the change point estimator
$\hat \theta^{(1)}$ in \refeq{theta1}. Setting
\[
M(n) := \Big\| \sum_{i=1}^n \frac{1}{N}\widehat{F}_i - \frac{n}{N^2} \sum_{i=1}^N \widehat{F}_i \Big\|^2
\]
and ${\hat n} := N \hat \theta^{(1)}$, we see that
\[
{\hat n} = \mathrm{argmax}_{n=1,...,N}[M(n)-M(n^*)].
\]
Here $n^* = \lfloor N \theta \rfloor$ and the above equality holds
because $M(n^*)$ is a constant. Let us define for ease of notation the terms
\begin{align*}
    E(n):=&\frac{1}{N}\sum_{i=1}^n \{\widehat{F}_i-\mathbb{E}
    [\widehat{F}_i]\}  - \frac{n}{N^2} \sum_{i=1}^N
    \{\widehat{F}_i-\mathbb{E}[\widehat{F}_i]\},  \\
    D(n):=&\frac{n}{N^2} \sum_{i=1}^N\mathbb{E}[\widehat{F}_i]
    -\frac{1}{N}\sum_{i=1}^n\mathbb{E}[\widehat{F}_i], \\
    \tilde D(n) := &\frac{n(n^*-N)}{N^2}  F_1+\frac{n(N-n^*)}{N^2} F_N.
\end{align*}
We now show that
\begin{align}\label{e:idRd}
&\lim_{M \to \infty}\limsup_N\mathbb{P}(\max_{n \le n^*-b_N, } M(n)-M(n^*) \ge 0) =0, \\
&\lim_{M \to \infty}\limsup_N\mathbb{P}(\max_{n > n^*+b_N, } M(n)-M(n^*)\ge 0) =0  \nonumber
\end{align}
where $b_N:= \max(M, M a_N^{-2}/K)$ and $M>0$. For simplicity,
we focus on the first identity.
Notice that we can rewrite the difference inside the maximum as
\begin{align} \label{e:diffR}
   & M(n)-M(n^*) =  \langle D(n)-D(n^*)-E(n)+E(n^*), D(n)+D(n^*)+E(n)+E(n^*)\rangle\\
    = & \{\|D(n)\|^2-\|D(n^*)\|^2\} + \{{-\|E(n)\|^2}+\|E(n^*)\|^2\}+\langle D(n)-D(n^*), E(n)+E(n^*)\rangle \nonumber\\&- \langle E(n)-E(n^*),D(n)+D(n^*)\rangle =: A_1(n)+A_2(n)+A_3(n)-A_4(n). \nonumber
\end{align}
It is  now enough to prove that
\[
\lim_{M \to \infty}\limsup_N\mathbb{P}(\max_{n \le n^*-b_N } A_1(n)/3+A_i(n) \ge 0)=0, \qquad \textnormal{for} \,\, i=2,3,4.
\]
To obtain these bounds,
we use the estimates from Lemma \ref{lem_cph}.
For $i=2,3$, we can simply show that $A_1(n)$ asymptotically
dominates $A_2(n), A_3(n)$.
More precisely, using Lemma \ref{lem_cph} parts a), b) shows that
\begin{align*}
    A_1(n)/3+A_2(n) \le -c a_N^{2}(n^*-n)/N +O_P(1/(NK)).
\end{align*}
Now, $|n-n^*|\ge b_N \ge M a_N^{-2}/K$, implies that
\[
c a_N^{2}(n^*-n)/N \ge c\frac{M}{NK},
\]
which goes to $\infty$ as $M$ does, whereas the part $O_P(1/(NK))$ is independent of $M$. \\
Similarly, we can use a) and c) of Lemma \ref{lem_cph}, to see that
\begin{align*}
   & A_1(n)/3+A_2(n) \le -c a_N^{2}(n^*-n)/N + c a_N(n-n^*)/N O_P(1/\sqrt{NK}) \\
    =& \frac{a_N(n^*-n)}{N}\big(-c a_N+O_P(1/\sqrt{NK})\big).
\end{align*}
Since $a_N$ is of larger order than $1/\sqrt{NK}$ by assumption, the part inside the brackets is negative, with probability converging to $1$, as $N \to \infty$.\\
Finally, we consider the sum $A_1(n)/3+A_4(n)$. Here a slightly more subtle argument is needed. We decompose the indices $n=1,...,n^*-b_N$ in blocks of size $b_N2^\ell$ for $\ell=1,2...$, with the the first block consisting of the indices $n^*-b_N, n^*-b_N-1,...,n^*-3b_N$, the second consisting of the indices $n^*-4b_N-1,...,n^*-7b_N$ and so on. We call these blocks $B_1,B_2,...$. Then, we observe that
\[
\max_{n \le n^*-b_N } A_1(n)/3+A_4(n) \ge \max_{\ell } \max_{n \in B_\ell} A_1(n)/3+A_4(n).
\]
Using estimate a) from Lemma \ref{lem_cph}, we have
\[
\max_{n \in B_\ell} A_1(n)/3 \ge -c a_N^2 (n^*-b_N-\ell b_N)/N.
\]
 We can then use the bound $d)$ from Lemma \ref{lem_cph} to see that
\begin{align*}
& \mathbb{P}(\max_{n \le n^*-b_N } A_1(n)/3+A_4(n)\ge 0)\\
\le & \sum_{\ell \ge 1} \mathbb{P}(\max_{n \in B_\ell} A_1(n)/3+A_4(n)\ge 0)\\
\le & \sum_{\ell \ge 1} \mathbb{P}( \max_{n \in B_\ell}A_4(n)\ge ca_N^2 b_N2^\ell/N)\\
\le &  c \sum_{\ell \ge 1} \frac{a_N^2  b_N 2^\ell/(N^2K)}{a_N^4b_N^22^{2\ell-2}/N^2}  \le   \sum_{\ell \ge 1} \frac{c }{M 2^{\ell-2}}
\end{align*}
The right side does not depend on $N$ anymore and converges to $0$,
 as  $M \to \infty$.
This completes the proof. \hfill \QED

\medskip

The following lemma was used in the proof of Theorem \ref{t:th1}.

\begin{lemma} \label{lem_cph}
Under the assumptions of Theorem \ref{t:loc1}
(and so of Theorem \ref{t:th1}),
    \begin{itemize}
        \item[a)] $A_1(n) <-c a_N^{2}(n^*-n)/N$.
        \item[b)] $\max_{n \le n^*} |A_2(n)| = O_P(1/(NK))$.
        \item[c)] $|A_3(n)| \le c a_N(n-n^*)/N O_P(1/\sqrt{NK})$.
        \item[d)] For any $\kappa =2,...,N$ it holds that $ \mathbb{E}\max_{|n-n^*|\le \kappa }|A_4(n)|^2 \le c a_N O_P(\sqrt{\kappa}/(N\sqrt{K}))$.
    \end{itemize}
    Here all Landau symbols and constants are independent of $M$.
\end{lemma}

\noindent{\sc Proof.}
We begin by considering $A_1(n)$. More precisely,
we investigate  first
$\tilde A_1(n):= \{\|\tilde D(n)\|^2-\|\tilde D(n^*)\|^2\} $.
By the definition of $\tilde D$,
    \begin{align} \label{e:A1}
        \tilde A_1(n) = \|F_N-F_1\|^2 \Big(\frac{n^2(N-n^*)^2-(n^*)^2(N-n^*)^2}{N^4}\Big).
    \end{align}
In the proof of Theorem \ref{t:loc1}, we have demonstrated
 that for a sufficiently small constant $c>0$ it holds that $\|F_N-F_1\|^2 >c a_N^2$. Next, we consider the second factor in \refeq{A1}. Recall that  $N-n^*\ge cN$ for some small enough $c>0$ and that $n^2-(n^*)^2=(n+n^*)(n-n^*)$, where again $n+n^* \ge n^* \ge cN$.
    Putting these results together yields
    \[
    \tilde A_1(n) <- \frac{c(n^*-n) a_N^2}{N}.
    \]
    Next, we consider the difference $|A_1(n)-\tilde A_1(n)|$, which can be expressed as
    \begin{align*}
        \big|\|F_N-F_1\|^2- \|\mathbb{E}\widehat{F}_N-\mathbb{E}\widehat{F}_1\|^2\big| \Big|\frac{n^2(N-n^*)^2-(n^*)^2(N-n^*)^2}{N^4}\Big|.
    \end{align*}
    By analogous arguments as before, the second factor is upper bounded by $c|n-n^*|/N $. The first factor can be expressed as
    \begin{align*}
        |2 \langle [\mathbb{E}\widehat{F}_N-\mathbb{E}\widehat{F}_1]-[F_N-F_1], F_N-F_1\rangle + \|[\mathbb{E}\widehat{F}_N-\mathbb{E}\widehat{F}_1]-[F_N-F_1]\|^2|,
    \end{align*}
    which follows by a version of the third binomial formula for Hilbert spaces. Recalling Lemma \ref{lemma_mean}, we observe that
    \[
    \|[\mathbb{E}\widehat{F}_N-\mathbb{E}\widehat{F}_1]-[F_N-F_1]\|^2 \le \frac{c}{K^2}
    \]
    and using additionally that $\|F_1-F_N\|\le c/a_N$
    \[
     |\langle [\mathbb{E}\widehat{F}_N-\mathbb{E}\widehat{F}_1]-[F_N-F_1], F_N-F_1\rangle| \le \frac{ca_N}{K}.
    \]
    Employing the assumption $a_N K \to \infty$, it follows that the term $(ca_N)/K$ dominates. Hence, we conclude, that
    \[
    |A_1(n)-\tilde A_1(n)| \le \frac{c (n^*-n)a_N}{NK}
    \]
    Since $a_N$ is asymptotically dominated by $1/K$, it follows that $\tilde A_1(n)$ dominates the remainder $|A_1(n)-\tilde A_1(n)|$, proving the rate a).

    To demonstrate b), it suffices to show the desired rate for $A_{2,1}(n):=\|E(n)\|^2$, since $\max_n\|E(n)\|^2 \ge \|E(n^*)\|^2 $. Next, we recall that by definition of $E(n)$ we can decompose
    \begin{align*}
    \|E(n)\|\le \frac{1}{N}\Big\|\sum_{i=1}^n \{\widehat{F}_i-\mathbb{E}[\widehat{F}_i]\}  \Big\|+\Big\| \frac{n}{N^2} \sum_{i=1}^N \{\widehat{F}_i-\mathbb{E}[\widehat{F}_i]\}\Big\| =: A_{2,1,1}(n)(n)+A_{2,1,2}(n).
    \end{align*}
    It suffices to show the rate $A_{2,1,1}(n), A_{2,1,2}(n) = O_P(1/\sqrt{NK})$ (uniformly in $n$) to get the desired result and we focus on the more difficult term $A_{2,1,1}(n)$ only. Notice that for all $i$ it holds that
    \[
    \mathbb{E} \|\widehat{F}_i-\mathbb{E}[\widehat{F}_i]\|^2 \le c/K
    \]
    (see Lemma \ref{lemma_mean}). Now, given the independence of the random variables across $i$ it holds for any two indices $1 \le n_1<n_2\le N$, that
    \begin{align}
    \label{e:moricz}
    \mathbb{E} \Big\|\sum_{i=n_1}^{n_2} \widehat{F}_i-\mathbb{E}[\widehat{F}_i]\Big\|^2 \le \frac{c(n_2-n_1)}{K}.
    \end{align}
    So, using Theorem 3.1 in \cite{moricz:serfling:stout:1982} implies that
    \[
    \mathbb{E}\big[\|\max_n A_{2,1,1}(n)\|^2\big] \le \frac{c}{KN},
    \]
    with a possibly larger constant $c$. Notice that the cited theorem was originally formulated for real valued random variables, but the proof carries over directly to random variables on a  Hilbert space. This concludes the proof of b)

    c) follows by similar techniques as before. We observe that
    \[
    A_3(n) \le \|D(n)-D(n^*)\| \|E(n)+E(n^*)\|.
    \]
    By the same bounds as in the last step, we conclude that $max_n \|E(n)+E(n^*)\| = O_P(1/\sqrt{NK})$. Turning to the first factor, we see that
    \begin{align}
    & \|D(n)-D(n^*)\| \le \|D(n)-D(n^*)-[\tilde D(n)-\tilde D(n^*)]\|
    +\|\tilde D(n)-\tilde D(n^*)\| \\
    =: & A_{3,1}(n)+ A_{3,2}(n). \nonumber
    \end{align}
    Using the analogous calculations as in a) shows that
    \begin{align*}
    A_{3,1}(n) = &\Big(\frac{(n^*-n)(N-n^*)}{N^2}\Big) \|F_1
    -F_N-[\mathbb{E}\widehat{F}_1-\mathbb{E}\widehat{F}_N]\| \\
    \le c \frac{n^*-n}{NK},
    \end{align*}
    where we have used Lemma \ref{lemma_mean}, to bound the second factor in the first line. Turning to $A_{3,2}(n)$, we observe (again with the same techniques  as in a)), that
    \[
    A_{3,2}(n) \le  \frac{ca_N(n^*-n)}{N}
.    \]
Putting these rates together and noticing that $a_N$ dominates $1/K$, we observe that
\[
A_3(n) \le \frac{ca_N(n^*-n)}{N}O_P\Big( \frac{1}{\sqrt{NK}}\Big),
\]
proving c).

Finally, we turn to d).  We can decompose $A_4(n)$ into
\[
A_4(n) \le  \|E(n)-E(n^*) \| \|D(n)+D(n^*)\| =:A_{4,1}(n)\cdot A_{4,2}(n).
\]
$A_{4,2}(n)$ is bounded by $\|D(n)\|+\|D(n^*)\|$. Again, using Lemma \ref{lemma_mean} we observe that
\[
\|D(n)\|+\|D(n^*)\|\le  \|\tilde D(n)\|+\|\tilde D(n^*)\| + \frac{c}{K}.
\]
Furthermore, by definition we have $\|\tilde D(n)\| \le \|F_1-F_N\| \le a_N$. Since $a_N$ dominates $1/K$ we have
\[
A_{4,2}(n) \le c a_N.
\]
Next, we turn to $A_{4,1}(n)$, which using the definition of $E$ can be upper bounded by
\[
\Big\|\frac{1}{N}\sum_{i=n}^{n^*} \{\widehat{F}_i-\mathbb{E}[\widehat{F}_i]\}\Big\|  + \Big\|\frac{n-n^*}{N^2} \sum_{i=1}^N \{\widehat{F}_i-\mathbb{E}[\widehat{F}_i]\} \Big\| =: A_{4,1,1}(n)+A_{4,1,2}(n).
\]
We confine our proof to the more difficult term $A_{4,1,1}(n)$. We employ Theorem 3.1 in \cite{moricz:serfling:stout:1982} (where the condition of the named theorem is satisfied due to \refeq{moricz}), which entails
\[
\mathbb{E}[\max_{\kappa \le n \le n^*}\|A_{4,1,1}(n)\|^2] \le \frac{\kappa}{N^2K}
\]
and hence that $\max_{\kappa \le n \le n^*} A_{4,1,1}(n)=O_P(\sqrt{\kappa}/(N\sqrt{K})). $ Combining the rates of $A_{4,1}$ and $A_{4,2}$ now yields the desired result d).\hfill \QED

\medskip

\noindent{\sc Proof of Theorems \ref{t:loc2} and \ref{t:th2}} The proofs of these theorems follow similar steps as those for Theorems \ref{t:loc1} and \ref{t:th1}. The details are, however, easier given that in this case we consider real-valued time series. We have therefore decided to omit a proof of these results to avoid redundancy.  \hfill \QED
\medskip

\noindent{\sc Proof of Proposition \ref{prop:th:fin}} We confine the proof to the more difficult case of the mixed alternative $\sigma_{(2)}=  (1+1/\sqrt{K})\sigma_{(1)}+\tilde \sigma/\sqrt{K}$, that violates both $H^{(1)}$ and $H^{(2)}$. We notice, we can rewrite
\[
\sigma_{(2)}=  \sigma_{(1)}+\big(\tilde \sigma+\sigma_{(1)}\big)/\sqrt{K}
\]
and by the proof of Theorem \ref{t:loc1} (see the convergence \refeq{S1k} in particular), we observe that
\[
\frac{\widehat{S}^{(1)}}{N} \overset{P}{\to} c_1>0,
\]
where $c_1>0$ is a constant depending on $\sigma_{(1)}$ and $\tilde \sigma$. By a similar line of argumentation, we can show that
\[
\frac{K \widehat{S}^{(2)}}{N} \overset{P}{\to} c_2>0,
\]
where  $c_2>0$ is a constant that depends on $\sigma_{(1)}$ and $\tilde \sigma$ as well. Now, notice that we can rewrite the following fraction of $p$-values in terms of the distribution functions
\begin{align} \label{e:ratio1}
\frac{ p^{(1)}}{ p^{(1)}+ p^{(2)}} = \frac{ 1-\Lambda^{(1)}[\widehat{S}^{(1)}]}{ (1-\Lambda^{(1)}[\widehat{S}^{(1)}])+ (1-\Lambda^{(2)}[\widehat{S}^{(2)}])}.
\end{align}
If we can prove for an arbitrarily small, but fixed $c_3>0$ that
\begin{align} \label{e:ratio2}
\frac{1-\Lambda^{(1)}[\widehat{S}^{(1)}]}{1-\Lambda^{(2)}[\widehat{S}^{(2)}]} = \mathcal{O}(\exp(-c_3 N)),
\end{align}
it follows that the ratio in \refeq{ratio1} is exponentially decaying and hence, by definition of $\hat \theta$ in \refeq{theta} that
\[
\hat \theta = \hat \theta^{(1)} + \mathcal{O}_P\bigg( \frac{1}{N}\bigg).
\]
This, together with the fact that
\[
 \hat \theta^{(1)} = \theta + \mathcal{O}_P\bigg( \frac{1}{N}\bigg),
\]
(Theorem \ref{t:th1}) would already imply the desired result.
Now, to prove \refeq{ratio2}, we notice that by the continuous mapping theorem
\begin{align*}
    \frac{1-\Lambda^{(1)}[\widehat{S}^{(1)}]}{1-\Lambda^{(2)}[\widehat{S}^{(2)}]} \frac{1-\Lambda^{(2)}[(N/K)c_2]}{1-\Lambda^{(1)}[Nc_1]}=\frac{1-\Lambda^{(1)}[N (\widehat{S}^{(1)}/N)]}{1-\Lambda^{(2)}[(N/K) (K\widehat{S}^{(2)}/N)]} \frac{1-\Lambda^{(2)}[(N/K)c_2]}{1-\Lambda^{(1)}[Nc_1]} \overset{P}{\to} 1.
\end{align*}
Consequently, it suffices to analyze the deterministic scaling ratio and show that
\[
\frac{1-\Lambda^{(1)}[Nc_1]}{1-\Lambda^{(2)}[(N/K)c_2]}  = \mathcal{O}(\exp(-c_3 N)),
\]
to prove \refeq{ratio2}. Since $K \to \infty$ as $N \to \infty$, this implies that for any arbitrarily small constant $c_L>0$ there exists an $N_0>0$ such that for all $N>N_0$ we have $c_2/K <c_1 c_L$. It follows by monotonicity of the denominator and Theorem \ref{thm:thp} that as $N \to \infty$
\[
\frac{1-\Lambda^{(1)}[Nc_1]}{1-\Lambda^{(2)}[(N/K)c_2]} \le \frac{1-\Lambda^{(1)}[Nc_1]}{1-\Lambda^{(2)}[N c_1 c_L]} \le \exp(-c_3N)
\]
for an adequate choice of $c_L$ and a sufficiently small constant $c_3>0$, which concludes the proof. \hfill \QED
%
%


\begin{theorem} \label{thm:thp}
    Under Assumption \ref{a:model}, there exist constants
    $0<c_L< c_U<\infty$, and $c_L'c_U'>0$ such that for sufficiently large $x$
\[
    \frac{1-\Lambda^{(1)}(x)}{1-\Lambda^{(2)}(c_L x)}\le \exp(-c_L' \,x), \qquad \frac{1-\Lambda^{(1)}(x)}{1-\Lambda^{(2)}(c_Ux)} \ge \exp(c_{U}'\,x).
\]
The constants in this Theorem
depend on the eigenvalues $\lambda_i$ in Theorem \ref{thm:S1}
 and the long--run variance $\lambda$ in Theorem \ref{thm:S2}.
\end{theorem}

\noindent{\sc Proof. }
We confine the proof to the second equation (the first one follows by the same reasoning).
Let us define $B:= \int_0^1 [\mathbb{B}(x)]^2 dx$ and let $B_1, B_2,...$ be i.i.d. copies of $B$.
Then
\[
S^{(1)} \overset{d}{=} \sum_{i=1}^\infty \lambda_i B_i, \qquad S^{(2)} \overset{d}{=} (4\lambda) B,
\]
where $\lambda_1 \ge \lambda_2 \ge ...$
are defined in Theorem \ref{thm:S1} and $\lambda$
in Theorem \ref{thm:S2}. For simplicity, we set $\lambda=1$.
According to eq. (49) in \cite{tolmatz:2002} we can write for large $x$
\begin{align} \label{e:exp:rate}
1-\Lambda^{(2)}(x) = \frac{c_1 \exp(-c_2  x) (1+O(1/x))}{\sqrt{x}} ,
\end{align}
for absolute constants $c_1, c_2$. Now, let us investigate $\Lambda^{(1)}$. Therefore, let us assume that there exists an $\eta>0$ such that $c_3:=\sum_{i \ge 1} \lambda_i i^{\eta} <\infty$ and set $c_L:=3c_3$. Now, it follows that
\begin{align*}
   & 1-\Lambda^{(1)}(c_Lx) = \mathbb{P}\Big(S^{(1)}>c_Lx\Big) = \mathbb{P}\Big(\sum_{i=1}^\infty \lambda_i B_i>(c_Lx/c_3)\sum_{i=1}^\infty \lambda_i i^{\eta}\Big)\\
   = & \mathbb{P}\Big(\sum_{i=1}^\infty \lambda_i (B_i-3xi^\eta)>0\Big) \le \sum_{i=1}^\infty \mathbb{P}\Big( \lambda_i B_i>3xi^\eta \lambda_i \Big)  \\
   =& \sum_{i=1}^\infty  \frac{c_1 \exp(-3c_2i^\eta x  ) (1+O(1/x))}{\sqrt{[3xi^\eta] }}\\
   =& \frac{\exp(-2c_2x) }{ \sqrt{x} }
   \sum_{i=1}^\infty  \frac{c_1 \exp(-c_2x [3i^\eta -2] ) (1+O(1/x))}{\sqrt{3i^\eta }} .
\end{align*}
The series on the right is of size $O(1)$ and the first factor decays
at a rate $\exp(-2c_2x)$, which comparing it to \refeq{exp:rate} yields the desired result.

Finally, we have to show that
the eigenvalues $(\lambda_i)_{i \in \mathbb{N}}$
have indeed the property that for some $\eta>0$
it holds that $\sum_{i \ge 1} \lambda_i i^{\eta} <\infty$.
This follows directly from the fact that the asymptotic
covariance kernel of $\widehat{F}_1$ is
Lipschitz continuous (defined in Theorem \ref{thm:cov:F};
the verification of the Lipschitz property is easy).
It thus satisfies the conditions of Theorem 3.2 in
\cite{griebel:li:2018} and hence there exists
a sufficiently large constant $c>0$ that $|\lambda_i| \le c i^{-2}$
\hfill \QED


\section{Implementation}\label{sec:imple}
In this section, we provide details on the implementation of our approach.
In the following, for a matrix $A \in \mathbb{R}^{L \times K}$,
we refer to the $\ell$--th row as $A[\ell,]$, to the
$k$--th column as $A[,k]$ and to the entry $A_{\ell,k}$ as $A[\ell, k]$.
When applying a function $f: \mathbb{R} \to \mathbb{R}$ to a vector
or matrix, we apply it entry-wise. The Euclidean norm and inner product
are denoted by $\|\cdot\|$ and $\langle \cdot, \cdot \rangle$, respectively.

We begin by transforming our raw data into estimates of the quadratic variation process, $\widehat{Q}$ (appearing in \eqref{e:Qhat})  in Algorithm \ref{alg:Q}.

\begin{algorithm}\small
\caption{Empirical Quadratic Variation  $\{\widehat{Q}_i({k/K}): i=1,\ldots,N, \,\, k=1,\ldots,K\} $}\label{alg:Q}
\begin{algorithmic}
\State \textbf{input:} $N$ (number of curves), $K$ (number of measurements), $R \in \mathbb{R}^{N\times (K+1)}$ (matrix of observations $\{R_i({k/K}): i=1,\ldots,N, \,\, k=0,\ldots,K\}$)
\State \textbf{output:} Quadratic variation  matrix $Q \in \mathbb{R}^{N\times K}$
\Function{Q-Functions}{$N$, $K$, $R$}
\State Define $SI \in \mathbb{R}^{N\times K}$ with $0$ entries 
\For{$l=1,\ldots,K$}

$SI[,l]:=\left(R[,l+1]-R[,l]\right)^2$
\EndFor
\State Define $Q \in \mathbb{R}^{N\times K}$ with
$Q[,1] = SI[,1]$, $0$ otherwise

\For{$l=2,\ldots,K$}
\State $Q[,l]:=Q[,l-1]+SI[,l]$
\EndFor \\
\Return $Q$
    \EndFunction
\end{algorithmic}
\end{algorithm}

\noindent Algorithm \ref{alg:F} below transforms the estimated quadratic variation into the normalized version $\widehat{F}_i$, appearing in \eqref{e:F}. This quantity is   used  as inputs of the test of hypothesis $H_0^{(1)}$.
\bigskip\bigskip\bigskip\bigskip

\begin{algorithm}\small
\caption{ $\{\widehat{F}_i({k/K}): i=1,\ldots,N, \,\, k=1,\ldots,K\} $ }\label{alg:F}
\begin{algorithmic}
\State \textbf{input:} $N$ (number of curves), $K$ (number of measurements), $Q \in \mathbb{R}^{N\times K}$ (output of Algorithm \ref{alg:Q})
\State \textbf{output:} The standardized quadratic variation matrix $F \in \mathbb{R}^{N\times K}$
\Function{F-Functions}{$N$, $K$,  $Q$}
\State Define $F \in \mathbb{R}^{N\times K}$ with $0$ entries
\For{$n=1,\ldots,N$}
\For{$k=1,\ldots,K$}

 \;\; \;\;\; $F[n,k]:=Q[n,k]/Q[n,K]$
\EndFor
\EndFor\\
\Return $F$
    \EndFunction
\end{algorithmic}
\end{algorithm}

\noindent Algorithms \ref{alg:S1}, \ref{alg:p1} and \ref{alg:t1}, respectively,  calculate the  CUSUM statistic $\widehat{S}^{(1)}$, the empirical $p$-value $\hat p^{(1)}$ and the estimator $\hat \theta^{(1)}$.  In the estimation of  $\hat p^{(1)}$ there exist two user-determined parameters: $r$, the number of simulations for the limiting distribution and $B$, a dimension reduction parameter that is used to truncate the infinite sum in the definition of $\widehat{S}^{(1)}$ (see \eqref{e:def:S1}). As defaults, we recommend using $r=1000$ and choosing $B$ large enough such that at least $95\%$ of the empirical variance is explained. More precisely, we pick $B$ as the smallest value s.t.\\[-1ex]
\[
\frac{\sum_{b=1}^B \hat \lambda_b}{Tr[\hat c_F]} \ge 95\%\\[-1ex]
\]
\noindent where $Tr$ refers to the trace of a covariance kernel $\hat c_F$, defined as $\int_0^1 \hat c_F(u, u) du$. The empirical eigenvalues and the estimator $\hat c_F$ are defined in Algorithm \ref{alg:p1}.

\begin{algorithm}\small
\caption{Test statistic $\widehat{S}^{(1)}$}\label{alg:S1}
\begin{algorithmic}
\State \textbf{input:} $N$ (number of curves), $K$ (number of measurements), $F \in \mathbb{R}^{N\times K}$ (output of Algorithm \ref{alg:F})
\State \textbf{output:} Statistic $\widehat{ S}^{(1)}$
\Function{F-CUSUM}{$N$, $K$, $F$}
\State Define $PS \in \mathbb{R}^{N\times K}$ with $PS[1,]:=F[1,]$ and otherwise $0$ entries
\For{$n=2,\ldots,N$}
\State $PS[n,]:=PS[n-1,]+F[n,]$
\EndFor
\State Define $S=0$
\For{$n=1,\ldots,N$}
\State $S=S+\|PS[n,]-(n/N)PS[N,]\|^2$
\EndFor
\State $\widehat{S}^{(1)}=S/N^2$\\
\Return $\widehat{S}^{(1)}$
    \EndFunction
\end{algorithmic}
\end{algorithm}

\begin{algorithm}\small
\caption{Empirical $p$-value  $\hat p^{(1)}$}\label{alg:p1}
\begin{algorithmic}
\State \textbf{input:} $N$ (number of curves), $K$ (number of measurements), $F \in \mathbb{R}^{N\times K}$ (output of Algorithm \ref{alg:F}), $B$ (dimension reduction parameter), $r$ (simulation number for $p$-values)
\State \textbf{output:} $\hat p^{(1)}$, empirical $p$-value for statistic $\widehat{ S}^{(1)}$
\Function{F-Cov}{$N$, $K$, $F$, $B$}
\State Define $C \in \mathbb{R}^{K\times K}$ with $0$-entries
\For{$n=2,\ldots,N$}
\State $C = C + (F[n,]-F[n-1,])\cdot (F[n,]-F[n-1,])^\top$.
\EndFor
\State Collect the $B$ largest eigenvalues of $\hat c_F:=C/(2NK)$ in the vector $v_\lambda:=(\hat \lambda_1,\ldots, \hat \lambda_B)$ \\
\Return $v_\lambda$
\EndFunction
\Function{Int-BB}{$B$, $N$}
\State Define $v_B \in \mathbb{R}^{B}$ with $0$-entries
\For{$rep=1,\ldots,B$}
    \State Define $v_{rep}=(B_{rep}(0),B_{rep}(1/N), B_{rep}(2/N),\ldots,B_{rep}(1))$ (where $B_1,\ldots, B_B$ are \\ $\qquad\quad \quad $ independent Brownian Bridges)
\State $v_B[rep] = \|v_{rep}\|^2/N$
\EndFor \\
\Return $v_B$
\EndFunction
\Function{pval-1}{$N$, $K$, $B$, $r$, $F$}
\State Define $v_r \in \mathbb{R}^{r}$ with $0$-entries
\State Calculate $v_\lambda = \textnormal{F-Cov}(N,K,F,  B)$
\For{$rep=1,\ldots,r$}
\State Generate fresh $v_{B}=$Int-BB$(B,N)$
  \State  $v_r[rep] = \langle v_\lambda, v_{B}\rangle$
\EndFor
\State Order $v_r$ with entries in decreasing order
\State Calculate $c =$F-CUSUM$(N,K,F)$
\State Determine $i^* = \textnormal{argmin}\{|c-v_r[i]|: i=1,\ldots,r\}$\State $\hat p^{(1)}:=i^*/r$\\
\Return  $\hat p^{(1)}$
\EndFunction
\end{algorithmic}
\end{algorithm}

\begin{algorithm}\small
\caption{Change point estimator $\hat \theta^{(1)}$}\label{alg:t1}
\begin{algorithmic}
\State \textbf{input:} $N$ (number of curves), $K$ (number of measurements), $F \in \mathbb{R}^{N \times K}$ (output of Algorithm \ref{alg:F})
\State \textbf{output:} Change point estimator $\hat \theta^{(1)}$
\Function{F-CP}{$N$, $K$, $F$}
\State Define $PS \in \mathbb{R}^{N\times K}$ with $PS[1,]:=F[1,]$ and otherwise $0$ entries
\For{$n=2,\ldots,N$}
\State $PS[n,]:=PS[n-1,]+F[n,]$
\EndFor
\State Define $v_N \in \mathbb{R}^{N}$ with $0$ entries
\For{$n=1,\ldots,N$}
\State Set $v_N[n]=\|PS[n,]-(n/N)PS[N,]\|^2$
\EndFor
\State  $n^*= \textnormal{argmax}\{v_N[n]:n=1,\ldots,N\}$
\State $\hat \theta^{(1)} = n^*/N$\\
\Return $\hat \theta^{(1)}$
    \EndFunction
\end{algorithmic}
\end{algorithm}

\noindent Next, we turn to the test of $H_0^{(2)}$, beginning with the calculation of the logarithmized total variation $\log(\widehat{Q}_i(1))$, appearing in \eqref{e:logQ(1)}, then calculating the CUSUM statistic $\widehat{S}^{(2)}$ in \eqref{e:S2}, and subsequently approximating the $p$-value. These are implemented in Algorithms \ref{alg:logQ}, \ref{alg:S2} and  \ref{alg:p2}. Algorithm \ref{alg:t2} entails the (potential) time change $\hat{\theta}^{(2)}$.
When approximating the $p$-value, we call the LAMBDA-function, that gives an estimate of the long-run variance $(4\lambda)$, where $\lambda$ is defined in   \eqref{e:long:cov:g}. There exist many preimplemented methods in statistical softwares to approximate the long--run variance and hence we do not select any specific method.

\begin{algorithm}\small
\caption{ 
$\{\log \widehat{Q}_i(1): i=1,\ldots,N\} $}\label{alg:logQ}
\begin{algorithmic}
\State \textbf{input:} $N$ (number of curves),  $Q \in \mathbb{R}^{N\times K}$ (output of Algorithm \ref{alg:Q})
\State \textbf{output:} Log total quadratic variation, $LTQ \in \mathbb{R}^{N}$
\Function{Log--TQV}{$N$,  $Q$}
\State Define $LTQ \in \mathbb{R}^{N}$ with
$LTQ  = \log Q[,K]$\\
\Return $LTQ$
    \EndFunction
\end{algorithmic}
\end{algorithm}

\begin{algorithm}\small
\caption{Test statistic $\widehat{S}^{(2)}$}\label{alg:S2}
\begin{algorithmic}
\State \textbf{input:} $N$ (number of curves),  $LTQ$ (output of Algorithm \ref{alg:logQ})
\State \textbf{output:} Statistic $\widehat{ S}^{(2)}$
\Function{logQ-CUSUM}{$N$,  $LTQ$}
\State Define $PS \in \mathbb{R}^{N}$ with $PS[1]=LTQ[1]$ and otherwise $0$ entries
\For{$n=2,\ldots,N$}
\State $PS[n]=PS[n-1]+LTQ[n]$
\EndFor
\State Define $S=0$
\For{$n=1,\ldots,N$}
\State $S=S+\left \vert PS[n]-(n/N)PS[N]\right \vert^2$
\EndFor
\State $\widehat{S}^{(2)}=S/N^2$\\
\Return $\widehat{S}^{(2)}$
    \EndFunction
\end{algorithmic}
\end{algorithm}

\begin{algorithm}\small
\caption{Empirical $p$-value  $\hat p^{(2)}$}\label{alg:p2}
\begin{algorithmic}
\State \textbf{input:} $N$ (number of curves),  $LTQ$ (output of Algorithm \ref{alg:logQ}), $r$ (simulation number for $p$-values)
\State \textbf{output:} $\hat p^{(2)}$, $p$-value for statistic $\widehat{ S}^{(2)}$






\Function{pval-2}{$N$,  $B$, $r$, $LTQ$}
\State Calculate $ \tilde \lambda = \textnormal{LAMBDA}(LTQ)$
\State Define $v_r \in \mathbb{R}^{r}$ with $0$-entries

\For{$rep=1,\ldots,r$}
\State Generate fresh $v_{B}:=$Int-BB$(1,N)$
  \State  $v_r[rep] =  \tilde \lambda \cdot v_{B}$
\EndFor
\State Order $v_r$ with entries in decreasing order
\State Calculate $c =$LOGQ-CUSUM$(N,K,F)$
\State Determine $i^* = \textnormal{argmin}\{|c-v_r[i]|: i=1,\ldots,r\}$
\State $\hat p^{(2)} = i^*/r$ \\
\Return $\hat p^{(2)}$
\EndFunction
\end{algorithmic}
\end{algorithm}

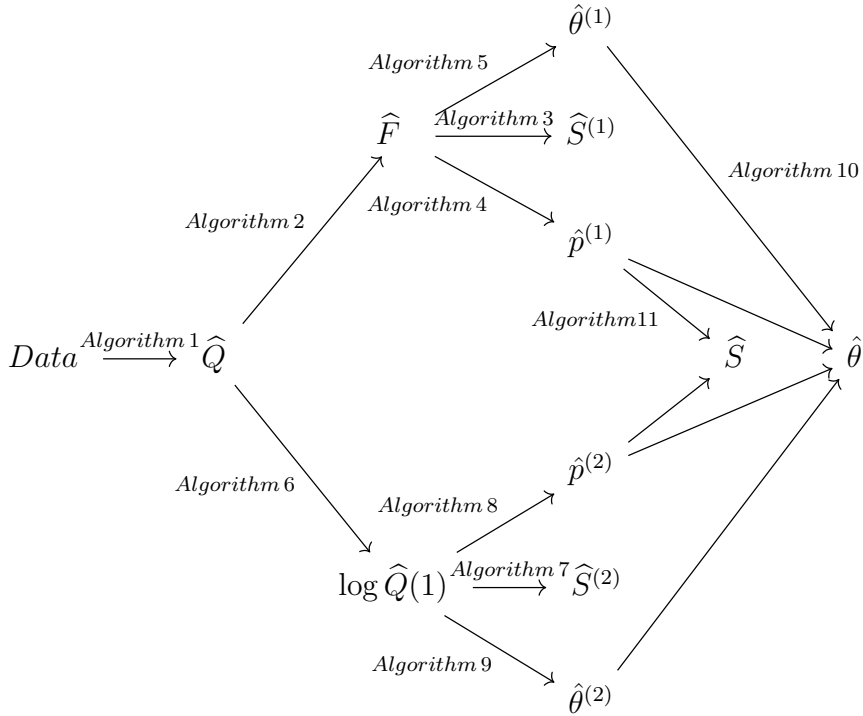
\begin{figure}
\begin{tikzcd}
&&&\hat \theta^{(1)}\ar[dddrr, "Algorithm \, \ref{alg:t}"] &&\\
&&\widehat{F} \,\,\,\,  \ar[r,"Algorithm  \,  \ref{alg:S1}"] \ar[ur,"Algorithm  \, \ref{alg:t1}"] \ar[dr,"Algorithm  \, \ref{alg:p1}"']& \widehat{S}^{(1)} &&\\
&&&\hat p^{(1)} \ar[dr, "Algorithm \ref{alg:S}"'] \ar[drr]&&
\\
\qquad Data\,\, \ar[r,"Algorithm  \, \ref{alg:Q}"] & \,\, \widehat{Q} \,\, \ar[uur,"Algorithm  \, \ref{alg:F}"]\ar[ddr,"Algorithm  \, \ref{alg:logQ}"']&& & \widehat{S} &\hat \theta  
\\
&&&\hat p^{(2)} \arrow{ur}  \arrow{urr}&&\\
&&\log \widehat{Q}(1)\,\, \ar[r,"Algorithm  \,  \ref{alg:S2}"] \ar[ur,"Algorithm  \, \ref{alg:p2}"] \ar[dr,"Algorithm  \, \ref{alg:t2}"']& \,\, \widehat{S}^{(2)} &&\\
&&&\hat \theta^{(2)}\ar[uuurr]&&
\end{tikzcd}
\caption{\textit{Flowchart of our inference procedure, with arrows indicating input relations.}}
\end{figure}

\begin{algorithm}\small
\caption{Change point estimator $\hat \theta^{(2)}$}\label{alg:t2}
\begin{algorithmic}
\State \textbf{input:} $N$ (number of curves),  $LTQ$ (output of Algorithm \ref{alg:logQ})
\State \textbf{output:} Change point estimator $\hat \theta^{(2)}$
\Function{logQ-CP}{$N$,  $LTQ$}
\State Define $PS \in \mathbb{R}^{N\times K}$ with $PS[1]=LTQ[1]$ and otherwise $0$ entries
\For{$n=2,\ldots,N$}
\State $PS[n]=PS[n-1]+LTQ[n]$
\EndFor
\State Define $v_N \in \mathbb{R}^{N}$ with $0$ entries
\For{$n=1,\ldots,N$}
\State Set $v_N[n]=|PS[n,]-(n/N)PS[N,]|^2$
\EndFor
\State  $n^*= \textnormal{argmax}\{v_N[n]:n=1,\ldots,N\}$
\State $\hat \theta^{(2)} = n^*/N$\\
\Return $\hat \theta^{(2)}$
    \EndFunction
\end{algorithmic}
\end{algorithm}

\newpage
\noindent Finally, in Algorithms \ref{alg:t} and \ref{alg:S}, we combine the $p$-values and change point estimators to the pooled change point estimator and the pooled test of $H_0$. As before, we call $q_{1-\alpha}$ the upper $\alpha$-quantile of the chi-squared distribution with four degrees of freedom.

\begin{algorithm}\small
\caption{Change point estimator $\hat \theta$}\label{alg:t}
\begin{algorithmic}
\State \textbf{input:} $\hat \theta^{(1)}$ (output of Algorithm \ref{alg:t1}),  $\hat \theta^{(2)}$ (output of Algorithm \ref{alg:t2}), $\hat p^{(1)}$ (output of Algorithm \ref{alg:p1}), $\hat p^{(2)}$ (output of Algorithm \ref{alg:p2})
\State \textbf{output:} Change point estimator $\hat \theta$
\Function{CP}{$\hat \theta^{(1)}$,$\hat \theta^{(2)}$,$\hat p^{(1)}$,$\hat p^{(2)}$}
\State Define $\hat\theta = \frac{ \hat p^{(1)}}{ \hat p^{(1)}+ \hat p^{(2)}} \hat \theta^{(2)}+\frac{ \hat p^{(2)}}{ \hat p^{(1)}+ \hat p^{(2)}} \hat \theta^{(1)}$

\Return $\hat \theta$
    \EndFunction
\end{algorithmic}
\end{algorithm}

\begin{algorithm}\small
\caption{The Statistic $\hat S$ and test decision}\label{alg:S}
\begin{algorithmic}
\State \textbf{input:}  $\hat p^{(1)}$ (output of Algorithm \ref{alg:p1}), $\hat p^{(2)}$ (output of Algorithm \ref{alg:p2}), $\alpha$ (nominal level)
\State \textbf{output:}  Statistic $\hat S$, and test decision
\Function{S}{$\hat p^{(1)}$,$\hat p^{(2)}$, $\alpha$}
\State $\widehat{S} = -2\{\log(\hat p^{(1)})+ \log(\hat p^{(2)})\}$
\If{$\widehat{S}>q_{1-\alpha}$}
\State Define $decision =1$
\Else{}
\State Define $decision =0$
\EndIf\\
\Return $(\widehat{S},decision)$
 \EndFunction
\end{algorithmic}
\end{algorithm}

\section{Details of the computation  of test statistics and critical
values in Section \ref{sec:fsp} }\label{s:dc}

\subsection{Additional information related to Section \ref{ss:es}}

\paragraph{Computation of  $\int_{0}^{t} \sigma (u) d W (u)$}
We can avoid the numerical integral method (e.g. Euler-Maruyama)
to calculate the integral $\int_{0}^{t} \sigma (u) d W (u)$.
According to Dambis-Dubins-Schwarz theorem, see e.g.
Section 5.3.2 in \cite{le2016brownian},
any continuous local martingale $M$ can be written as a ``time-changed'' Brownian motion. In particular, if $W$ is a Brownian motion and $\sigma$ is $W$-integrable then the result can be applied to $X(t) = \int_{0}^{t} \sigma (u) d W_i (u)$. This gives
\[
\forall t, \quad \int_{0}^{t} \sigma(u) d W(u)
= W\left( \int_{0}^{t} \sigma^2(u)du\right) \ \ \ a.s.
\]

Thus, generating from $X(t) = \int_{0}^{t} \sigma (u) d W (u)$ reduces to generating from Brownian motion. Additionally, Brownian motion admits independent and normally distributed increments. It is enough to generate independent normal variables (precise) and then sum them up (fast) to obtain discrete observations of Brownian paths. Specifically, the increments are in the form
$$
	 \int_{0}^{t} \sigma (u) d W (u) - \int_{0}^{s} \sigma (u) d W (u) = W\left( \int_{0}^{t} \sigma^2 (u) d u \right) - W\left( \int_{0}^{s} \sigma^2 (u) d u \right), \quad 0 \leq s < t <1,
$$
that are normally distributed with mean zero and variance $\int_{s}^{t} \sigma^2 (u) d u$, and non-overlapping increments are independent.\\

Based on our discretization on $t$, the increment can be further spelled out as
\begin{align*}
	X(t_k) - X\left( t_{k-1}\right)  &= W\left( \int_{0}^{t_k} \sigma^2 (u) d u \right) - W\left( \int_{0}^{ t_{k-1}} \sigma^2 (u) d u \right)\\
	& = W(G(t_k)) -  W(G(t_{k-1})), \quad k=1,...,K.
\end{align*}
Denote such increment as $d(t_k) = X\left( t_{k}\right) - X(t_{k-1})$. We can simulate $d(t_k)$ by independent normal variables, i.e. $d(t_k) \sim \mathcal{N}(0, G(t_k) - G(t_{k-1}))$. Therefore, the trajectory of $X(t_k)$ is the summation of those independent normal variables
$$
	X(t_k) = \sum_{s=1}^{k} d(t_k), \quad k=1,...,K.
$$

\paragraph{Computation of $\int_{0}^{1} B^2 (u) d u$}
Following \cite{horvath:kokoszka:rice:2014},
we use the expansion discussed in
\cite{shorack:wellner:1986}, pp 210--211,
to approximate the squared integral of Brownian bridge,
$$
	\int_{0}^{1} B^2 (u) d u
\approx \sum_{j=1}^{J} \frac{Z_j^2}{j^2 \pi^2},
$$
where $\left\lbrace Z_j\right\rbrace_{j=1}^{\infty} $
are i.i.d. standard normal random variables.
There is thus no need to simulate
the  trajectories and to perform numerical integration.
In our work, we used $J = 500$.

\paragraph{Simulation procedure for testing a shape change
 $H_0^{(1)}$}
\begin{enumerate}
	\item Simulate the data $\left\lbrace R_i(t_k), i = 1,...,N,  \ k = 0,...,K\right\rbrace $ based on the DGP.
	\item Calculate the realized quadratic variation processes
	$$
	\widehat{Q}_i (t) = \sum_{k=1}^{K} \left| R_i (t_k) - R_i (t_{k-1}) \right|^2 \mathbb{I}\left\lbrace t_k \leq t\right\rbrace, \qquad t \in[0,1], \qquad i = 1,2,...,N.
	$$
	\item Calculate the empirical standardized  quadratic variation
	$$
		\widehat{F}_i (t) = \frac{\widehat{Q}_i (t)}{\widehat{Q}_i (1)}, \qquad t \in[0,1], \qquad i = 1,2,...,N.
	$$
	\item Calculate the test statistic $\widehat{S}^{(1)}$
	$$
		\widehat{S}^{(1)} := \frac{K}{N^2}\sum_{n=1}^N \int_0^1\Big(\sum_{i=1}^n \widehat{F}_i(u) - \frac{n}{N} \sum_{i=1}^N \widehat{F}_i(u) \Big)^2 du.
	$$
	\item We use the following first-order difference estimator(FDE)\footnote{I think here we can divide by $2(N-1)$, rather than $2N$. This is because there are only $N-1$ of first-order difference observations. Based on a small-scale simulation (not reported), I notice that dividing by $2(N-1)$ gives slightly better size than dividing by $2N$.}:
		$$
		c_F (u, v) = \frac{1}{2(N-1)} \sum_{i=2}^{N} \left\lbrace F_i (u) - F_{i-1} (u)\right\rbrace \left\lbrace  F_i (v) - F_{i-1} (v)\right\rbrace.
		$$	
	\item Collect the first largest $B$ eigenvalues $\left\lbrace \lambda_1,...,\lambda_B \right\rbrace $ of $c_F (u, v)$ so that
	$$
		 \frac{\sum_{\ell=1}^{B}\lambda_\ell}{\sum_{\ell=1}^{K}\lambda_\ell} \geq 95\%.
	$$
	\item Approximate the limit distribution of $\widehat{S}^{(1)}$ by simulating the following quantity by $r=5000$ times
	$$
		\sum_{\ell=1}^{B} \lambda_\ell  \sum_{j=1}^{J} \frac{Z_j^2}{j^2 \pi^2},
	$$
	and use its empirical distribution  to obtain the $\hat{p}^{(1)}$.
	\item If reject  $H_0^{(1)}$, perform the change point estimator
	$$
		\hat \theta^{(1)} = \frac{1}{N}\underset{n\in \{1,\ldots,N\}}{\textnormal{argmax}}\int_0^1\Big(\sum_{i=1}^n \widehat{F}_i(u) - \frac{n}{N} \sum_{i=1}^N \widehat{F}_i(u) \Big)^2 du.
	$$
	\item Repeat Steps (1)--(8) for $M=5000$ times to obtain the empirical rejection rate.
\end{enumerate}

\paragraph{Simulation procedure for testing a change
in total volatility $H_0^{(2)}$}
\begin{enumerate}
	\item Use $\left\lbrace \widehat{Q}_i (t)\right\rbrace_{i=1}^N $
calculated above to obtain
$\left\lbrace\log \widehat{Q}_i (1)\right\rbrace_{i=1}^N$.
\item Calculate the test statistic $\widehat{S}^{(2)}$
	$$
		\widehat{S}^{(2)}:= \frac{1}{N^2} \sum_{n=1}^N \Big(\sum_{i=1}^n \log(\widehat{Q}_i(1)) - \frac{n}{N}\sum_{i=1}^N \log(\widehat{Q}_i(1))\Big)^2.
	$$
	\item Calculate the long-run variance of $\left\lbrace\log \widehat{Q}_i (1)\right\rbrace_{i=1}^N$ by
	$$
	\hat{\lambda} = \hat{\gamma} (0) + 2\sum_{h=1}^{N-1} K\left( \frac{h}{H} \right) \hat{\gamma} (h),
	$$
	where
	$$
	\hat{\gamma} (h) = \frac{1}{N} \sum_{i=1}^{N-h} \left( \log \widehat{Q}_i (1) - \frac{1}{N} \sum_{i=1}^{N} \log \widehat{Q}_i (1)\right) \left( \log \widehat{Q}_{i+h} (1) - \frac{1}{N} \sum_{i=1}^{N} \log \widehat{Q}_i (1)\right).
	$$
	\item Approximate the limit distribution of $\widehat{S}^{(2)}$ by simulating the following quantity by $r=5000$ times
	$$
		\hat{\lambda}  \sum_{j=1}^{J} \frac{Z_j^2}{j^2 \pi^2},
	$$
	and use its empirical distribution  to obtain the $\hat{p}^{(2)}$.
	\item If reject  $H_0^{(2)}$, perform the change point estimator
	$$
	\hat \theta^{(2)} =  \frac{1}{N}  \underset{n\in \{1,\ldots,N\}}{\textnormal{argmax}}\Big(\sum_{i=1}^n \log(\widehat{Q}_i(1)) - \frac{n}{N}\sum_{i=1}^N \log(\widehat{Q}_i(1))\Big)^2,
	$$
	\item Repeat Steps (1)--(5) for $M=5000$ times to obtain the empirical rejection rate.
\end{enumerate}

\paragraph{Simulation procedure for testing the global
null hypothesis $H_0$}
\begin{enumerate}
	\item Calculate the test statistic $\widehat{S}$ by
	$$
		\widehat{S} = -2\{\log(\hat{p}^{(1)})+ \log(\hat{p}^{(2)})\}.
	$$
	\item Use its limit distribution $\chi_4^2$ to obtain the $p$-value of the global test.
	\item If reject  $H_0$, perform the pooled change point estimator
	$$
		 \hat\theta = \frac{ \hat{p}^{(1)}}{ \hat{p}^{(1)}+ \hat{p}^{(2)}} \hat \theta^{(2)}+\frac{ \hat{p}^{(2)}}{ \hat{p}^{(1)}+ \hat{p}^{(2)}} \hat \theta^{(1)}.
	$$
	\item Repeat Steps (1)--(3) for $M=5000$ times to obtain the empirical rejection rate.
\end{enumerate}

\subsection{Distribution of change point estimators}\label{ss:est}
Now we validate the convergence of the change point estimators, $\hat{\theta}_1, \hat{\theta}_2, \hat{\theta}$, under different alternative hypotheses. Figure \ref{fig:est_HA1} provides the violin plot (with included boxplot) for $\hat{\theta}_1$ under $H_{A,1}$ with $\theta=0.5$. We can observe that $\hat{\theta}_1$ converges to the true $\theta=0.5$ as $N$ and $K$ increases. Figure \ref{fig:est_HA2} provides the violin plot (with included boxplot) for $\hat{\theta}_2$ under $H_{A,2}$ with $\theta=0.5$. As $K$ is not relevant for $H_{A,2}$, we only present $\hat{\theta}_2$ in terms of different $N$, and there is a converge in $\hat{\theta}_2$ to the true $\theta=0.5$. Figure \ref{fig:est_HA3} provides the violin plot (with included boxplot) for $\hat{\theta}$ under $H_{A,3}$ with $\theta=0.5$. We can observe that $\hat{\theta}$ is already very close to $\theta=0.5$, even under small $N$ and small $K$. The density is very concentrated around  $\theta=0.5$, and get more concentrated for larger $N$ small $K$.

\begin{figure}[H]
	\centering
	\includegraphics[width=0.45\linewidth]{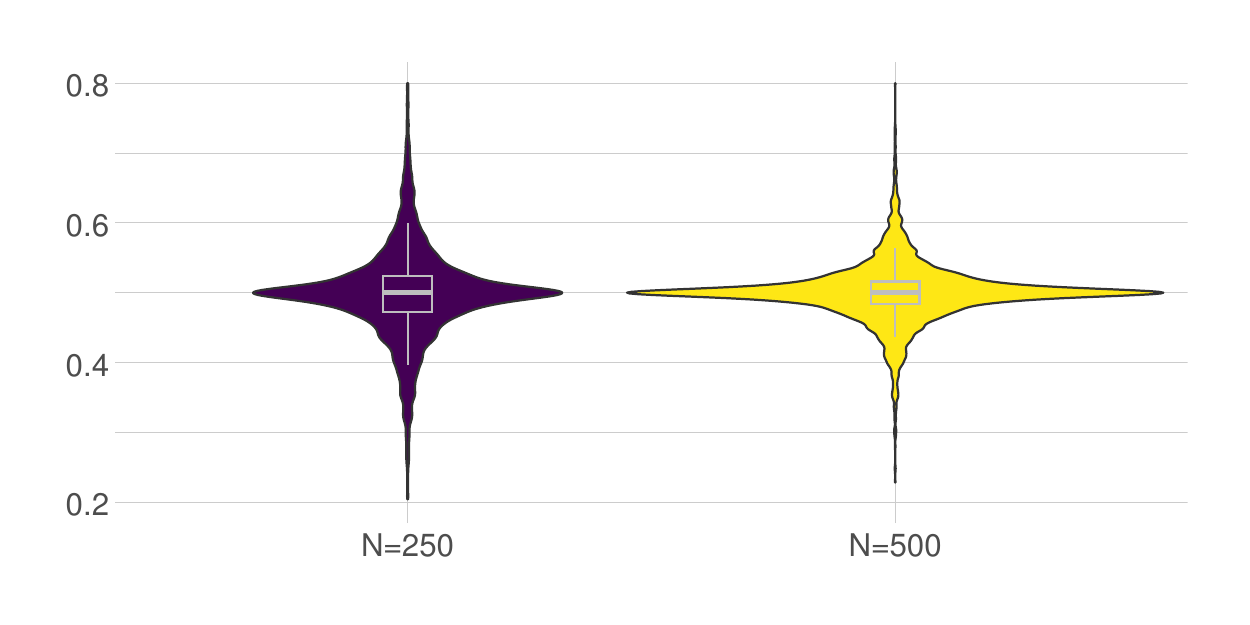} \\
	 \includegraphics[width=0.45\linewidth]{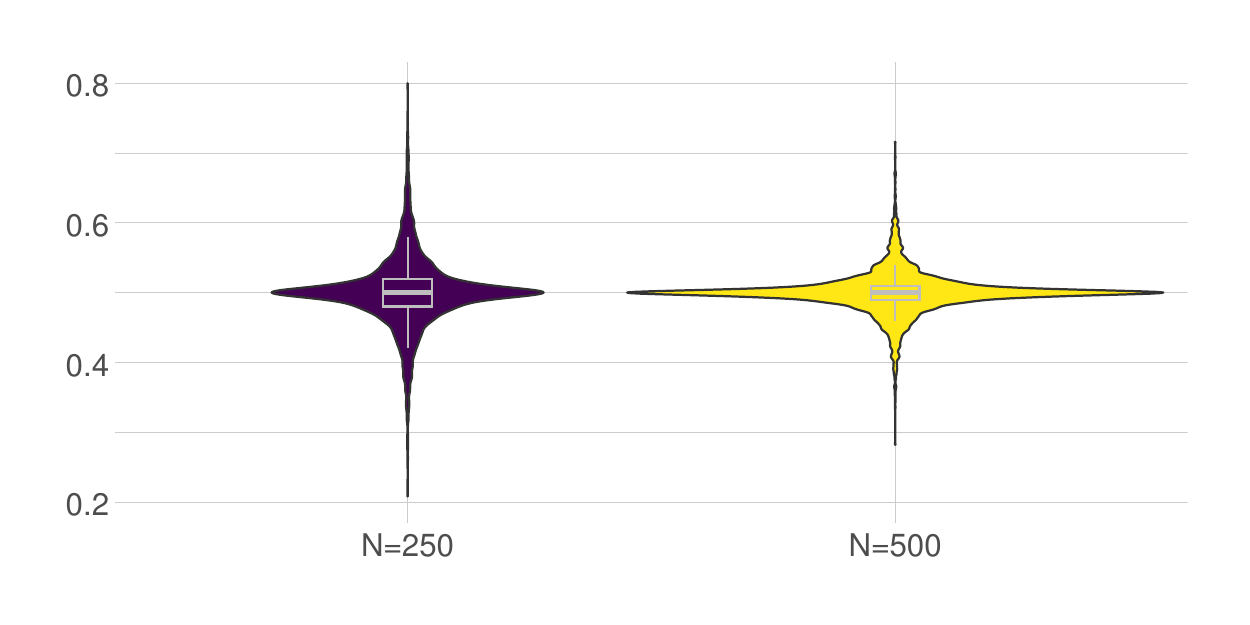}
	 \includegraphics[width=0.45\linewidth]{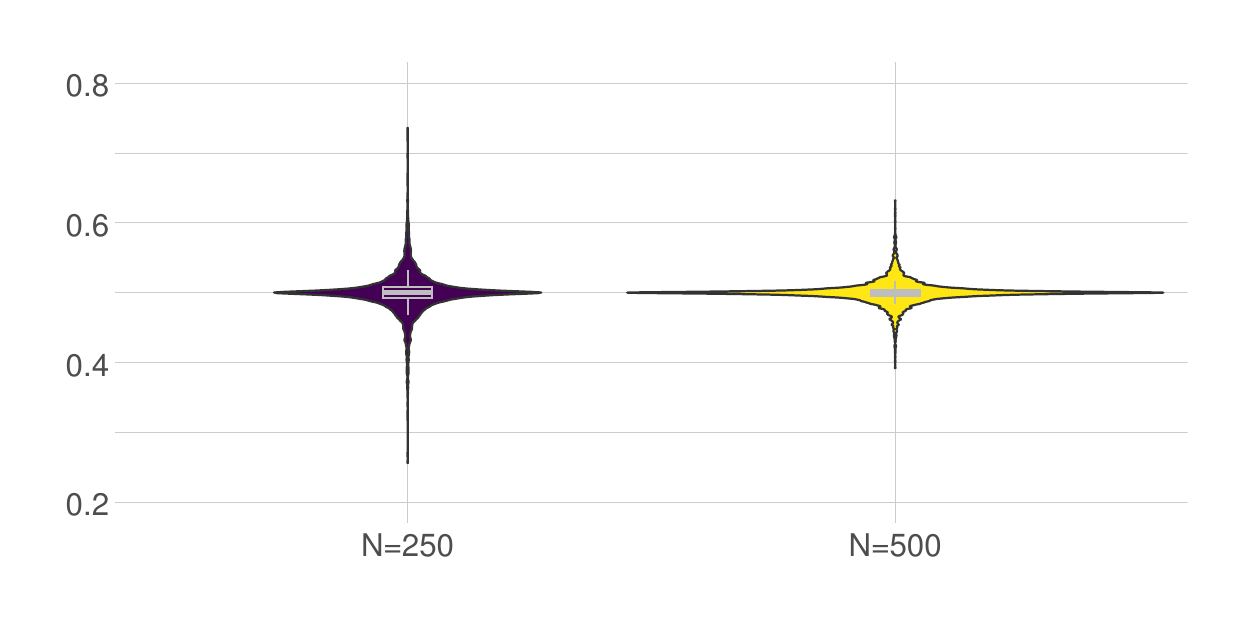}
	\caption{Distribution of $\hat{\theta}_1$ under $H_{A,1}$ with $\theta=0.5$. Top panel: $K= 26$; Bottom left panel: $K= 39$; Bottom right panel: $K= 78$. }
	\label{fig:est_HA1}
\end{figure}

\begin{figure}[H]
	\centering
	\includegraphics[width=0.7\linewidth]{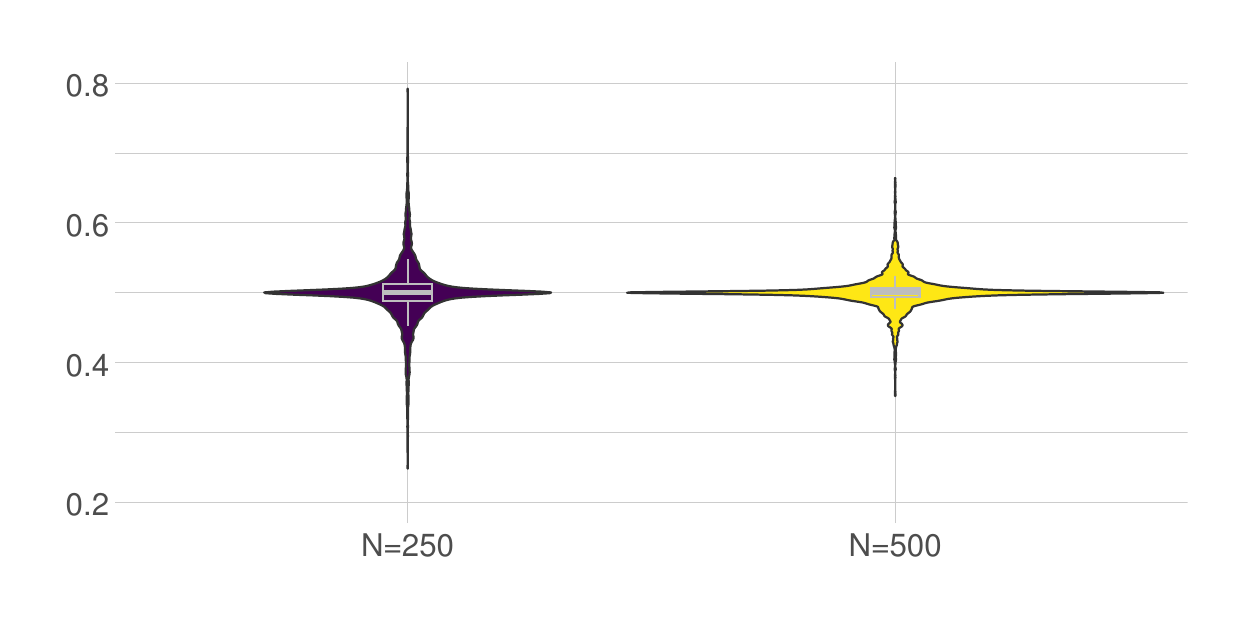}
	\caption{Distribution of $\hat{\theta}_2$ under $H_{A,2}$ with $\theta=0.5$.}
	\label{fig:est_HA2}
\end{figure}

\begin{figure}[H]
	\centering
	\includegraphics[width=0.45\linewidth]{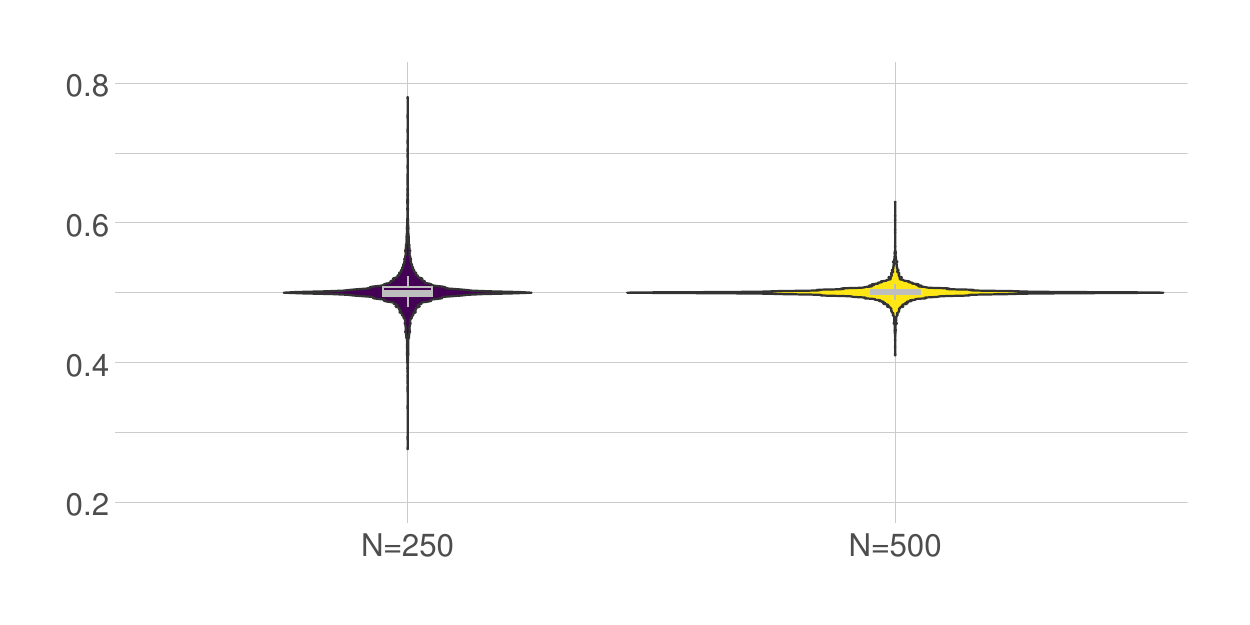} \\
	 \includegraphics[width=0.45\linewidth]{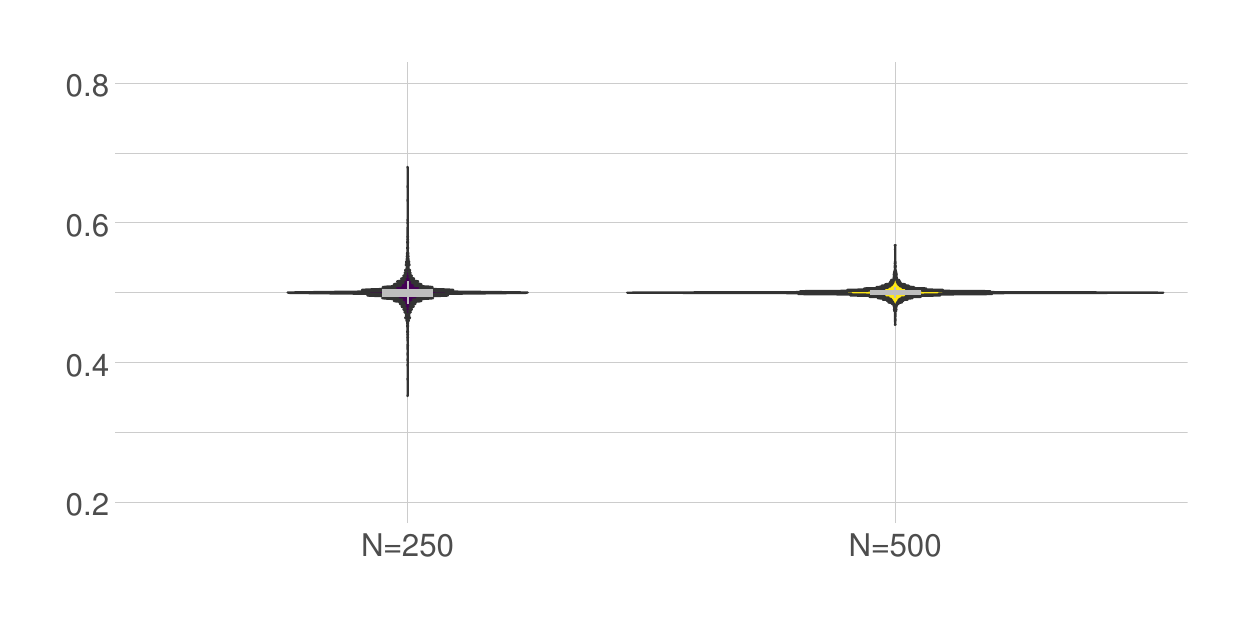}
	 \includegraphics[width=0.45\linewidth]{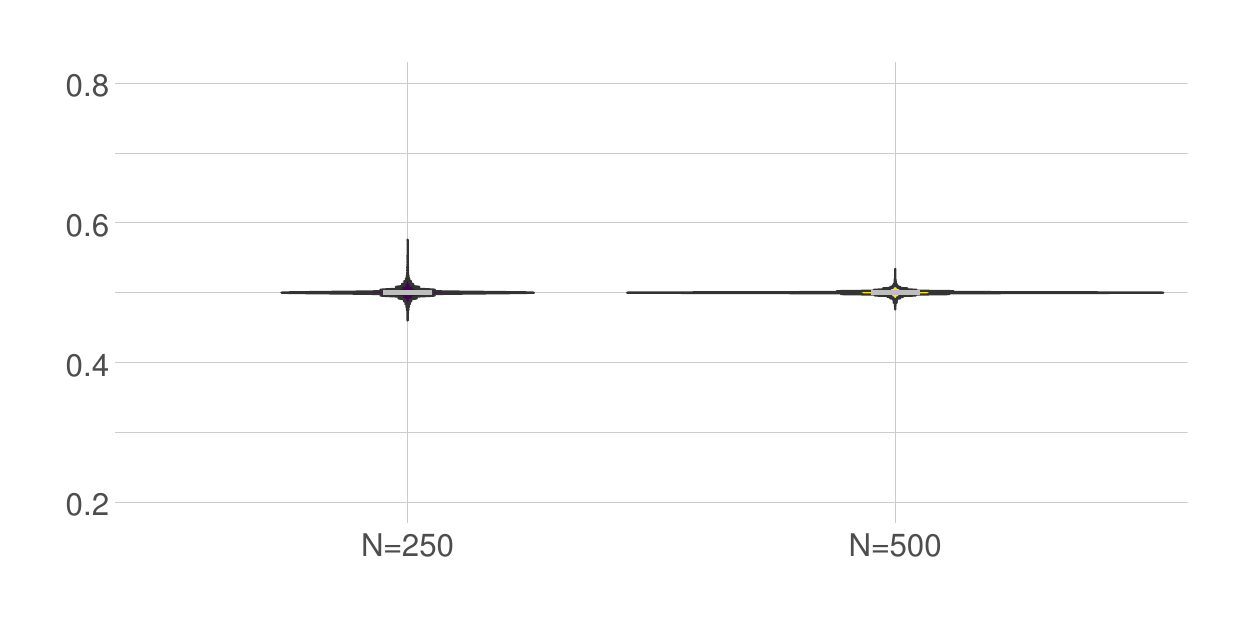}
	\caption{Distribution of $\hat{\theta}$ under $H_{A,3}$ with $\theta=0.5$. Top panel: $K= 26$; Bottom left panel: $K= 39$; Bottom right panel: $K= 78$. }
	\label{fig:est_HA3}
\end{figure}
\clearpage

\end{document}